\documentclass[11pt]{article}

\usepackage[english]{babel}
\usepackage[a4paper,margin=0.89in]{geometry}
\usepackage{amsmath,amssymb}
\usepackage{graphicx}
\usepackage[colorlinks=true, allcolors=blue]{hyperref}
\usepackage{wrapfig}
\usepackage[numbers,comma,sort&compress]{natbib}
\usepackage{titling}
\usepackage{floatrow}
\usepackage{enumitem}
\usepackage{float}

\usepackage{authblk}

\title{A Spin-Based Pathway to Testing the Quantum Nature of Gravity\vspace{0em}}

\begin{document}
\date{} 

\author[1]{Sougato Bose\thanks{s.bose@ucl.ac.uk}}
\author[2]{Anupam Mazumdar}
\author[3]{Roger Penrose}
\author[4,5]{Ivette Fuentes}
\author[6]{Marko Toro\v{s}}
\author[7]{Ron Folman}
\author[8]{Gerard J. Milburn}
\author[9]{Myungshik Kim}
\author[10,11]{Adrian Kent}
\author[12]{A. T. M. Anishur Rahman}
\author[13]{Cyril Laplane}
\author[14]{Aaron Markowitz}
\author[1,15]{Debarshi Das}
\author[4]{Ethan Campos-Méndez} 
\author[1]{Eva Kilian}
\author[7]{David Groswasser}
\author[7]{Menachem Givon}
\author[7]{Or Dobkowski}
\author[7]{Peter Skakunenko}
\author[7]{Maria Muretova}
\author[7]{Yonathan Japha}
\author[7]{Naor Levi}
\author[7]{Omer Feldman}
\author[10]{Dami\'an Pital\'ua-Garc\'ia}
\author[1]{Jonathan M.H. Gosling}
\author[16]{Ka-Di Zhu}
\author[17]{Marco Genovese}
\author[4]{Kia Romero-Hojjati}
\author[1]{Ryan J. Marshman}
\author[1]{Markus Rademacher}
\author[18]{Martine Schut}
\author[19]{Melanie Bautista-Cruz}
\author[2]{Qian Xiang}
\author[12]{Stuart M. Graham}
\author[12]{James E. March}
\author[12]{William J. Fairbairn}
\author[12]{Karishma S. Gokani}
\author[20]{Joseph Aziz}
\author[20]{Richard Howl}
\author[2]{Run Zhou}
\author[2]{Ryan Rizaldy}
\author[21]{Thiago Guerreiro}
\author[2]{Tian Zhou}
\author[22]{Jason Twamley}
\author[19]{Chiara Marletto}
\author[19]{Vlatko Vedral}
\author[1]{Jonathan Oppenheim}
\author[23]{Mauro Paternostro}
\author[4]{Hendrik Ulbricht}
\author[1]{Peter F. Barker}
\author[24]{Thomas P. Purdy}
\author[24]{M. V. Gurudev Dutt}
\author[25]{Andrew A. Geraci}
\author[14,26]{David C. Moore}
\author[12]{Gavin W. Morley\thanks{gavin.morley@warwick.ac.uk}}

\affil[1]{Department of Physics and Astronomy, University College London, Gower Street, WC1E 6BT London, UK}
\affil[2]{Van Swinderen Institute, University of Groningen, 9747 AG, The Netherlands}
\affil[3]{Mathematical Institute, Andrew Wiles Building, University of Oxford, Radcliffe Observatory Quarter, Woodstock Road, Oxford, OX2 6GG, UK}
\affil[4] {School of Physics and Astronomy, University of Southampton, Southampton SO17 1BJ, UK}
\affil[5] {Keble College, University of Oxford, Oxford OX1 3PG, UK}
\affil[6]{Faculty of Mathematics and Physics, University of Ljubljana, Jadranska 19, SI-1000 Ljubljana, Slovenia}
\affil[7]{Department of Physics, Ben-Gurion University of the Negev, Be’er Sheva 84105, Israel}
\affil[8]{Department of Physics and Astronomy, University of Sussex, Sussex House, Brighton, BN1 9RH UK, and National Quantum Computing Centre, Rutherford Appleton Laboratory, Harwell Campus, Oxfordshire OX11 0QX, UK}
\affil[9]{Korea Institute for Advanced Study, Seoul, Korea, and Blackett Laboratory, Imperial College London, London, UK}
\affil[10]{Centre for Quantum Information and Foundations, DAMTP, Centre for Mathematical Sciences, University of Cambridge, Wilberforce Road, Cambridge CB3 0WA, UK} \affil[11] {Perimeter Institute for Theoretical Physics, 31 Caroline Street North, Waterloo, ON N2L 2Y5, Canada}
\affil[12]{Department of Physics, University of Warwick, Gibbet Hill Road, Coventry CV4 7AL, UK}
\affil[13]{School of Physics, The University of Sydney, NSW 2006, Australia}
\affil[14]{Wright Laboratory, Department of Physics, Yale University, New Haven, Connecticut 06520, USA}
\affil[15]{Department of Physics, Shiv Nadar Institution of Eminence, Gautam Buddha Nagar, Uttar Pradesh 201314, India}
\affil[16]{Key Laboratory of Artificial Structures and Quantum Control (Ministry of Education), School of Physics and Astronomy, Shanghai Jiao Tong University, DongChuan Road, Shanghai 200240, China}
\affil[17]{INRIM, strada delle cacce 91, 10135 Turin Italy}
\affil[18]{Centre for Quantum Technologies, National University of Singapore, 3 Science Drive 2, 117543, Singapore}
\affil[19]{Clarendon Laboratory, University of Oxford, Parks Road, Oxford OX1 3PU, UK}
\affil[20]{Department of Physics, Royal Holloway, University of London, Egham, Surrey, TW20 0EX, UK}
\affil[21]{Department of Physics, Pontifical Catholic University of Rio de Janeiro, Rio de Janeiro 22451-900, Brazil}
\affil[22] {Quantum Machines Unit, Okinawa Institute of Science and Technology Graduate University, Onna, Okinawa 904-0495, Japan}
\affil[23] {Quantum Theory Group, Dipartimento di Fisica e Chimica Emilio Segrè, Università degli Studi di Palermo, via Archirafi 36, I-90123 Palermo, Italy, and Centre for Quantum Materials and Technologies, School of Mathematics and Physics, Queen's University Belfast, Belfast, BT7 1NN, UK}
\affil[24] {Department of Physics and Astronomy, University of Pittsburgh, Pittsburgh, Pennsylvania 15260, USA}
\affil[25]{Center for Fundamental Physics, Department of Physics and Astronomy, Northwestern University, Evanston, Illinois 60208, USA and Center for Interdisciplinary Exploration and Research in Astrophysics (CIERA), Northwestern University, Evanston, Illinois 60208, USA}
\affil[26]{Yale Quantum Institute, Yale University, New Haven, Connecticut 06520, USA}

\setcounter{Maxaffil}{0}
\renewcommand\Affilfont{\itshape\small}

\begin{titlepage}
\clearpage

\maketitle

\thispagestyle{empty}

\vspace{-3em}

\begin{abstract}
A key open problem in physics is the correct way to combine gravity (described by general relativity) with everything else (described by quantum mechanics). This problem suggests that general relativity and possibly also quantum mechanics need fundamental corrections. Most physicists expect that gravity should be quantum in character, but gravity is fundamentally different to the other forces because it alone is described by spacetime geometry. Experiments are needed to test whether gravity, and hence space-time, is quantum or classical. We propose an experiment to test the quantum nature of gravity by checking whether gravity can entangle two micron-sized crystals. A pathway to this is to create macroscopic quantum superpositions of each crystal first using embedded spins and Stern-Gerlach forces. These crystals could be nanodiamonds containing nitrogen-vacancy (NV) centres. The spins can subsequently be measured to witness the gravitationally generated entanglement. This is based on extensive theoretical feasibility studies and experimental progress in quantum technology. The eventual experiment will require a medium-sized consortium with excellent suppression of decoherence including vibrations and gravitational noise. In this white paper, we review the progress and plans towards realizing this. While implementing these plans, we will further explore the most macroscopic superpositions that are possible, which will test theories that predict a limit to this.
\end{abstract}
\end{titlepage}
\pagebreak

\section{Executive Summary}

Experiments with levitated nanoparticles could lead to a test of the quantum nature of gravity. Recent quantum-technology developments bring this within reach. To advance this we should prioritize: 
\begin{itemize}
  
\item
Building instruments to create a spatial superposition of a nanoparticle. 
\item 
Fabrication of high-purity crystals with embedded qubits, such as micron-sized diamonds containing a single nitrogen-vacancy (NV) centre in the middle. 
\item 
Reducing the electromagnetic and other background forces so that the mutual gravitational interaction between the two nanoparticles dominates.
\end{itemize} 

\section{Scientific context}

General relativity and quantum mechanics reign supreme in their domains, but we do not yet know the correct way to combine them in a way that describes our universe. Although a century of theoretical work has gone into solving this fundamental challenge, experiments are needed to guide the way. We aim to establish the key technologies needed to realize a laboratory experiment that will answer the core question: ``is gravity classical or quantum?"

Protocols to test the quantum nature of spacetime have recently been proposed \cite{Bose:2017nin,ICTS, marletto2017gravitationally,bose2025massive,hanif2024testing,krisnanda2020observable,howl2021non,haine2021searching,dec_Vs_diff,lami2024testing,kryhin2025distinguishable,carney2019tabletop, Belenchia:2018szb,Guerreiro2020,parikh2021signatures,tobar2024detecting,vermeulen2025photon, Vermeulen2021, Marletto2024, Genovese2024, MarlettoVedralRMP, MarlettoVedralPRD2020, MarlettoVedralPRD2018, Miki2025, Mari2025, Hogan2012, Hogan2017, Holometer2016, Pradyumna2020, Pikovski2012, Howl2023}. The test in Refs.  \cite{Bose:2017nin,ICTS, marletto2017gravitationally,bose2025massive} is based on the {\em necessity} for gravity to be quantum for it to entangle masses (assuming that gravity is not an action at a distance)~\cite{Bose:2017nin,Marshman:2019sne,Bose:2022uxe}. Here we refer to this as QGEM (Quantum-Gravity-Induced Entanglement of Masses). About 60 years ago, Feynman colloquially suggested~\cite{DeWitt:1957obj} that to describe a process where a mass in a quantum superposition moves another mass, one required gravity to be quantum. However, it was not clear exactly ``what" to measure, ``why" that would evidence the quantum character of gravity and ``how" to make an effective experiment. QGEM has recently provided answers to the above questions inspired by experiments on increasingly massive objects in the quantum regime \cite{delic2020cooling,tebbenjohanns2020motional}, precision gravity measurements \cite{westphal2021measurement,fuchs2024measuring}, matter wave interferomentry \cite{Folman2019_GM,SGI_experiment} and highly coherent spins in nanocrystals \cite{March23_GM}.
 
\begin{figure}[t]
\floatbox[{\capbeside\thisfloatsetup{capbesideposition={right,top},capbesidewidth=10cm}}]{figure}[\FBwidth]
{\caption{Schematic of two spatial quantum superpositions interacting via gravitational and EM interactions. The spatial superposition distance is denoted by $\Delta x$, while the two matter-wave interferometers are separated by distance $d$. The duration of the experiment, $\tau$, should be similar to the coherence time. The entanglement witness is read by spin measurements after the matter-wave interferometry, using Bell-type measurements on the spins.}\label{fig:1}}
{\includegraphics[width=6cm]{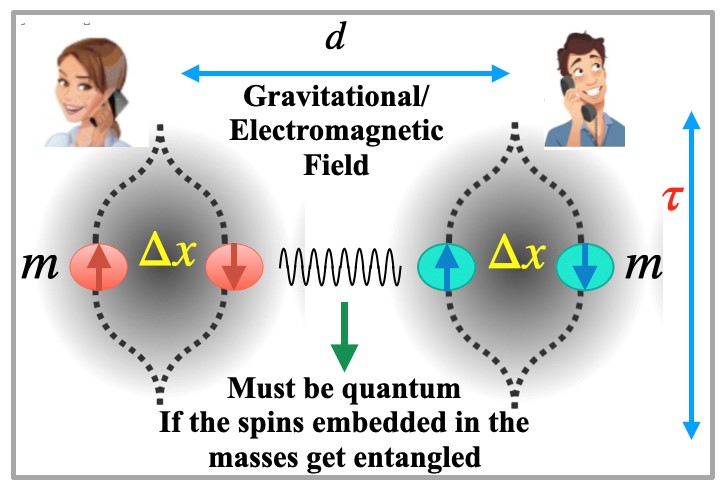}}
\end{figure}

These techniques could allow the preparation of two 1-$\mu$m-sized masses, each in a spatial superpositon, and enable measurement of whether they become entangled through their gravitational interaction, as shown in Fig.~\ref{fig:1}. If gravity can be in a quantum superposition, the two quantum systems will become entangled~\cite{Marshman:2019sne,Bose:2022uxe,christodoulou2019possibility}.
Entanglement is a quantum entity that cannot be mimicked in a classical world. When the combined state of two quantum systems cannot be written as a product of their individual states, they are entangled. Conversely, if gravity were classical and acted as a mediator between masses \cite{Bose:2017nin,Marshman:2019sne}, it cannot entangle two masses as two initially unentangled quantum systems cannot become entangled via local operations and classical communication (LOCC) alone~\cite{Bennett_1996}. Indeed, any theory in which gravity is classical and in which the quantum interactions are local or screened, cannot generate entanglement~\cite{Oppenheim23}. Thus, observing the correlations evidencing entanglement between masses is a key fundamental test for quantum natured gravity. If gravity is quantum, it would be possible to entangle only through the weak gravitational forces accessible in the lab \cite{Bose:2017nin}, where general relativity can be approximated by the Newtonian potential. Even at low energies, when we are prone to the lowest order contribution in the gravitational potential, such as the Newtonian potential, the QGEM protocol suggests~\cite{Marshman:2019sne,Bose:2022uxe, Carney_2019,Belenchia:2018szb,Danielson:2021egj,Christodoulou:2022vte,Rufo:2024ulr} that we would require the quantum nature of gravity to entangle the two masses, despite the fact that there is no $\hslash$ contribution at this level.

To get a levitated microdiamond into a spatial superposition, we aim to use the Stern-Gerlach effect on the spin of a nitrogen-vacancy (NV) centre inside the diamond~\cite{Scala13_GM, Duan2013_GM}. We will put the NV spin into a spin superposition, and then an inhomogeneous magnetic field will provide a superposition of forces on the diamond. To demonstrate matter-wave interferometry will then require that we do a closed-loop Stern-Gerlach experiment~\cite{Scala13_GM, Duan2013_GM, Folman2019_GM, SGI_experiment}. Single atoms in Bose-Einstein condensates have been used to demonstrate matter-wave interferometry~\cite{Kasevich2015_GM}, including experiments using the Stern-Gerlach effect~\cite{Folman2019_GM}. Quantum superposition has been demonstrated with molecules made up of 2000 atoms ($4\times10^{-23}$~kg), with a superposition distance of 266~nm~\cite{arndt}. The superposition of single atoms ($10^{-25}$~kg) has been demonstrated with a superposition distance that is close to one meter~\cite{Kasevich2015_GM}. Our diamonds are made up of around $10^{12}$ atoms with a mass of around $10^{-14}$~kg. The gravitational coupling has been measured between two 1-mm gold spheres with mass below $10^{-4}$~kg~\cite{westphal2021measurement}. Gravitational entanglement could be tested without using the Stern-Gerlach effect, such as by using the free expansion of quantum wavepackets with or without inverted potentials~\cite{StateExpansion2024, StateExpansion2021, KieselPNAS, OriolDoubleWell2024, TalbotLau2014, DattaMiao2021, Plenio2021}. A superposition of rotational states~\cite{KuhnRotation2017, SticklerRotors2022} is not likely to provide enough superposition distance to test for gravitational entanglement. 

The electromagnetic (EM) interaction could entangle the two masses also, and minimizing this unwanted piece of known physics is a key challenge that we will solve experimentally. Micron-sized particles have been levitated in vacuum with zero net charge previously~\cite{Moore14_mcp}, but the electric multipole effects such as electric dipoles may have to be reduced as well. If the two objects did possess net charges, the leading-order contribution would be the Coulomb interaction mediated by the Standard-Model photon, but with neutral objects the dipole-dipole induced interaction, e.g. Casimir's potential mediated by two photons, could also entangle the two masses~\cite{vandeKamp:2020rqh}. Minimizing this EM-induced entanglement will be a key part of the developments needed. The magnetic spin-spin interaction between the two NV spins will be weak enough that it can be neglected.  

NV spins have been controlled inside nanodiamonds that have been levitated both optically~\cite{VamivakasNV, TongcangLiNV2016} and in ion traps~\cite{BensonNV2014,DelordNV2018, DelordNature2020,TongcangLiNV2024, Conangla2018}. However, optical trapping pumps heat into the diamond~\cite{RahmanBurning2016,Frangeskou18_GM} which may prevent having the diamond at an internal temperature of $1$~K to avoid blackbody radiation causing decoherence. Ion traps will not trap particles with no net charge.  

The implications of QGEM for combining general relativity with quantum theory are substantial: QGEM will test the common prediction of a broad set of theories of quantum gravity at extremely low energies, i.e., that gravity produces an entangling interaction between masses. While it is still hotly debated how we should quantize gravity or treat spacetime microscopically, in the low-energy limit where laboratory experiments will take place, the predictions of these theories are equivalent to a quantum-natured mediator of the gravitational interaction: massless quantized spin-2 fluctuations around flat spacetime convey interactions between masses. However, without any experiment, there is no empirical evidence for this fundamental picture of gravity and, in fact, there are stochastic semi-classical theories in which sources (masses) are quantum, while the gravitational interaction between them would not be entangling~\cite{diosi1998coupling, kafri2014classical,tilloy2016sourcing,oppenheim2022constraints,OppenheimPRX_GM}. No existing experiment can distinguish the above theories from one in which gravity is truly quantum in character, while QGEM will~\cite{Marshman:2019sne,Bose:2022uxe,Carney_2019,Belenchia:2018szb,Danielson:2021egj,Christodoulou:2022vte,Rufo:2024ulr, Christodoulou2023}. Setups similar to QGEM will also test other independent aspects of gravity as a quantum entity such as its behaviour when measured \cite{hanif2024testing}.

There are theories in which a fundamental mechanism would cause the wavefunction of a macroscopic superposition to collapse within some short time~\cite{GRW86, Diosi1987, diosi1998coupling, Penrose96, Karolyhazy66, kafri2014classical, bassireview, MacroscopicRMP2018}. Even without positing additional collapse dynamics, any theory in which gravity is fundamentally classical necessarily causes wavefunction collapse~\cite{dec_Vs_diff, OppenheimPRX_GM}, since any classical–quantum coupling acts as an entanglement-breaking channel. This collapse rate serves as a bound for the experimental proposal of \cite{dec_Vs_diff} based on a {\it decoherence-diffusion trade-off}.

Penrose and Di\`{o}si have argued that gravity should render quantum superpositions unstable. Penrose’s reasoning is based on fundamental conflicts between quantum theory and general relativity. From this, he derives an energy uncertainty relation that sets a timescale for the reduction of a spatial superposition of a massive object to a definite state of one location or the other. This relation predicts that superpositions involving masses of $5.7 \times 10^{-16}$~kg would collapse within one second \cite{howl2019exploring}. Independently, Di\`{o}si arrived at essentially the same timescale through classical arguments. He developed a stochastic model, sometimes referred to as the ``Di\`{o}si-Penrose model” which predicts a spontaneous heating effect, basically ruled out in a recent experiment~\cite{Donadi2021}. Penrose’s actual proposal~\cite{Penrose2022} does not predict such heating, and its implications are currently being explored. If gravity does collapse the wave function, demonstrating a specific quantum effect of the gravitational field, this is very relevant to the entanglement issues of this paper, and will require even more demanding conditions. In any case, preparing massive quantum superpositions remains an essential requirement for understanding the interplay between quantum theory and gravity, this being a primary aspect of the current project.

\section{Readiness and Expected Challenges}
Like any ambitious experiment, QGEM presents significant technical challenges. Among them:

\noindent {\textbf{(C1) Initial state preparation}}: Preparing an initial quantum state of the nanoparticle about which we have nearly full knowledge (a “pure” state) is critical. For creating the spatial superposition and for the convenience of evidencing the entanglement, there should be a spin qubit (shown by arrows in Fig.~\ref{fig:1}) embedded in this crystal as well. This pure state requires a high degree of translational and rotational~\cite{KuhnRotation2017,Zhou2025,Rizaldy2024} cooling and/or intricate measurements. The initial amplitude of the motion of the nanoparticles will need to be very low. It has been demonstrated that the translational motion can be reduced to the quantum ground state for optically trapped silica nanoparticles~\cite{delic2020cooling, MagriniCooling2021, TebbenjohannsCooling2021, Kamba2022, Piotrowski2023, Dania2025}.  

\noindent
{\textbf{(C2) Creating the superposition}}: Generating a quantum superposition of a nanoparticle in two places at once is key to this protocol. For the configuration shown in Fig. \ref{fig:1}, the superposition distance, $\Delta x$, between the two components of the nanoparticle's trajectory is required to be $\sim 100~{\rm \mu m}$, for a nanoparticle with mass of order $10^{-15}-10^{-14}$~kg. To achieve this, we propose to put an embedded electron spin in the nanoparticle into a quantum superposition of spin states. Applying an inhomogeneous magnetic field then leads to Stern-Gerlach forces that create a spatially separated superposition by accelerating $|\uparrow\rangle$ to the left and $|\downarrow\rangle$ to the right. Applying spin flips at appropriate times can then complete an interferometer sequence, as shown in Fig.~\ref{fig:WanDrop}. These spins will also be used to witness the entanglement. Alternative protocols such as wavefunction expansion followed by an optical grating to create multi-component superpositions are also possible. The successful use of a screen for electromagnetic (EM) backgrounds may lessen the superposition size requirement dramatically as discussed in Refs.  \cite{vandeKamp:2020rqh,Schut:2021svd,Schut:2023hsy,Schut:2025blz}.

\noindent
{\textbf{(C3) Coherence}}: We need to maintain the superpositions for long enough to observe entanglement, requiring coherence times of order 1~second for both the spin and the matter wave. This requires isolation of the extremely fragile quantum superpositions from coupling to their environment: a process called decoherence that destroys the states needed for QGEM. While rapid progress has been made in the past decades in creating superpositions of ever-larger objects~\cite{arndt}, QGEM will require extreme isolation from vibrations and gas collisions as well as thermal, electromagnetic, and gravitational noise~\cite{Toros:2020dbf, Schut2025, Moorthy2025, ORI11_GM, Nair2023}.  

\noindent
{\textbf{(C4) Two adjacent gravitationally interacting Schr\"odinger cat states}}: Two such matter-wave interferometers must be created next to each other separated by a distance $d$ as shown in Fig.~\ref{fig:1}, and the EM interactions between the particles must be minimized so their mutual gravitational attraction dominates. For this, we will need to characterize and greatly reduce other sources of entanglement generation which come from EM interactions, such as the Coulomb and Casimir interactions. The value of $d$ may then need to be around 400~$\mu$m but this depends on several things including the protocol used and the screening that is possible. 

\noindent
{\textbf{(C5) Complete quantum interference and read out}}: Maintaining the necessary interference contrast requires reducing noise-induced variations between the two spatially separated trajectories in the superposition (dotted lines in Fig.~\ref{fig:1}). The final measurement will read out the single electron spins in each nanoparticle to calculate an entanglement witness, similarly to a Bell-inequality violation \cite{BellReviewRMP2014}. It would be valuable, but more difficult, to do this while closing the standard loopholes in Bell-violation tests~\cite{Kent2021}. Recent realizations of the Stern-Gerlach interferometer with atoms have shown that spin is a good observable~\cite{Folman2019_GM}. 

A number of technical risks remain, which must ultimately be addressed to realize QGEM. These include achieving the required vacuum pressures, particle temperatures, and vibration isolation needed to realize the full experiment. The experimental apparatus must also be housed at a site with sufficiently low gravity gradient noise to avoid coupling to the environment, which appears feasible at an underground location~\cite{Toros:2020dbf}. The best current work on reducing the dangling bonds on the surface of diamond~\cite{Rosskopf14_GM} measures 0.01 electron spins per nm$^2$ which suggests that there will be over 10,000 unpaired spins on the surface of the nanodiamond~\cite{Pedernales20_GM}. This would cause a large Stern-Gerlach force which is different for each run of the interferometry experiment because the dangling bond spins will randomize between each run, leading to noise and loss of contrast. A 10~T magnet could be used to prepolarize these dangling bonds at 1~K using Boltzmann thermalization before dropping the diamond out of this high magnetic field for the matter-wave interferometry protocol~\cite{WoodPRA22_GM}. While the above issues represent formidable engineering challenges in their own right, they appear to be within the capabilities of state-of-the-art systems.



\section{Objectives}

\begin{wrapfigure}{R}{0.40\textwidth}
    \begin{center}
 \vspace{-10mm}
    \includegraphics[width=0.9\textwidth]{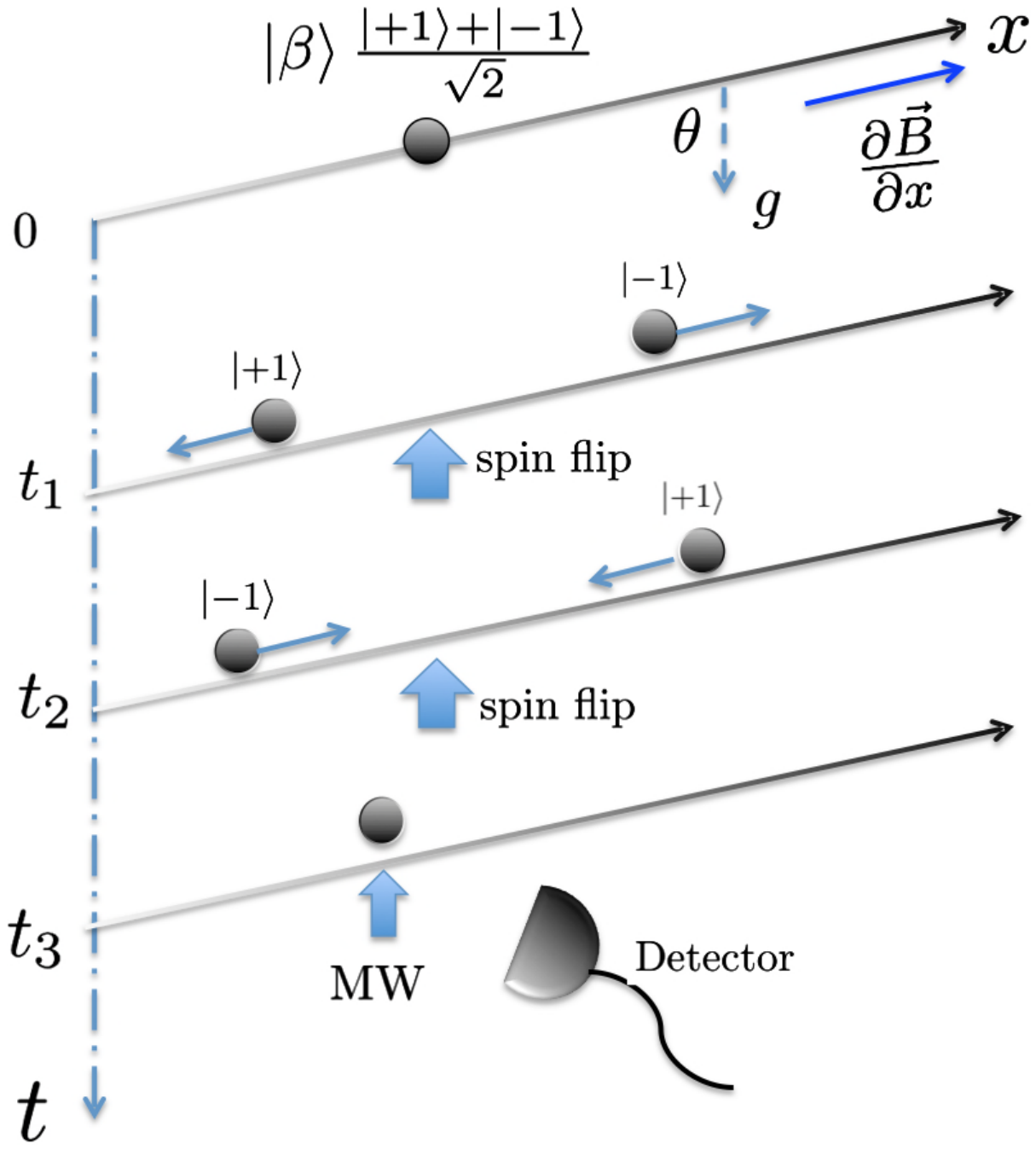}
    \caption{Schematic of the protocol from Ref.~\cite{Wan16_GM} for creating a spatial superposition using the Stern-Gerlach effect with a single spin.
    \label{fig:WanDrop}}
    \end{center}
    \vspace{-10mm}
\end{wrapfigure}

Work to date has provided roadmaps to realize the experiment in Fig.~\ref{fig:1}, overcoming the challenges C1-C5~\cite{Bose:2017nin, WoodPRA22_GM,Pedernales20_GM,vandeKamp:2020rqh, Zhou:2022epb, Zhou:2022frl, Zhou:2022jug,Zhou:2024voj, Braccini:2024fey,SGI_experiment,Henkel:2021wmj,Japha:2022xyg,Zhou2025,Rizaldy2024, Genovese2024}. The next step is a series of pathfinder experiments to optimize the experimental implementation for a future QGEM experiment. A key milestone will be the first nanoparticle spatial superposition, which could be achieved by following our protocol~\cite{Wan16_GM} shown in Fig.~\ref{fig:WanDrop}. For this, we will optically polarize a single nitrogen-vacancy (NV) spin and then put it into a spin superposition using a $\pi/2$ pulse. A strong inhomogeneous magnetic field of 10$^4$~T/m~\cite{WoodPRA22_GM, DUrso16_GM} or more would be applied. The Stern-Gerlach effect will then provide a spatial superposition, before further spin flips will bring the two superposition components back together for interferometry. Stern-Gerlach interferometry has very recently proved to be exceptionally agile in enabling complex configurations leading to new discoveries, such as the observation of the quantum equivalence principle~\cite{QequivalenceFolman}. Atom-chips~\cite{AtomChipReview2016} could be used (as in Ref. \cite{QequivalenceFolman}) as they can allow the position of the diamond and the magnetic gradients to be controlled with extremely high precision, including the realization of complex diamagnetic traps and guides. 

As is clear from the description of the challenges above, a collaboration between researchers with complementary expertise is needed to develop the techniques needed to implement QGEM. The relevant areas include low-noise optical levitation of nanoparticles in a vacuum, and applications of these techniques to measurements of gravity and other fundamental interactions. While optical traps work with pure nanodiamonds at an internal temperature of 300~K~\cite{Frangeskou18_GM}, these traps will increase the internal temperature of the nanodiamonds too much to be compatible with the 1~K internal temperature that will ultimately be required. Diamagnetic traps avoid this heating and the photon scattering present in optical traps, and have been demonstrated with microdiamonds~\cite{DUrso16_GM, Twamley19_GM}. Achieving the Stern-Gerlach interferometry with good spin coherence requires expertise in the manipulation and readout of single NV centers, including spin dynamic decoupling. Reaching sufficiently large superposition distances will require free flight for a period of around 1~s because the trapping potential works against the Stern-Gerlach forces~\cite{Wan16_GM, WoodPRA22_GM}. Screening, gravity sensing, and close-range force measurement methods are also of crucial importance here. 

{
\parfillskip=0pt
\parskip=0pt
\par}
 
Improved theoretical proposals could make the experimental challenges easier, including consideration of gravity gradient noise and random acceleration noise. Futher theory work is also needed to model the experiments and direct them, particularly toward optimizing the integrated experiment. 


\section{Methodology}

\subsection{Electrostatic background forces}
A key technical challenge to realizing the ambitious experiments described above is to ensure the nanoparticles interact only through their mutual gravitational attraction, and no other forces. This challenge is extreme-- the intrinsic strength of gravitational interactions is nearly 40~orders of magnitude smaller than EM interactions. However, since gravitational interactions are always attractive while electromagnetic effects can be canceled in macroscopic objects by precisely balancing positive and negative electric charges, it is possible to enhance the relative strength of gravity compared to other forces by working with the largest objects possible, as long as they maintain sufficient isolation from their environment. This tradeoff leads to the most viable proposals for such tests occurring with nanoscale objects of the type envisioned here: smaller systems such as atoms cannot interact with a large enough force from gravity compared to stray EM effects, while larger systems cannot be sufficiently isolated from the environment and maintained in the required quantum states.

A significant component of the work required in the short term aims to address the dominant expected EM backgrounds, with the goal of proving the feasibility of experiments where the only significant interactions between nanoparticles are through gravity. Electrical neutralization of optically trapped particles has been demonstrated in vacuum with single $e^-$ precision~\cite{Moore14_mcp}, ensuring that such objects have exactly the same number of protons and electrons and that the leading-order EM interactions of the particle with its environment or other particles are canceled. However, once this monopole charge is zero, higher-order multipole moments in the electric charge distribution provide the dominant backgrounds. Measurements indicate that 10~$\mu$m-sized silica particles typically have permanent electric dipole moments $p \sim 10^{-2}\ e$~cm~\cite{Afek21_dipoles,Rider16_chameleons}, which scale approximately with the volume of the particle~\cite{Afek21_dipoles} indicating that 10~pg particles may have dipole moments $\sim 10^{-4}\ e$~cm. If not mitigated, the expected dipole-dipole interaction energy between two 2-micron-sized diamond particles separated by $\sim$100~$\mu$m would then be:
\begin{equation}
    \frac{U_{dipole}}{U_{gravity}} \approx 3\times 10^6 \left(\frac{\mathrm{10\ pg}}{m} \right)^2 \left(\frac{100\ \mu\mathrm{m}}{d} \right)^2 \left(\frac{p}{10^{-4}\ e\ \mathrm{cm}} \right)^2.
\end{equation}
These estimates, while uncertain, indicate that we expect to need to reduce this interaction by a factor $\gg 10^6$ to ensure it is subdominant to gravitational interactions between the particles. Here we focus on electric dipole moments, rather than magnetic dipole-dipole interactions between the particles, since the latter are estimated to be negligible (much smaller than gravitational interactions) for the relevant experimental parameters.

An initial goal will be to use the same methods to characterize the intrinsic dipole moments of nanodiamonds suitable for QGEM. Characterizing these forces will further constrain the required nanoparticle masses and spatial superposition sizes, as well as the optimal separation of the interferometers. 

Once these dipole moments are measured, techniques to demonstrate the required mitigation will be developed. Five primary methods are envisioned to mitigate these interactions:
\begin{itemize}[noitemsep,topsep=0pt]
\item {\em Distance:} Since the dipole-dipole interaction energy scales as $d^{-3}$ for separation $d$, while the gravitational potential between the particles scales as $d^{-1}$, larger separations minimize the effect of these backgrounds, provided sufficient gravitational interaction can be maintained in achievable decoherence times~\cite{Fragolino:2023agd,Schut:2023tce}.
\item {\em Averaging:} While the objects themselves are expected to have intrinsic charge disorder that may not be possible to eliminate sufficiently with materials engineering alone, control of the rotational degrees of freedom allows averaging away of higher-order multipole moments in the charge distribution, as in Ref. ~\cite{Afek21_dipoles}.
\item {\em In-situ annealing:} 
We will measure the nanodiamond electric multipole moments in our traps before and after various treatments such as UV excitation, in-situ thermal annealing, and hydrogen atom passivation. These treatments could reduce the multipole moments in a reproducible way, or in a random way. If they have a random effect, then we can use a `repeat-until-success' protocol. 
\item {\em Shielding:} A conductive shield can be placed between the particles to attenuate EM forces while not shielding gravitational interactions. However, such a surface will typically deviate from a true, perfectly conductive equipotential, and avoiding decoherence due to coupling of the electric dipole moments in the spheres to ``patch potentials'' on the shield itself must be avoided.
\item {\em Materials development:} The nanodiamonds will need to be made from diamond with very low concentrations of crystal imperfections. It is reasonable to aim for zero dislocations in each 1-micron diamond and <100 nitrogen atoms. 
\end{itemize}
Given the technical challenge to eliminate these background forces, it is likely that a combination of all five methods will be required to reach the needed suppression, and the proposed work aims to characterize and substantially advance the maturity of all these options. Diamond fabrication is discussed more below as it is also important for the spin coherence times. 

While this work should be performed in parallel to development of an EM shield, it is essential to optimize the experiment and assess the viability of a shield compared to reduction of EM backgrounds only through materials engineering, annealing, rotation and separation. In particular, any electric multipole moments present in the particles will induce particle-shield interactions, even if the particle-particle interactions are fully screened. These interactions can arise, e.g., due to the presence of patch potentials on the surfaces of even high-quality conductors in practical implementations to date~\cite{Garrett20_patches}. However, even for a perfectly ideal conductive shield, the induced charge on the shield due to a nearby dipole (i.e., an image dipole) would lead to a force between the particle and shield of a similar order of magnitude to the dipole-dipole interaction. Since this interaction could lead to decoherence of the required state, it is possible that the shield cannot by itself be used to mitigate electric dipole interactions, and techniques to reduce the effective multipole moments in the spheres at a similar level are required regardless of the presence of the shield. Motional dynamic decoupling~\cite{Pedernales20_GM} could partially overcome the decoherence from image dipoles due to a uniform shield.

\subsection{Test of Casimir screen}
\label{sec:wp2}

It will be important to investigate the interactions between a trapped nanodiamond and a conducting screen. This includes the temperature dependence of the Casimir effect and patch effects with the surface as a function of the distance the nanodiamond is trapped from the surface. Calibrated force sensing has been demonstrated at the zeptonewton scale ($10^{-21}$ N) with nanoparticles trapped in an optical standing wave trap \cite{Ranjit:2016} as well as the ability to trap a nanoparticle at micron-range distances from a conducting surface and perform scanning force sensing in the plane perpendcular to the surface while the particle remains trapped in a standing wave formed by retroreflecting the optical tweezer beam from the surface \cite{montoya2022}.   

The Casimir force between a dielectric sphere of radius a, and a metal plane with distance z from the centre of the sphere can be written using the proximity force approximation (PFA) as \cite{casimirpolder} $F_{\rm{c}}=- \eta \frac{\pi^3
a \hslash c} {360 (z-a)^3}$, where $c$ is the speed of light in vacuum, in the limit that $(z-a) \ll a$.   The
prefactor $\eta$ characterizes the reduction in the force compared
with that between two perfect conductors \cite{lambrecht}.  For $z
\gg a$, the force takes the Casimir-Polder \cite{casimirpolder} form
$F_{\rm{cp}}=-\frac{3\hslash c \alpha_V}{8\pi^2\epsilon_0}
\frac{1}{z^5}$, where $\alpha_V=3\epsilon_0 V
\frac{\epsilon-1}{\epsilon+2}$ is the electric polarizability, $\epsilon$ is the relative permittivity, $\epsilon_0$ is the permittivity of free space and $V$ is the volume of the sphere. 

The gradient of the Casimir force near the static mirror surface produces a fractional shift in the resonance
frequency of the sphere given by $|\delta \omega_0 / \omega_0| =
\frac{|\partial{F_{\rm{c}}}/\partial{z}|}{2k}$ where $k$ is the spring constant. A similar frequency
shift has been measured due to the Casimir-Polder force on an atomic
sample trapped near a surface \cite{harber}. 
The minimum detectable frequency shift due to thermal
noise is given by $|\delta \omega_0 /
\omega_0|_{\rm{min}}=\sqrt{\frac{k_BT_{\rm{CM}}
b}{k\omega_0 Q z_{\rm{rms}}^2}}$ in a measurement bandwidth $b$ for a particle with an effective center-of-mass temperature $T_{\rm{CM}}$ and quality factor $Q$, driven with amplitude $z_{\rm{rms}}$ where $k_B$ is the Boltzmann constant.  Other sources of systematic fractional
frequency shifts near the surface, for example from diffuse scattered light on
the gold surface, should be experimentally characterized. We expect the Casimir force to be detectable using a nanoparticle of diameter $170$~nm at distances z below approximately $10$~microns from the surface, depending on the sensitivity achieved in the cryogenic vacuum~\cite{Elahi:2024dbb}.

Feasibility studies have led to constraints on both the masses and the sizes of the superposition, including the effects of the Casimir screening on the diamonds of the two interferometers proposed for QGEM~\cite{vandeKamp:2020rqh,Elahi:2024dbb,Schut:2023eux,Schut:2023hsy}. For example, this work showed that with an ambient temperature of $1$~K and a Young's modulus of $E=137$~GPa, a copper shield that is well clamped will not decohere a superposition of size $\sim20$~${\rm \mu m}$ for mass $10^{-15}$~kg. There will be more constraints from an actual experiment, including the patch potentials that will modify these numbers for the blueprint of the QGEM experiment.

A laser grating could be used to study the effect of the grating on the trajectory of the wave packet of a freely falling laser-cooled nanoparticle along a drop path, with the aim of generating an interference pattern in the measured position of the particle, generated by dropping one particle at a time. The required vibrational stability at the level of $10$ nm and pressure in a cryogenic vacuum below $10^{-12}$ mbar is achievable. One challenge that must be overcome while building statistics when measuring the final position of the center of mass of the dropped particle is that there can be variation in the size, shape, and mass of the nanodiamonds. To overcome this technical hurdle, a recapture beam and optical ``elevator'' scheme could be used upon particle detection to allow re-using the same diamond. To help mitigate risk for these measurements, nanospheres of silica are significantly monodisperse as verified by SEM imaging and can be used as a backup to demonstrate the desired matter wave interference phenomena for objects of similar mass, where distinct but similar particles can be more easily loaded for each instance of the experiment. While the long-term goals of the QGEM experiment involve magnetic trapping and state-splitting, the optical setup provides a testbed to establish the feasibility of creating superpositions with particles as large as 10$^{-20}$~kg, which would represent an improvement by three orders of magnitude with respect to the state of the art~\cite{arndt}, an important milestone along the way for future experiments with even larger macroscopic quantum superpositions. 

Gravitational acceleration (g and G) could then be observed on the interference pattern from such a particle. Tilting of the laser diffraction grating could be used to measure the effect of the Earth's gravitational field on the interference pattern of the particle. 

Schemes for large spatial superpositions are presented in Refs.~\cite{vandeKamp:2020rqh, Nair2023}, and there are proposals for even more ambitious mass-independent superposition schemes~\cite{Marshman:2021wyk,
Zhou:2022frl,Zhou:2022jug,Zhou:2022epb,Zhou:2024voj,Braccini:2024fey}.

Based on previous sub-mm gravity tests and measuring forces near surfaces \cite{Geraci:2008,montoya2022},  measuring acceleration from a nearby surface adjacent to the drop path of the interferometer would be of interest. This will be relevant for determining background forces due to induced dipoles, the Casimir effect, and patch potentials, and will help establish the feasibility of measuring the gravitational field sourced by a nanoparticle in a quantum superposition.

Advances in sensitivity made possible by pushing the sensitivity of these sensors into the quantum regime along with improved understanding and mitigation of systematic effects due to background electromagnetic interactions such as the Casimir effect and patch potentials will enable several orders of magnitude of improvement in the search for new physics beyond the Standard Model in addition to paving the way for tests of the role of gravity in quantum entanglement. Schemes based on the QGEM setup could be used to probe entanglement between two particles due to the Casimir-Polder interaction, as well as searches for axion dark matter and fifth forces~\cite{Barker:2022mdz}. Spin-off applications may include inertial sensors, accelerometers, or for appropriately functionalized nanoparticles, ultrasensitive detectors of electric or magnetic fields.

\subsection{Spin Control to create the spatial superpositions}

Using an NV center to create a spatial superposition will be an important milestone. This is a key step towards performing the test of the quantum nature of gravity because a large superposition distance of around $100$ $\mu$m will be required, which could be achieved with the Stern-Gerlach effect~\cite{Bose:2017nin}. The Stern-Gerlach effect requires an inhomogeneous magnetic field, which is already present in the diamagnetic traps that have been used to levitate microdiamonds in vacuum~\cite{DUrso16_GM, Twamley19_GM}. The diamagnetic trapping is due to the strong inhomogeneous magnetic field from sharp magnets,  but could also be engineered by utilizing current-carrying wires on atom chips~\cite{Folman2019_GM}. 

The spatial superposition experiment is limited in duration by a number of noise sources including the NV spin coherence time. The longest nanodiamond spin coherence times are $780$~$\mu$s with dynamic decoupling at room temperature~\cite{March23_GM, WoodPRB22_GM, Frangeskou18_GM}. This time may be limited by the nitrogen concentration of around $100$~ppb, or by unpaired electron spins on the surface of the $^{12}$C nanodiamond. Further reducing the nitrogen concentration and increasing the size of the diamond to one micron should improve this which could be achieved by fabricating diamond nanopillars from electronic-grade $^{12}$C diamond with a nitrogen concentration of under $5$~ppb~\cite{Awschalom, Englund}. Laser-writing ~\cite{Stephen19_GM} or ion implantation~\cite{TylerCMP2025} could be used to put one NV into each microdiamond. The cylindrical shape may be useful as a sharp electrode could be used to prevent rotations except around the cylinder axis. [111]-oriented diamond should be used so that the flat faces of the cylinders will be [111] faces. Then one in every four NV will be aligned along the cylinder axis as desired. Spinning the diamond fast about the cylinder axis would partially average away the electric multipole moments. If the density of unpaired electrons on the diamond surface cannot be sufficiently reduced, materials other than diamond may be needed such as pentacene-doped naphthalene~\cite{SteinerPentacene}.  

The $1$~ms coherence time matches with typical diamagnetic trap periods of around $10$~ms. If the spatial superposition experiment lasts for 1~ms then the diamond will effectively be in free flight: the trapping potential will not significantly prevent the expansion of the superposition distance. This will allow use of the free-flight proposal~\cite{Wan16_GM} which achieves larger superposition distances than earlier proposals~\cite{Scala13_GM} which assumed a much shorter trap period. This may only permit a small superposition distance of $5$~pm for a $2\times 10^{-15}$~kg diamond, but it would demonstrate many of the techniques needed for the test of the quantum nature of gravity. 

To go to larger superpostion distances using NV centres, the internal temperature of the diamond should be cryogenically reduced. The NV spin coherence has been shown to extend to over 1~s at low temperatures~\cite{Taminiau_GM} and  we can expect similar gains from cooling large high-purity nanodiamonds and microdiamonds. High-purity nanodiamonds have another key benefit which is to make them much more transparent than commercially available nanodiamonds. This means that the weak green light which excites the NV and the weak IR light  used for monitoring the diamond position are not absorbed so much, reducing the heat load on the diamond. This will be particularly important when the diamond is cryogenically cooled to a low internal temperature. 

Initial spatial superposition experiments could follow the free-flight protocol~\cite{Wan16_GM} which is shown schematically in Fig.~\ref{fig:WanDrop}. A microdiamond containing a single NV would be feedback cooled and then the position monitoring laser and the feedback would be turned off. A $3$~$\mu$s green pulse would polarize the NV spin to $m_s=0$. A $100$-ns microwave $\pi/2$ pulse would create a spin superposition which would cause a spatial superposition to grow with constant acceleration in the inhomogeneous magnetic field. Fig. \ref{fig:WanDrop} starts with a spin superposition of $m_s=+1$ and $m_s=-1$ which is more experimentally demanding~\cite{WoodPRA22_GM, Japha:2022xyg} but could lead to a larger spatial superposition than the spin superposition of $m_s=-1$ and $m_s=0$. After waiting a time $t_1$, a microwave $\pi$ pulse would be used to flip the spin states so the two superposition components slow down and then move back towards each other. After a further time $2t_1$, another microwave $\pi$ pulse would be applied so that the two superposition components meet for interference with no relative velocity after a further period of $t_1$. The NV spin state would then be detected optically using another green pulse. The total time, $4t_1$, would be similar to the 1~ms spin coherence time and much shorter than the trap period. This sequence would be repeated $10^4$ times to build up signal-to-noise. This averaged measurement sequence will then be repeated for a series of tilts of the optical table with respect to gravity to detect interference fringes due to a gravitational phase. These fringes would be clear evidence that a spatial superposition was created. A tilt of $3.5$~mrad = $0.2$ degrees would be enough to see one full fringe. This is a convenient regime as the tilt noise will not be overwhelming, but it is simple to controllably tilt an optical table by $0.5$~degrees which should allow more than one fully fringe. For future cryogenic versions of this experiment, the tilt noise should be greatly reduced by the use of the motional dynamic decoupling~\cite{Pedernales20_GM, WoodPRA22_GM}. This will be needed anyway to access longer experiments. 

\section{Timeline}

We are currently building different traps for levitating micron-sized diamonds and nanodiamonds as a step towards putting a diamond into a superposition of being in two places at once. This will be a key result, although the initial superposition distance will be very small: perhaps on the order of 10 to 1000~fm. Such a result will already test quantum mechanics in a new regime of macroscopicity. We will then work to increase this distance to the 10~$\mu$m to 100~$\mu$m scale required by lowering the temperatures and increasing the durations of the experiments. This will require that we reduce the decoherence. Theories like the Di\'osi-Penrose gravitationally-induced-collapse conjecture would predict that such a state could not be created. In parallel, we are setting up experiments featuring two diamonds that are close enough to interact with each other. In addition, we are investigating the implementation of and requirements on electromagnetic screening for nanodiamonds trapped at few micron range from a surface. Going forward, moving these experiments into underground labs may become important to reduce background vibrations and the gravity gradient noise. With sufficient resources and suitable facilities, significant steps towards the realization of the QGEM experiment are possible within the next decade.  

\section{Impact of research}

This research program is focused on what is arguably the most fundamental open problem in physics: finding experimental input to test the correct way to combine quantum mechanics with general relativity. 

Beyond this central objective, by establishing methods to access the quantum regime of matter-wave interferometry with levitated objects, a variety of precision sensing applications emerge. Methods of realizing matter wave interference along with improved understanding and mitigation of systematic effects due to background electromagnetic interactions such as the Casimir effect and patch potentials will enable several orders of magnitude of improvement in the search for new interactions between massive particles (i.e., ``fifth forces''), which appear in  many models of physics beyond the Standard Model~\cite{Geraci:2010}. Prior to reaching the full sensitivity needed for QGEM, these techniques will also allow stringent tests of quantum collapse models~\cite{bassireview}. 

Generating and reading out long-lived macroscopic superpositions would also permit searches for dark matter not possible with other techniques~\cite{Carney20_DM, Afek22_DM, Moore20_BSM,Riedel13_DM,Barker:2022mdz,Rufo:2025rps} as well as dark energy searches~\cite{Liu2018}. In particular, if sufficiently strong interactions occur between the particles and dark matter, the presence of dark matter could lead to unavoidable decoherence during the measurement protocol~\cite{Riedel13_DM}, allowing dark matter models to be searched for with smaller masses or couplings than can be detected with existing technologies~\cite{Carney20_DM}, or directly via entanglement like in the QGEM protocol~\cite{Barker:2022mdz,Rufo:2025rps}. If such dark matter interactions were to provide the dominant decoherence mechanism, the experimental protocols envisioned here would become more challenging to implement. However, observing such decoherence---and definitively linking it to dark matter through the inherent momentum sensitivity of these techniques~\cite{Afek22_DM,Monteiro20_DM,kilian2023requirements,kilian2024optimal,kilian2024dark}---would provide a major discovery in its own right.

Other relevant spin-off applications in sensing technologies include inertial sensors \cite{andyhart2015,Marshman20_GM,Wu:2024bzd, Rademacher2019}, gravitational wave detectors \cite{Marshman20_GM}, high-bandwidth accelerometers \cite{Monteiro20_force,winstone_2022}, and, using nanoparticles that are intentionally functionalized with an electric charge or magnetic moment, ultrasensitive detectors of electric or magnetic fields. Let's entangle things using gravity! 

\section*{Acknowledgments}
G.M. is supported by grants from the UK Research and Innovation (UKRI) Engineering and Physical Sciences Research Council (EPSRC): grants number EP/T001062/1 (Quantum Computation and Simulation Hub) and number EP/V056778/1 (Prosperity Partnership). G.M. and S.B. are supported by the UKRI Science and Technologies Facilities Council (STFC) grant number ST/W006561/1. M.T. acknowledges funding from the Slovenian Research and Innovation Agency (ARIS) under Contracts No. N1-0392, No. P1-0416, No. SN-ZRD/22-27/0510. M.P. acknowledges the European Union’s Horizon Europe EIC-Pathfinder project QuCoM (101046973), the UK EPSRC (EP/T028424/1), the Department for the Economy Northern Ireland under the US-Ireland R\&D Partnership Programme. MS is supported by the National Research Foundation, Singapore through the National Quantum Office, hosted in A*STAR, under its Centre for Quantum Technologies Funding Initiative (S24Q2d0009) and by the Ministry of Education, Singapore under the Academic Research Fund Tier 1 (FY2022, A-8000988-00-00. G.D. and T.P. acknowledge support from the John F. Templeton Foundation Grant ID 63121 and Alfred P. Sloan Foundation Grant No. 2023-21131. R.F. wishes to thank the Israel Science foundation (grants no. 856/18, 1314/19, 3515/20, and 3470/21), and the German-Israeli DIP project (Hybrid devices: FO 703/2-1) supported by the DFG, as well as the Gordon and Betty Moore (doi.org/10.37807/GBMF11936), and Simons (MP-TMPS-00005477) foundations. The research of S.B., A.M., A.G., D.M and G.M.  is funded in part by the Gordon and Betty Moore Foundation through Grant GBMF12328, DOI 10.37807/GBMF12328, and by the Alfred P. Sloan Foundation under Grant No. G-2023-21130.

\bibliography{CERNQG}

\begin{thebibliography}{158}%
\makeatletter
\providecommand \@ifxundefined [1]{%
 \@ifx{#1\undefined}
}%
\providecommand \@ifnum [1]{%
 \ifnum #1\expandafter \@firstoftwo
 \else \expandafter \@secondoftwo
 \fi
}%
\providecommand \@ifx [1]{%
 \ifx #1\expandafter \@firstoftwo
 \else \expandafter \@secondoftwo
 \fi
}%
\providecommand \natexlab [1]{#1}%
\providecommand \enquote  [1]{``#1''}%
\providecommand \bibnamefont  [1]{#1}%
\providecommand \bibfnamefont [1]{#1}%
\providecommand \citenamefont [1]{#1}%
\providecommand \href@noop [0]{\@secondoftwo}%
\providecommand \href [0]{\begingroup \@sanitize@url \@href}%
\providecommand \@href[1]{\@@startlink{#1}\@@href}%
\providecommand \@@href[1]{\endgroup#1\@@endlink}%
\providecommand \@sanitize@url [0]{\catcode `\\12\catcode `\$12\catcode `\&12\catcode `\#12\catcode `\^12\catcode `\_12\catcode `\%12\relax}%
\providecommand \@@startlink[1]{}%
\providecommand \@@endlink[0]{}%
\providecommand \url  [0]{\begingroup\@sanitize@url \@url }%
\providecommand \@url [1]{\endgroup\@href {#1}{\urlprefix }}%
\providecommand \urlprefix  [0]{URL }%
\providecommand \Eprint [0]{\href }%
\providecommand \doibase [0]{http://dx.doi.org/}%
\providecommand \selectlanguage [0]{\@gobble}%
\providecommand \bibinfo  [0]{\@secondoftwo}%
\providecommand \bibfield  [0]{\@secondoftwo}%
\providecommand \translation [1]{[#1]}%
\providecommand \BibitemOpen [0]{}%
\providecommand \bibitemStop [0]{}%
\providecommand \bibitemNoStop [0]{.\EOS\space}%
\providecommand \EOS [0]{\spacefactor3000\relax}%
\providecommand \BibitemShut  [1]{\csname bibitem#1\endcsname}%
\let\auto@bib@innerbib\@empty
\bibitem [{\citenamefont {Bose}\ \emph {et~al.}(2017)\citenamefont {Bose}, \citenamefont {Mazumdar}, \citenamefont {Morley}, \citenamefont {Ulbricht}, \citenamefont {Toro\v{s}}, \citenamefont {Paternostro}, \citenamefont {Geraci}, \citenamefont {Barker}, \citenamefont {Kim},\ and\ \citenamefont {Milburn}}]{Bose:2017nin}%
  \BibitemOpen
  \bibfield  {author} {\bibinfo {author} {\bibfnamefont {S.}~\bibnamefont {Bose}}, \bibinfo {author} {\bibfnamefont {A.}~\bibnamefont {Mazumdar}}, \bibinfo {author} {\bibfnamefont {G.~W.}\ \bibnamefont {Morley}}, \bibinfo {author} {\bibfnamefont {H.}~\bibnamefont {Ulbricht}}, \bibinfo {author} {\bibfnamefont {M.}~\bibnamefont {Toro\v{s}}}, \bibinfo {author} {\bibfnamefont {M.}~\bibnamefont {Paternostro}}, \bibinfo {author} {\bibfnamefont {A.}~\bibnamefont {Geraci}}, \bibinfo {author} {\bibfnamefont {P.}~\bibnamefont {Barker}}, \bibinfo {author} {\bibfnamefont {M.~S.}\ \bibnamefont {Kim}}, \ and\ \bibinfo {author} {\bibfnamefont {G.}~\bibnamefont {Milburn}},\ }\href {\doibase 10.1103/PhysRevLett.119.240401} {\bibfield  {journal} {\bibinfo  {journal} {Phys. Rev. Lett.}\ }\textbf {\bibinfo {volume} {119}},\ \bibinfo {pages} {240401} (\bibinfo {year} {2017})}\BibitemShut {NoStop}%
\bibitem [{ICT(2016)}]{ICTS}%
  \BibitemOpen
  \href@noop {} {}\bibinfo {howpublished} {\url{https://www.youtube.com/watch?v=0Fv-0k13s_k}} (\bibinfo {year} {2016}),\ \bibinfo {note} {accessed 31st March 2025}\BibitemShut {NoStop}%
\bibitem [{\citenamefont {Marletto}\ and\ \citenamefont {Vedral}(2017)}]{marletto2017gravitationally}%
  \BibitemOpen
  \bibfield  {author} {\bibinfo {author} {\bibfnamefont {C.}~\bibnamefont {Marletto}}\ and\ \bibinfo {author} {\bibfnamefont {V.}~\bibnamefont {Vedral}},\ }\href {\doibase 10.1103/PhysRevLett.119.240402} {\bibfield  {journal} {\bibinfo  {journal} {Phys. Rev. Lett.}\ }\textbf {\bibinfo {volume} {119}},\ \bibinfo {pages} {240402} (\bibinfo {year} {2017})}\BibitemShut {NoStop}%
\bibitem [{\citenamefont {Bose}\ \emph {et~al.}(2025)\citenamefont {Bose}, \citenamefont {Fuentes}, \citenamefont {Geraci}, \citenamefont {Khan}, \citenamefont {Qvarfort}, \citenamefont {Rademacher}, \citenamefont {Rashid}, \citenamefont {Toro{\v{s}}}, \citenamefont {Ulbricht},\ and\ \citenamefont {Wanjura}}]{bose2025massive}%
  \BibitemOpen
  \bibfield  {author} {\bibinfo {author} {\bibfnamefont {S.}~\bibnamefont {Bose}}, \bibinfo {author} {\bibfnamefont {I.}~\bibnamefont {Fuentes}}, \bibinfo {author} {\bibfnamefont {A.~A.}\ \bibnamefont {Geraci}}, \bibinfo {author} {\bibfnamefont {S.~M.}\ \bibnamefont {Khan}}, \bibinfo {author} {\bibfnamefont {S.}~\bibnamefont {Qvarfort}}, \bibinfo {author} {\bibfnamefont {M.}~\bibnamefont {Rademacher}}, \bibinfo {author} {\bibfnamefont {M.}~\bibnamefont {Rashid}}, \bibinfo {author} {\bibfnamefont {M.}~\bibnamefont {Toro{\v{s}}}}, \bibinfo {author} {\bibfnamefont {H.}~\bibnamefont {Ulbricht}}, \ and\ \bibinfo {author} {\bibfnamefont {C.~C.}\ \bibnamefont {Wanjura}},\ }\href {\doibase 10.1103/RevModPhys.97.015003} {\bibfield  {journal} {\bibinfo  {journal} {Reviews of Modern Physics}\ }\textbf {\bibinfo {volume} {97}},\ \bibinfo {pages} {015003} (\bibinfo {year} {2025})}\BibitemShut {NoStop}%
\bibitem [{\citenamefont {Hanif}\ \emph {et~al.}(2024)\citenamefont {Hanif}, \citenamefont {Das}, \citenamefont {Halliwell}, \citenamefont {Home}, \citenamefont {Mazumdar}, \citenamefont {Ulbricht},\ and\ \citenamefont {Bose}}]{hanif2024testing}%
  \BibitemOpen
  \bibfield  {author} {\bibinfo {author} {\bibfnamefont {F.}~\bibnamefont {Hanif}}, \bibinfo {author} {\bibfnamefont {D.}~\bibnamefont {Das}}, \bibinfo {author} {\bibfnamefont {J.}~\bibnamefont {Halliwell}}, \bibinfo {author} {\bibfnamefont {D.}~\bibnamefont {Home}}, \bibinfo {author} {\bibfnamefont {A.}~\bibnamefont {Mazumdar}}, \bibinfo {author} {\bibfnamefont {H.}~\bibnamefont {Ulbricht}}, \ and\ \bibinfo {author} {\bibfnamefont {S.}~\bibnamefont {Bose}},\ }\href {https://doi.org/10.1103/PhysRevLett.133.180201} {\bibfield  {journal} {\bibinfo  {journal} {Physical Review Letters}\ }\textbf {\bibinfo {volume} {133}},\ \bibinfo {pages} {180201} (\bibinfo {year} {2024})}\BibitemShut {NoStop}%
\bibitem [{\citenamefont {Krisnanda}\ \emph {et~al.}(2020)\citenamefont {Krisnanda}, \citenamefont {Tham}, \citenamefont {Paternostro},\ and\ \citenamefont {Paterek}}]{krisnanda2020observable}%
  \BibitemOpen
  \bibfield  {author} {\bibinfo {author} {\bibfnamefont {T.}~\bibnamefont {Krisnanda}}, \bibinfo {author} {\bibfnamefont {G.~Y.}\ \bibnamefont {Tham}}, \bibinfo {author} {\bibfnamefont {M.}~\bibnamefont {Paternostro}}, \ and\ \bibinfo {author} {\bibfnamefont {T.}~\bibnamefont {Paterek}},\ }\href {https://doi.org/10.1038/s41534-020-0243-y} {\bibfield  {journal} {\bibinfo  {journal} {npj Quantum Information}\ }\textbf {\bibinfo {volume} {6}},\ \bibinfo {pages} {12} (\bibinfo {year} {2020})}\BibitemShut {NoStop}%
\bibitem [{\citenamefont {Howl}\ \emph {et~al.}(2021)\citenamefont {Howl}, \citenamefont {Vedral}, \citenamefont {Naik}, \citenamefont {Christodoulou}, \citenamefont {Rovelli},\ and\ \citenamefont {Iyer}}]{howl2021non}%
  \BibitemOpen
  \bibfield  {author} {\bibinfo {author} {\bibfnamefont {R.}~\bibnamefont {Howl}}, \bibinfo {author} {\bibfnamefont {V.}~\bibnamefont {Vedral}}, \bibinfo {author} {\bibfnamefont {D.}~\bibnamefont {Naik}}, \bibinfo {author} {\bibfnamefont {M.}~\bibnamefont {Christodoulou}}, \bibinfo {author} {\bibfnamefont {C.}~\bibnamefont {Rovelli}}, \ and\ \bibinfo {author} {\bibfnamefont {A.}~\bibnamefont {Iyer}},\ }\href {https://doi.org/10.1103/PRXQuantum.2.010325} {\bibfield  {journal} {\bibinfo  {journal} {PRX Quantum}\ }\textbf {\bibinfo {volume} {2}},\ \bibinfo {pages} {010325} (\bibinfo {year} {2021})}\BibitemShut {NoStop}%
\bibitem [{\citenamefont {Haine}(2021)}]{haine2021searching}%
  \BibitemOpen
  \bibfield  {author} {\bibinfo {author} {\bibfnamefont {S.~A.}\ \bibnamefont {Haine}},\ }\href {https://dx.doi.org/10.1088/1367-2630/abd97d} {\bibfield  {journal} {\bibinfo  {journal} {New Journal of Physics}\ }\textbf {\bibinfo {volume} {23}},\ \bibinfo {pages} {033020} (\bibinfo {year} {2021})}\BibitemShut {NoStop}%
\bibitem [{\citenamefont {Oppenheim}\ \emph {et~al.}(2023)\citenamefont {Oppenheim}, \citenamefont {Sparaciari}, \citenamefont {{\v{S}}oda},\ and\ \citenamefont {Weller-Davies}}]{dec_Vs_diff}%
  \BibitemOpen
  \bibfield  {author} {\bibinfo {author} {\bibfnamefont {J.}~\bibnamefont {Oppenheim}}, \bibinfo {author} {\bibfnamefont {C.}~\bibnamefont {Sparaciari}}, \bibinfo {author} {\bibfnamefont {B.}~\bibnamefont {{\v{S}}oda}}, \ and\ \bibinfo {author} {\bibfnamefont {Z.}~\bibnamefont {Weller-Davies}},\ }\href {https://doi.org/10.1038/s41467-023-43348-2} {\bibfield  {journal} {\bibinfo  {journal} {Nature Communications}\ }\textbf {\bibinfo {volume} {14}},\ \bibinfo {pages} {7910} (\bibinfo {year} {2023})}\BibitemShut {NoStop}%
\bibitem [{\citenamefont {Lami}\ \emph {et~al.}(2024)\citenamefont {Lami}, \citenamefont {Pedernales},\ and\ \citenamefont {Plenio}}]{lami2024testing}%
  \BibitemOpen
  \bibfield  {author} {\bibinfo {author} {\bibfnamefont {L.}~\bibnamefont {Lami}}, \bibinfo {author} {\bibfnamefont {J.~S.}\ \bibnamefont {Pedernales}}, \ and\ \bibinfo {author} {\bibfnamefont {M.~B.}\ \bibnamefont {Plenio}},\ }\href {\doibase 10.1103/PhysRevX.14.021022} {\bibfield  {journal} {\bibinfo  {journal} {Physical Review X}\ }\textbf {\bibinfo {volume} {14}},\ \bibinfo {pages} {021022} (\bibinfo {year} {2024})}\BibitemShut {NoStop}%
\bibitem [{\citenamefont {Kryhin}\ and\ \citenamefont {Sudhir}(2025)}]{kryhin2025distinguishable}%
  \BibitemOpen
  \bibfield  {author} {\bibinfo {author} {\bibfnamefont {S.}~\bibnamefont {Kryhin}}\ and\ \bibinfo {author} {\bibfnamefont {V.}~\bibnamefont {Sudhir}},\ }\href {https://doi.org/10.1103/PhysRevLett.134.061501} {\bibfield  {journal} {\bibinfo  {journal} {Physical Review Letters}\ }\textbf {\bibinfo {volume} {134}},\ \bibinfo {pages} {061501} (\bibinfo {year} {2025})}\BibitemShut {NoStop}%
\bibitem [{\citenamefont {Carney}\ \emph {et~al.}(2019{\natexlab{a}})\citenamefont {Carney}, \citenamefont {Stamp},\ and\ \citenamefont {Taylor}}]{carney2019tabletop}%
  \BibitemOpen
  \bibfield  {author} {\bibinfo {author} {\bibfnamefont {D.}~\bibnamefont {Carney}}, \bibinfo {author} {\bibfnamefont {P.~C.}\ \bibnamefont {Stamp}}, \ and\ \bibinfo {author} {\bibfnamefont {J.~M.}\ \bibnamefont {Taylor}},\ }\href {\doibase 10.1088/1361-6382/aaf9ca} {\bibfield  {journal} {\bibinfo  {journal} {Classical and Quantum Gravity}\ }\textbf {\bibinfo {volume} {36}},\ \bibinfo {pages} {034001} (\bibinfo {year} {2019}{\natexlab{a}})}\BibitemShut {NoStop}%
\bibitem [{\citenamefont {Belenchia}\ \emph {et~al.}(2018)\citenamefont {Belenchia}, \citenamefont {Wald}, \citenamefont {Giacomini}, \citenamefont {Castro-Ruiz}, \citenamefont {Brukner},\ and\ \citenamefont {Aspelmeyer}}]{Belenchia:2018szb}%
  \BibitemOpen
  \bibfield  {author} {\bibinfo {author} {\bibfnamefont {A.}~\bibnamefont {Belenchia}}, \bibinfo {author} {\bibfnamefont {R.~M.}\ \bibnamefont {Wald}}, \bibinfo {author} {\bibfnamefont {F.}~\bibnamefont {Giacomini}}, \bibinfo {author} {\bibfnamefont {E.}~\bibnamefont {Castro-Ruiz}}, \bibinfo {author} {\bibfnamefont {{\v{C}}.}~\bibnamefont {Brukner}}, \ and\ \bibinfo {author} {\bibfnamefont {M.}~\bibnamefont {Aspelmeyer}},\ }\href {\doibase 10.1103/PhysRevD.98.126009} {\bibfield  {journal} {\bibinfo  {journal} {Phys. Rev. D}\ }\textbf {\bibinfo {volume} {98}},\ \bibinfo {pages} {126009} (\bibinfo {year} {2018})}\BibitemShut {NoStop}%
\bibitem [{\citenamefont {Guerreiro}(2020)}]{Guerreiro2020}%
  \BibitemOpen
  \bibfield  {author} {\bibinfo {author} {\bibfnamefont {T.}~\bibnamefont {Guerreiro}},\ }\href {\doibase 10.1088/1361-6382/ab9d5d} {\bibfield  {journal} {\bibinfo  {journal} {Classical and Quantum Gravity}\ }\textbf {\bibinfo {volume} {37}},\ \bibinfo {pages} {155001} (\bibinfo {year} {2020})}\BibitemShut {NoStop}%
\bibitem [{\citenamefont {Parikh}\ \emph {et~al.}(2021)\citenamefont {Parikh}, \citenamefont {Wilczek},\ and\ \citenamefont {Zahariade}}]{parikh2021signatures}%
  \BibitemOpen
  \bibfield  {author} {\bibinfo {author} {\bibfnamefont {M.}~\bibnamefont {Parikh}}, \bibinfo {author} {\bibfnamefont {F.}~\bibnamefont {Wilczek}}, \ and\ \bibinfo {author} {\bibfnamefont {G.}~\bibnamefont {Zahariade}},\ }\href {https://doi.org/10.1103/PhysRevD.104.046021} {\bibfield  {journal} {\bibinfo  {journal} {Physical Review D}\ }\textbf {\bibinfo {volume} {104}},\ \bibinfo {pages} {046021} (\bibinfo {year} {2021})}\BibitemShut {NoStop}%
\bibitem [{\citenamefont {Tobar}\ \emph {et~al.}(2024)\citenamefont {Tobar}, \citenamefont {Manikandan}, \citenamefont {Beitel},\ and\ \citenamefont {Pikovski}}]{tobar2024detecting}%
  \BibitemOpen
  \bibfield  {author} {\bibinfo {author} {\bibfnamefont {G.}~\bibnamefont {Tobar}}, \bibinfo {author} {\bibfnamefont {S.~K.}\ \bibnamefont {Manikandan}}, \bibinfo {author} {\bibfnamefont {T.}~\bibnamefont {Beitel}}, \ and\ \bibinfo {author} {\bibfnamefont {I.}~\bibnamefont {Pikovski}},\ }\href {https://doi.org/10.1038/s41467-024-51420-8} {\bibfield  {journal} {\bibinfo  {journal} {Nature Communications}\ }\textbf {\bibinfo {volume} {15}},\ \bibinfo {pages} {7229} (\bibinfo {year} {2024})}\BibitemShut {NoStop}%
\bibitem [{\citenamefont {Vermeulen}\ \emph {et~al.}(2025)\citenamefont {Vermeulen}, \citenamefont {Cullen}, \citenamefont {Grass}, \citenamefont {MacMillan}, \citenamefont {Ramirez}, \citenamefont {Wack}, \citenamefont {Korzh}, \citenamefont {Lee}, \citenamefont {Zurek}, \citenamefont {Stoughton},\ and\ \citenamefont {McCuller}}]{vermeulen2025photon}%
  \BibitemOpen
  \bibfield  {author} {\bibinfo {author} {\bibfnamefont {S.~M.}\ \bibnamefont {Vermeulen}}, \bibinfo {author} {\bibfnamefont {T.}~\bibnamefont {Cullen}}, \bibinfo {author} {\bibfnamefont {D.}~\bibnamefont {Grass}}, \bibinfo {author} {\bibfnamefont {I.~A.~O.}\ \bibnamefont {MacMillan}}, \bibinfo {author} {\bibfnamefont {A.~J.}\ \bibnamefont {Ramirez}}, \bibinfo {author} {\bibfnamefont {J.}~\bibnamefont {Wack}}, \bibinfo {author} {\bibfnamefont {B.}~\bibnamefont {Korzh}}, \bibinfo {author} {\bibfnamefont {V.~S.~H.}\ \bibnamefont {Lee}}, \bibinfo {author} {\bibfnamefont {K.~M.}\ \bibnamefont {Zurek}}, \bibinfo {author} {\bibfnamefont {C.}~\bibnamefont {Stoughton}}, \ and\ \bibinfo {author} {\bibfnamefont {L.}~\bibnamefont {McCuller}},\ }\href {https://doi.org/10.1103/PhysRevX.15.011034} {\bibfield  {journal} {\bibinfo  {journal} {Physical Review X}\ }\textbf {\bibinfo {volume} {15}},\ \bibinfo {pages} {011034} (\bibinfo {year} {2025})}\BibitemShut {NoStop}%
\bibitem [{\citenamefont {Vermeulen}\ \emph {et~al.}(2021)\citenamefont {Vermeulen}, \citenamefont {Aiello}, \citenamefont {Ejlli}, \citenamefont {Griffiths}, \citenamefont {James}, \citenamefont {Dooley},\ and\ \citenamefont {Grote}}]{Vermeulen2021}%
  \BibitemOpen
  \bibfield  {author} {\bibinfo {author} {\bibfnamefont {S.~M.}\ \bibnamefont {Vermeulen}}, \bibinfo {author} {\bibfnamefont {L.}~\bibnamefont {Aiello}}, \bibinfo {author} {\bibfnamefont {A.}~\bibnamefont {Ejlli}}, \bibinfo {author} {\bibfnamefont {W.~L.}\ \bibnamefont {Griffiths}}, \bibinfo {author} {\bibfnamefont {A.~L.}\ \bibnamefont {James}}, \bibinfo {author} {\bibfnamefont {K.~L.}\ \bibnamefont {Dooley}}, \ and\ \bibinfo {author} {\bibfnamefont {H.}~\bibnamefont {Grote}},\ }\href {\doibase 10.1088/1361-6382/abe757} {\bibfield  {journal} {\bibinfo  {journal} {Classical and Quantum Gravity}\ }\textbf {\bibinfo {volume} {38}},\ \bibinfo {pages} {085008} (\bibinfo {year} {2021})}\BibitemShut {NoStop}%
\bibitem [{\citenamefont {Pietra}\ \emph {et~al.}(2024)\citenamefont {Pietra}, \citenamefont {Piacentini}, \citenamefont {Bernardi}, \citenamefont {Moreva}, \citenamefont {Napoli}, \citenamefont {Degiovanni}, \citenamefont {Genovese}, \citenamefont {Vedral},\ and\ \citenamefont {Marletto}}]{Marletto2024}%
  \BibitemOpen
  \bibfield  {author} {\bibinfo {author} {\bibfnamefont {G.~D.}\ \bibnamefont {Pietra}}, \bibinfo {author} {\bibfnamefont {F.}~\bibnamefont {Piacentini}}, \bibinfo {author} {\bibfnamefont {E.}~\bibnamefont {Bernardi}}, \bibinfo {author} {\bibfnamefont {E.}~\bibnamefont {Moreva}}, \bibinfo {author} {\bibfnamefont {C.}~\bibnamefont {Napoli}}, \bibinfo {author} {\bibfnamefont {I.~P.}\ \bibnamefont {Degiovanni}}, \bibinfo {author} {\bibfnamefont {M.}~\bibnamefont {Genovese}}, \bibinfo {author} {\bibfnamefont {V.}~\bibnamefont {Vedral}}, \ and\ \bibinfo {author} {\bibfnamefont {C.}~\bibnamefont {Marletto}},\ }\href@noop {} {\  (\bibinfo {year} {2024})},\ \Eprint {http://arxiv.org/abs/arXiv:2410.19601} {arXiv:2410.19601} \BibitemShut {NoStop}%
\bibitem [{\citenamefont {Vicentini}\ \emph {et~al.}(2024)\citenamefont {Vicentini}, \citenamefont {Bernardi}, \citenamefont {Moreva}, \citenamefont {Piacentini}, \citenamefont {Napoli}, \citenamefont {Degiovanni}, \citenamefont {Manzin},\ and\ \citenamefont {Genovese}}]{Genovese2024}%
  \BibitemOpen
  \bibfield  {author} {\bibinfo {author} {\bibfnamefont {M.}~\bibnamefont {Vicentini}}, \bibinfo {author} {\bibfnamefont {E.}~\bibnamefont {Bernardi}}, \bibinfo {author} {\bibfnamefont {E.}~\bibnamefont {Moreva}}, \bibinfo {author} {\bibfnamefont {F.}~\bibnamefont {Piacentini}}, \bibinfo {author} {\bibfnamefont {C.}~\bibnamefont {Napoli}}, \bibinfo {author} {\bibfnamefont {I.~P.}\ \bibnamefont {Degiovanni}}, \bibinfo {author} {\bibfnamefont {A.}~\bibnamefont {Manzin}}, \ and\ \bibinfo {author} {\bibfnamefont {M.}~\bibnamefont {Genovese}},\ }\href@noop {} {\  (\bibinfo {year} {2024})},\ \Eprint {http://arxiv.org/abs/arXiv:2405.21029} {arXiv:2405.21029} \BibitemShut {NoStop}%
\bibitem [{\citenamefont {Marletto}\ and\ \citenamefont {Vedral}(2025)}]{MarlettoVedralRMP}%
  \BibitemOpen
  \bibfield  {author} {\bibinfo {author} {\bibfnamefont {C.}~\bibnamefont {Marletto}}\ and\ \bibinfo {author} {\bibfnamefont {V.}~\bibnamefont {Vedral}},\ }\href {\doibase 10.1103/RevModPhys.97.015006} {\bibfield  {journal} {\bibinfo  {journal} {Rev. Mod. Phys.}\ }\textbf {\bibinfo {volume} {97}},\ \bibinfo {pages} {015006} (\bibinfo {year} {2025})}\BibitemShut {NoStop}%
\bibitem [{\citenamefont {Marletto}\ and\ \citenamefont {Vedral}(2020)}]{MarlettoVedralPRD2020}%
  \BibitemOpen
  \bibfield  {author} {\bibinfo {author} {\bibfnamefont {C.}~\bibnamefont {Marletto}}\ and\ \bibinfo {author} {\bibfnamefont {V.}~\bibnamefont {Vedral}},\ }\href {\doibase 10.1103/PhysRevD.102.086012} {\bibfield  {journal} {\bibinfo  {journal} {Phys. Rev. D}\ }\textbf {\bibinfo {volume} {102}},\ \bibinfo {pages} {086012} (\bibinfo {year} {2020})}\BibitemShut {NoStop}%
\bibitem [{\citenamefont {Marletto}\ and\ \citenamefont {Vedral}(2018)}]{MarlettoVedralPRD2018}%
  \BibitemOpen
  \bibfield  {author} {\bibinfo {author} {\bibfnamefont {C.}~\bibnamefont {Marletto}}\ and\ \bibinfo {author} {\bibfnamefont {V.}~\bibnamefont {Vedral}},\ }\href {\doibase 10.1103/PhysRevD.98.046001} {\bibfield  {journal} {\bibinfo  {journal} {Phys. Rev. D}\ }\textbf {\bibinfo {volume} {98}},\ \bibinfo {pages} {046001} (\bibinfo {year} {2018})}\BibitemShut {NoStop}%
\bibitem [{\citenamefont {Miki}\ \emph {et~al.}(2025)\citenamefont {Miki}, \citenamefont {Kaku}, \citenamefont {Liu}, \citenamefont {Ma},\ and\ \citenamefont {Chen}}]{Miki2025}%
  \BibitemOpen
  \bibfield  {author} {\bibinfo {author} {\bibfnamefont {D.}~\bibnamefont {Miki}}, \bibinfo {author} {\bibfnamefont {Y.}~\bibnamefont {Kaku}}, \bibinfo {author} {\bibfnamefont {Y.}~\bibnamefont {Liu}}, \bibinfo {author} {\bibfnamefont {Y.}~\bibnamefont {Ma}}, \ and\ \bibinfo {author} {\bibfnamefont {Y.}~\bibnamefont {Chen}},\ }\href@noop {} {\  (\bibinfo {year} {2025})},\ \Eprint {http://arxiv.org/abs/2503.11882} {2503.11882 [quant-ph]} \BibitemShut {NoStop}%
\bibitem [{\citenamefont {Mari}\ \emph {et~al.}(2025)\citenamefont {Mari}, \citenamefont {Zippilli},\ and\ \citenamefont {Vitali}}]{Mari2025}%
  \BibitemOpen
  \bibfield  {author} {\bibinfo {author} {\bibfnamefont {A.}~\bibnamefont {Mari}}, \bibinfo {author} {\bibfnamefont {S.}~\bibnamefont {Zippilli}}, \ and\ \bibinfo {author} {\bibfnamefont {D.}~\bibnamefont {Vitali}},\ }\href@noop {} {\  (\bibinfo {year} {2025})},\ \Eprint {http://arxiv.org/abs/2504.05998} {2504.05998 [quant-ph]} \BibitemShut {NoStop}%
\bibitem [{\citenamefont {Hogan}(2012)}]{Hogan2012}%
  \BibitemOpen
  \bibfield  {author} {\bibinfo {author} {\bibfnamefont {C.~J.}\ \bibnamefont {Hogan}},\ }\href {\doibase 10.1103/PhysRevD.85.064007} {\bibfield  {journal} {\bibinfo  {journal} {Phys. Rev. D}\ }\textbf {\bibinfo {volume} {85}},\ \bibinfo {pages} {064007} (\bibinfo {year} {2012})}\BibitemShut {NoStop}%
\bibitem [{\citenamefont {Hogan}\ \emph {et~al.}(2017)\citenamefont {Hogan}, \citenamefont {Kwon},\ and\ \citenamefont {Richardson}}]{Hogan2017}%
  \BibitemOpen
  \bibfield  {author} {\bibinfo {author} {\bibfnamefont {C.}~\bibnamefont {Hogan}}, \bibinfo {author} {\bibfnamefont {O.}~\bibnamefont {Kwon}}, \ and\ \bibinfo {author} {\bibfnamefont {J.}~\bibnamefont {Richardson}},\ }\href {\doibase 10.1088/1361-6382/aa73c0} {\bibfield  {journal} {\bibinfo  {journal} {Classical and Quantum Gravity}\ }\textbf {\bibinfo {volume} {34}},\ \bibinfo {pages} {135006} (\bibinfo {year} {2017})}\BibitemShut {NoStop}%
\bibitem [{\citenamefont {Chou}\ \emph {et~al.}(2016)\citenamefont {Chou}, \citenamefont {Gustafson}, \citenamefont {Hogan}, \citenamefont {Kamai}, \citenamefont {Kwon}, \citenamefont {Lanza}, \citenamefont {McCuller}, \citenamefont {Meyer}, \citenamefont {Richardson}, \citenamefont {Stoughton}, \citenamefont {Tomlin}, \citenamefont {Waldman},\ and\ \citenamefont {Weiss}}]{Holometer2016}%
  \BibitemOpen
  \bibfield  {author} {\bibinfo {author} {\bibfnamefont {A.~S.}\ \bibnamefont {Chou}}, \bibinfo {author} {\bibfnamefont {R.}~\bibnamefont {Gustafson}}, \bibinfo {author} {\bibfnamefont {C.}~\bibnamefont {Hogan}}, \bibinfo {author} {\bibfnamefont {B.}~\bibnamefont {Kamai}}, \bibinfo {author} {\bibfnamefont {O.}~\bibnamefont {Kwon}}, \bibinfo {author} {\bibfnamefont {R.}~\bibnamefont {Lanza}}, \bibinfo {author} {\bibfnamefont {L.}~\bibnamefont {McCuller}}, \bibinfo {author} {\bibfnamefont {S.~S.}\ \bibnamefont {Meyer}}, \bibinfo {author} {\bibfnamefont {J.}~\bibnamefont {Richardson}}, \bibinfo {author} {\bibfnamefont {C.}~\bibnamefont {Stoughton}}, \bibinfo {author} {\bibfnamefont {R.}~\bibnamefont {Tomlin}}, \bibinfo {author} {\bibfnamefont {S.}~\bibnamefont {Waldman}}, \ and\ \bibinfo {author} {\bibfnamefont {R.}~\bibnamefont {Weiss}} (\bibinfo {collaboration} {Holometer Collaboration}),\ }\href {\doibase 10.1103/PhysRevLett.117.111102} {\bibfield  {journal} {\bibinfo  {journal} {Phys. Rev. Lett.}\ }\textbf
  {\bibinfo {volume} {117}},\ \bibinfo {pages} {111102} (\bibinfo {year} {2016})}\BibitemShut {NoStop}%
\bibitem [{\citenamefont {Pradyumna}\ \emph {et~al.}(2020)\citenamefont {Pradyumna}, \citenamefont {Losero}, \citenamefont {Ruo-Berchera}, \citenamefont {Traina}, \citenamefont {Zucco}, \citenamefont {Jacobsen}, \citenamefont {Andersen}, \citenamefont {Degiovanni}, \citenamefont {Genovese},\ and\ \citenamefont {Gehring}}]{Pradyumna2020}%
  \BibitemOpen
  \bibfield  {author} {\bibinfo {author} {\bibfnamefont {S.~T.}\ \bibnamefont {Pradyumna}}, \bibinfo {author} {\bibfnamefont {E.}~\bibnamefont {Losero}}, \bibinfo {author} {\bibfnamefont {I.}~\bibnamefont {Ruo-Berchera}}, \bibinfo {author} {\bibfnamefont {P.}~\bibnamefont {Traina}}, \bibinfo {author} {\bibfnamefont {M.}~\bibnamefont {Zucco}}, \bibinfo {author} {\bibfnamefont {C.~S.}\ \bibnamefont {Jacobsen}}, \bibinfo {author} {\bibfnamefont {U.~L.}\ \bibnamefont {Andersen}}, \bibinfo {author} {\bibfnamefont {I.~P.}\ \bibnamefont {Degiovanni}}, \bibinfo {author} {\bibfnamefont {M.}~\bibnamefont {Genovese}}, \ and\ \bibinfo {author} {\bibfnamefont {T.}~\bibnamefont {Gehring}},\ }\href@noop {} {\bibfield  {journal} {\bibinfo  {journal} {Commun. Phys.}\ }\textbf {\bibinfo {volume} {3}} (\bibinfo {year} {2020})}\BibitemShut {NoStop}%
\bibitem [{\citenamefont {Pikovski}\ \emph {et~al.}(2012)\citenamefont {Pikovski}, \citenamefont {Vanner}, \citenamefont {Aspelmeyer}, \citenamefont {Kim},\ and\ \citenamefont {Brukner}}]{Pikovski2012}%
  \BibitemOpen
  \bibfield  {author} {\bibinfo {author} {\bibfnamefont {I.}~\bibnamefont {Pikovski}}, \bibinfo {author} {\bibfnamefont {M.~R.}\ \bibnamefont {Vanner}}, \bibinfo {author} {\bibfnamefont {M.}~\bibnamefont {Aspelmeyer}}, \bibinfo {author} {\bibfnamefont {M.~S.}\ \bibnamefont {Kim}}, \ and\ \bibinfo {author} {\bibfnamefont {{\v C}.}~\bibnamefont {Brukner}},\ }\href@noop {} {\bibfield  {journal} {\bibinfo  {journal} {Nat. Phys.}\ }\textbf {\bibinfo {volume} {8}},\ \bibinfo {pages} {393} (\bibinfo {year} {2012})}\BibitemShut {NoStop}%
\bibitem [{\citenamefont {Howl}\ \emph {et~al.}(2023)\citenamefont {Howl}, \citenamefont {Cooper},\ and\ \citenamefont {Hackerm{\"u}ller}}]{Howl2023}%
  \BibitemOpen
  \bibfield  {author} {\bibinfo {author} {\bibfnamefont {R.}~\bibnamefont {Howl}}, \bibinfo {author} {\bibfnamefont {N.}~\bibnamefont {Cooper}}, \ and\ \bibinfo {author} {\bibfnamefont {L.}~\bibnamefont {Hackerm{\"u}ller}},\ }\href@noop {} {\  (\bibinfo {year} {2023})},\ \Eprint {http://arxiv.org/abs/2304.00734} {2304.00734 [quant-ph]} \BibitemShut {NoStop}%
\bibitem [{\citenamefont {Marshman}\ \emph {et~al.}(2020{\natexlab{a}})\citenamefont {Marshman}, \citenamefont {Mazumdar},\ and\ \citenamefont {Bose}}]{Marshman:2019sne}%
  \BibitemOpen
  \bibfield  {author} {\bibinfo {author} {\bibfnamefont {R.~J.}\ \bibnamefont {Marshman}}, \bibinfo {author} {\bibfnamefont {A.}~\bibnamefont {Mazumdar}}, \ and\ \bibinfo {author} {\bibfnamefont {S.}~\bibnamefont {Bose}},\ }\href {\doibase 10.1103/PhysRevA.101.052110} {\bibfield  {journal} {\bibinfo  {journal} {Phys. Rev. A}\ }\textbf {\bibinfo {volume} {101}},\ \bibinfo {pages} {052110} (\bibinfo {year} {2020}{\natexlab{a}})}\BibitemShut {NoStop}%
\bibitem [{\citenamefont {Bose}\ \emph {et~al.}(2022)\citenamefont {Bose}, \citenamefont {Mazumdar}, \citenamefont {Schut},\ and\ \citenamefont {Toro\v{s}}}]{Bose:2022uxe}%
  \BibitemOpen
  \bibfield  {author} {\bibinfo {author} {\bibfnamefont {S.}~\bibnamefont {Bose}}, \bibinfo {author} {\bibfnamefont {A.}~\bibnamefont {Mazumdar}}, \bibinfo {author} {\bibfnamefont {M.}~\bibnamefont {Schut}}, \ and\ \bibinfo {author} {\bibfnamefont {M.}~\bibnamefont {Toro\v{s}}},\ }\href {\doibase 10.1103/PhysRevD.105.106028} {\bibfield  {journal} {\bibinfo  {journal} {Phys. Rev. D}\ }\textbf {\bibinfo {volume} {105}},\ \bibinfo {pages} {106028} (\bibinfo {year} {2022})}\BibitemShut {NoStop}%
\bibitem [{\citenamefont {De~Witt}(1957)}]{DeWitt:1957obj}%
  \BibitemOpen
  \bibinfo {editor} {\bibfnamefont {C.~M.}\ \bibnamefont {De~Witt}},\ ed.,\ \href {\doibase 10.34663/9783945561294-00} {\emph {\bibinfo {title} {{Proceedings: Conference on the Role of Gravitation in Physics, Chapel Hill, North Carolina, Jan 18-23, 1957}}}}\ (\bibinfo {year} {1957})\BibitemShut {NoStop}%
\bibitem [{\citenamefont {Deli{\'c}}\ \emph {et~al.}(2020)\citenamefont {Deli{\'c}}, \citenamefont {Reisenbauer}, \citenamefont {Dare}, \citenamefont {Grass}, \citenamefont {Vuleti{\'c}}, \citenamefont {Kiesel},\ and\ \citenamefont {Aspelmeyer}}]{delic2020cooling}%
  \BibitemOpen
  \bibfield  {author} {\bibinfo {author} {\bibfnamefont {U.}~\bibnamefont {Deli{\'c}}}, \bibinfo {author} {\bibfnamefont {M.}~\bibnamefont {Reisenbauer}}, \bibinfo {author} {\bibfnamefont {K.}~\bibnamefont {Dare}}, \bibinfo {author} {\bibfnamefont {D.}~\bibnamefont {Grass}}, \bibinfo {author} {\bibfnamefont {V.}~\bibnamefont {Vuleti{\'c}}}, \bibinfo {author} {\bibfnamefont {N.}~\bibnamefont {Kiesel}}, \ and\ \bibinfo {author} {\bibfnamefont {M.}~\bibnamefont {Aspelmeyer}},\ }\href {https://doi.org/10.1126/science.aba3993} {\bibfield  {journal} {\bibinfo  {journal} {Science}\ }\textbf {\bibinfo {volume} {367}},\ \bibinfo {pages} {892} (\bibinfo {year} {2020})}\BibitemShut {NoStop}%
\bibitem [{\citenamefont {Tebbenjohanns}\ \emph {et~al.}(2020)\citenamefont {Tebbenjohanns}, \citenamefont {Frimmer}, \citenamefont {Jain}, \citenamefont {Windey},\ and\ \citenamefont {Novotny}}]{tebbenjohanns2020motional}%
  \BibitemOpen
  \bibfield  {author} {\bibinfo {author} {\bibfnamefont {F.}~\bibnamefont {Tebbenjohanns}}, \bibinfo {author} {\bibfnamefont {M.}~\bibnamefont {Frimmer}}, \bibinfo {author} {\bibfnamefont {V.}~\bibnamefont {Jain}}, \bibinfo {author} {\bibfnamefont {D.}~\bibnamefont {Windey}}, \ and\ \bibinfo {author} {\bibfnamefont {L.}~\bibnamefont {Novotny}},\ }\href {https://doi.org/10.1103/PhysRevLett.124.013603} {\bibfield  {journal} {\bibinfo  {journal} {Physical Review Letters}\ }\textbf {\bibinfo {volume} {124}},\ \bibinfo {pages} {013603} (\bibinfo {year} {2020})}\BibitemShut {NoStop}%
\bibitem [{\citenamefont {Westphal}\ \emph {et~al.}(2021)\citenamefont {Westphal}, \citenamefont {Hepach}, \citenamefont {Pfaff},\ and\ \citenamefont {Aspelmeyer}}]{westphal2021measurement}%
  \BibitemOpen
  \bibfield  {author} {\bibinfo {author} {\bibfnamefont {T.}~\bibnamefont {Westphal}}, \bibinfo {author} {\bibfnamefont {H.}~\bibnamefont {Hepach}}, \bibinfo {author} {\bibfnamefont {J.}~\bibnamefont {Pfaff}}, \ and\ \bibinfo {author} {\bibfnamefont {M.}~\bibnamefont {Aspelmeyer}},\ }\href {https://doi.org/10.1038/s41586-021-03250-7} {\bibfield  {journal} {\bibinfo  {journal} {Nature}\ }\textbf {\bibinfo {volume} {591}},\ \bibinfo {pages} {225} (\bibinfo {year} {2021})}\BibitemShut {NoStop}%
\bibitem [{\citenamefont {Fuchs}\ \emph {et~al.}(2024)\citenamefont {Fuchs}, \citenamefont {Uitenbroek}, \citenamefont {Plugge}, \citenamefont {van Halteren}, \citenamefont {van Soest}, \citenamefont {Vinante}, \citenamefont {Ulbricht},\ and\ \citenamefont {Oosterkamp}}]{fuchs2024measuring}%
  \BibitemOpen
  \bibfield  {author} {\bibinfo {author} {\bibfnamefont {T.~M.}\ \bibnamefont {Fuchs}}, \bibinfo {author} {\bibfnamefont {D.~G.}\ \bibnamefont {Uitenbroek}}, \bibinfo {author} {\bibfnamefont {J.}~\bibnamefont {Plugge}}, \bibinfo {author} {\bibfnamefont {N.}~\bibnamefont {van Halteren}}, \bibinfo {author} {\bibfnamefont {J.-P.}\ \bibnamefont {van Soest}}, \bibinfo {author} {\bibfnamefont {A.}~\bibnamefont {Vinante}}, \bibinfo {author} {\bibfnamefont {H.}~\bibnamefont {Ulbricht}}, \ and\ \bibinfo {author} {\bibfnamefont {T.~H.}\ \bibnamefont {Oosterkamp}},\ }\href {https://doi.org/10.1126/sciadv.adk2949} {\bibfield  {journal} {\bibinfo  {journal} {Science Advances}\ }\textbf {\bibinfo {volume} {10}},\ \bibinfo {pages} {eadk2949} (\bibinfo {year} {2024})}\BibitemShut {NoStop}%
\bibitem [{\citenamefont {Amit}\ \emph {et~al.}(2019)\citenamefont {Amit}, \citenamefont {Margalit}, \citenamefont {Dobkowski}, \citenamefont {Zhou}, \citenamefont {Japha}, \citenamefont {Zimmermann}, \citenamefont {Efremov}, \citenamefont {Narducci}, \citenamefont {Rasel}, \citenamefont {Schleich},\ and\ \citenamefont {Folman}}]{Folman2019_GM}%
  \BibitemOpen
  \bibfield  {author} {\bibinfo {author} {\bibfnamefont {O.}~\bibnamefont {Amit}}, \bibinfo {author} {\bibfnamefont {Y.}~\bibnamefont {Margalit}}, \bibinfo {author} {\bibfnamefont {O.}~\bibnamefont {Dobkowski}}, \bibinfo {author} {\bibfnamefont {Z.}~\bibnamefont {Zhou}}, \bibinfo {author} {\bibfnamefont {Y.}~\bibnamefont {Japha}}, \bibinfo {author} {\bibfnamefont {M.}~\bibnamefont {Zimmermann}}, \bibinfo {author} {\bibfnamefont {M.~A.}\ \bibnamefont {Efremov}}, \bibinfo {author} {\bibfnamefont {F.~A.}\ \bibnamefont {Narducci}}, \bibinfo {author} {\bibfnamefont {E.~M.}\ \bibnamefont {Rasel}}, \bibinfo {author} {\bibfnamefont {W.~P.}\ \bibnamefont {Schleich}}, \ and\ \bibinfo {author} {\bibfnamefont {R.}~\bibnamefont {Folman}},\ }\href {\doibase 10.1103/PhysRevLett.123.083601} {\bibfield  {journal} {\bibinfo  {journal} {Phys. Rev. Lett.}\ }\textbf {\bibinfo {volume} {123}},\ \bibinfo {pages} {083601} (\bibinfo {year} {2019})}\BibitemShut {NoStop}%
\bibitem [{\citenamefont {Margalit}\ \emph {et~al.}(2021)\citenamefont {Margalit}, \citenamefont {Dobkowski}, \citenamefont {Zhou}, \citenamefont {Amit}, \citenamefont {Japha}, \citenamefont {Moukouri}, \citenamefont {Rohrlich}, \citenamefont {Mazumdar}, \citenamefont {Bose}, \citenamefont {Henkel},\ and\ \citenamefont {Folman}}]{SGI_experiment}%
  \BibitemOpen
  \bibfield  {author} {\bibinfo {author} {\bibfnamefont {Y.}~\bibnamefont {Margalit}}, \bibinfo {author} {\bibfnamefont {O.}~\bibnamefont {Dobkowski}}, \bibinfo {author} {\bibfnamefont {Z.}~\bibnamefont {Zhou}}, \bibinfo {author} {\bibfnamefont {O.}~\bibnamefont {Amit}}, \bibinfo {author} {\bibfnamefont {Y.}~\bibnamefont {Japha}}, \bibinfo {author} {\bibfnamefont {S.}~\bibnamefont {Moukouri}}, \bibinfo {author} {\bibfnamefont {D.}~\bibnamefont {Rohrlich}}, \bibinfo {author} {\bibfnamefont {A.}~\bibnamefont {Mazumdar}}, \bibinfo {author} {\bibfnamefont {S.}~\bibnamefont {Bose}}, \bibinfo {author} {\bibfnamefont {C.}~\bibnamefont {Henkel}}, \ and\ \bibinfo {author} {\bibfnamefont {R.}~\bibnamefont {Folman}},\ }\href {https://www.science.org/doi/abs/10.1126/sciadv.abg2879} {\bibfield  {journal} {\bibinfo  {journal} {Science Advances}\ }\textbf {\bibinfo {volume} {7}},\ \bibinfo {pages} {eabg2879} (\bibinfo {year} {2021})}\BibitemShut {NoStop}%
\bibitem [{\citenamefont {March}\ \emph {et~al.}(2023)\citenamefont {March}, \citenamefont {Wood}, \citenamefont {Stephen}, \citenamefont {Fervenza}, \citenamefont {Breeze}, \citenamefont {Mandal}, \citenamefont {Edmonds}, \citenamefont {Twitchen}, \citenamefont {Markham}, \citenamefont {Williams},\ and\ \citenamefont {Morley}}]{March23_GM}%
  \BibitemOpen
  \bibfield  {author} {\bibinfo {author} {\bibfnamefont {J.~E.}\ \bibnamefont {March}}, \bibinfo {author} {\bibfnamefont {B.~D.}\ \bibnamefont {Wood}}, \bibinfo {author} {\bibfnamefont {C.~J.}\ \bibnamefont {Stephen}}, \bibinfo {author} {\bibfnamefont {L.~D.}\ \bibnamefont {Fervenza}}, \bibinfo {author} {\bibfnamefont {B.~G.}\ \bibnamefont {Breeze}}, \bibinfo {author} {\bibfnamefont {S.}~\bibnamefont {Mandal}}, \bibinfo {author} {\bibfnamefont {A.~M.}\ \bibnamefont {Edmonds}}, \bibinfo {author} {\bibfnamefont {D.~J.}\ \bibnamefont {Twitchen}}, \bibinfo {author} {\bibfnamefont {M.~L.}\ \bibnamefont {Markham}}, \bibinfo {author} {\bibfnamefont {O.~A.}\ \bibnamefont {Williams}}, \ and\ \bibinfo {author} {\bibfnamefont {G.~W.}\ \bibnamefont {Morley}},\ }\href {\doibase 10.1103/PhysRevApplied.20.044045} {\bibfield  {journal} {\bibinfo  {journal} {Phys. Rev. Appl.}\ }\textbf {\bibinfo {volume} {20}},\ \bibinfo {pages} {044045} (\bibinfo {year} {2023})}\BibitemShut {NoStop}%
\bibitem [{\citenamefont {Christodoulou}\ and\ \citenamefont {Rovelli}(2019)}]{christodoulou2019possibility}%
  \BibitemOpen
  \bibfield  {author} {\bibinfo {author} {\bibfnamefont {M.}~\bibnamefont {Christodoulou}}\ and\ \bibinfo {author} {\bibfnamefont {C.}~\bibnamefont {Rovelli}},\ }\href {https://doi.org/10.1016/j.physletb.2019.03.015} {\bibfield  {journal} {\bibinfo  {journal} {Physics Letters B}\ }\textbf {\bibinfo {volume} {792}},\ \bibinfo {pages} {64} (\bibinfo {year} {2019})}\BibitemShut {NoStop}%
\bibitem [{\citenamefont {Bennett}\ \emph {et~al.}(1996)\citenamefont {Bennett}, \citenamefont {DiVincenzo}, \citenamefont {Smolin},\ and\ \citenamefont {Wootters}}]{Bennett_1996}%
  \BibitemOpen
  \bibfield  {author} {\bibinfo {author} {\bibfnamefont {C.~H.}\ \bibnamefont {Bennett}}, \bibinfo {author} {\bibfnamefont {D.~P.}\ \bibnamefont {DiVincenzo}}, \bibinfo {author} {\bibfnamefont {J.~A.}\ \bibnamefont {Smolin}}, \ and\ \bibinfo {author} {\bibfnamefont {W.~K.}\ \bibnamefont {Wootters}},\ }\href {\doibase 10.1103/PhysRevA.54.3824} {\bibfield  {journal} {\bibinfo  {journal} {Phys. Rev. A}\ }\textbf {\bibinfo {volume} {54}},\ \bibinfo {pages} {3824} (\bibinfo {year} {1996})}\BibitemShut {NoStop}%
\bibitem [{\citenamefont {Oppenheim}\ and\ \citenamefont {Weller-Davies}(2023)}]{Oppenheim23}%
  \BibitemOpen
  \bibfield  {author} {\bibinfo {author} {\bibfnamefont {J.}~\bibnamefont {Oppenheim}}\ and\ \bibinfo {author} {\bibfnamefont {Z.}~\bibnamefont {Weller-Davies}},\ }\href@noop {} {\  (\bibinfo {year} {2023})},\ \Eprint {http://arxiv.org/abs/2302.07283} {2302.07283 [gr-qc]} \BibitemShut {NoStop}%
\bibitem [{\citenamefont {Carney}\ \emph {et~al.}(2019{\natexlab{b}})\citenamefont {Carney}, \citenamefont {Stamp},\ and\ \citenamefont {Taylor}}]{Carney_2019}%
  \BibitemOpen
  \bibfield  {author} {\bibinfo {author} {\bibfnamefont {D.}~\bibnamefont {Carney}}, \bibinfo {author} {\bibfnamefont {P.~C.~E.}\ \bibnamefont {Stamp}}, \ and\ \bibinfo {author} {\bibfnamefont {J.~M.}\ \bibnamefont {Taylor}},\ }\href {\doibase 10.1088/1361-6382/aaf9ca} {\bibfield  {journal} {\bibinfo  {journal} {Class. Quant. Grav.}\ }\textbf {\bibinfo {volume} {36}},\ \bibinfo {pages} {034001} (\bibinfo {year} {2019}{\natexlab{b}})}\BibitemShut {NoStop}%
\bibitem [{\citenamefont {Danielson}\ \emph {et~al.}(2022)\citenamefont {Danielson}, \citenamefont {Satishchandran},\ and\ \citenamefont {Wald}}]{Danielson:2021egj}%
  \BibitemOpen
  \bibfield  {author} {\bibinfo {author} {\bibfnamefont {D.~L.}\ \bibnamefont {Danielson}}, \bibinfo {author} {\bibfnamefont {G.}~\bibnamefont {Satishchandran}}, \ and\ \bibinfo {author} {\bibfnamefont {R.~M.}\ \bibnamefont {Wald}},\ }\href {\doibase 10.1103/PhysRevD.105.086001} {\bibfield  {journal} {\bibinfo  {journal} {Phys. Rev. D}\ }\textbf {\bibinfo {volume} {105}},\ \bibinfo {pages} {086001} (\bibinfo {year} {2022})}\BibitemShut {NoStop}%
\bibitem [{\citenamefont {Christodoulou}\ \emph {et~al.}(2023{\natexlab{a}})\citenamefont {Christodoulou}, \citenamefont {Di~Biagio}, \citenamefont {Aspelmeyer}, \citenamefont {Brukner}, \citenamefont {Rovelli},\ and\ \citenamefont {Howl}}]{Christodoulou:2022vte}%
  \BibitemOpen
  \bibfield  {author} {\bibinfo {author} {\bibfnamefont {M.}~\bibnamefont {Christodoulou}}, \bibinfo {author} {\bibfnamefont {A.}~\bibnamefont {Di~Biagio}}, \bibinfo {author} {\bibfnamefont {M.}~\bibnamefont {Aspelmeyer}}, \bibinfo {author} {\bibfnamefont {{\v{C}}.}~\bibnamefont {Brukner}}, \bibinfo {author} {\bibfnamefont {C.}~\bibnamefont {Rovelli}}, \ and\ \bibinfo {author} {\bibfnamefont {R.}~\bibnamefont {Howl}},\ }\href {\doibase 10.1103/PhysRevLett.130.100202} {\bibfield  {journal} {\bibinfo  {journal} {Phys. Rev. Lett.}\ }\textbf {\bibinfo {volume} {130}},\ \bibinfo {pages} {100202} (\bibinfo {year} {2023}{\natexlab{a}})}\BibitemShut {NoStop}%
\bibitem [{\citenamefont {Rufo}\ \emph {et~al.}(2025)\citenamefont {Rufo}, \citenamefont {Mazumdar},\ and\ \citenamefont {Sab\'\i{}n}}]{Rufo:2024ulr}%
  \BibitemOpen
  \bibfield  {author} {\bibinfo {author} {\bibfnamefont {P.~G.~C.}\ \bibnamefont {Rufo}}, \bibinfo {author} {\bibfnamefont {A.}~\bibnamefont {Mazumdar}}, \ and\ \bibinfo {author} {\bibfnamefont {C.}~\bibnamefont {Sab\'\i{}n}},\ }\href {\doibase 10.1103/PhysRevA.111.022444} {\bibfield  {journal} {\bibinfo  {journal} {Phys. Rev. A}\ }\textbf {\bibinfo {volume} {111}},\ \bibinfo {pages} {022444} (\bibinfo {year} {2025})}\BibitemShut {NoStop}%
\bibitem [{\citenamefont {Scala}\ \emph {et~al.}(2013)\citenamefont {Scala}, \citenamefont {Kim}, \citenamefont {Morley}, \citenamefont {Barker},\ and\ \citenamefont {Bose}}]{Scala13_GM}%
  \BibitemOpen
  \bibfield  {author} {\bibinfo {author} {\bibfnamefont {M.}~\bibnamefont {Scala}}, \bibinfo {author} {\bibfnamefont {M.~S.}\ \bibnamefont {Kim}}, \bibinfo {author} {\bibfnamefont {G.~W.}\ \bibnamefont {Morley}}, \bibinfo {author} {\bibfnamefont {P.~F.}\ \bibnamefont {Barker}}, \ and\ \bibinfo {author} {\bibfnamefont {S.}~\bibnamefont {Bose}},\ }\href {http://link.aps.org/doi/10.1103/PhysRevLett.111.180403} {\bibfield  {journal} {\bibinfo  {journal} {Phys. Rev. Lett.}\ }\textbf {\bibinfo {volume} {111}},\ \bibinfo {pages} {180403} (\bibinfo {year} {2013})}\BibitemShut {NoStop}%
\bibitem [{\citenamefont {Yin}\ \emph {et~al.}(2013)\citenamefont {Yin}, \citenamefont {Li}, \citenamefont {Zhang},\ and\ \citenamefont {Duan}}]{Duan2013_GM}%
  \BibitemOpen
  \bibfield  {author} {\bibinfo {author} {\bibfnamefont {Z.-q.}\ \bibnamefont {Yin}}, \bibinfo {author} {\bibfnamefont {T.}~\bibnamefont {Li}}, \bibinfo {author} {\bibfnamefont {X.}~\bibnamefont {Zhang}}, \ and\ \bibinfo {author} {\bibfnamefont {L.~M.}\ \bibnamefont {Duan}},\ }\href {\doibase 10.1103/PhysRevA.88.033614} {\bibfield  {journal} {\bibinfo  {journal} {Phys. Rev. A}\ }\textbf {\bibinfo {volume} {88}},\ \bibinfo {pages} {033614} (\bibinfo {year} {2013})}\BibitemShut {NoStop}%
\bibitem [{\citenamefont {Kovachy}\ \emph {et~al.}(2015)\citenamefont {Kovachy}, \citenamefont {Asenbaum}, \citenamefont {Overstreet}, \citenamefont {Donnelly}, \citenamefont {Dickerson}, \citenamefont {Sugarbaker}, \citenamefont {M.},\ and\ \citenamefont {Kasevich}}]{Kasevich2015_GM}%
  \BibitemOpen
  \bibfield  {author} {\bibinfo {author} {\bibfnamefont {T.}~\bibnamefont {Kovachy}}, \bibinfo {author} {\bibfnamefont {P.}~\bibnamefont {Asenbaum}}, \bibinfo {author} {\bibfnamefont {C.}~\bibnamefont {Overstreet}}, \bibinfo {author} {\bibfnamefont {C.~A.}\ \bibnamefont {Donnelly}}, \bibinfo {author} {\bibfnamefont {S.~M.}\ \bibnamefont {Dickerson}}, \bibinfo {author} {\bibfnamefont {A.}~\bibnamefont {Sugarbaker}}, \bibinfo {author} {\bibfnamefont {H.~J.}\ \bibnamefont {M.}}, \ and\ \bibinfo {author} {\bibfnamefont {M.~A.}\ \bibnamefont {Kasevich}},\ }\href {\doibase 10.1038/nature16155} {\bibfield  {journal} {\bibinfo  {journal} {Nature}\ }\textbf {\bibinfo {volume} {528}},\ \bibinfo {pages} {530} (\bibinfo {year} {2015})}\BibitemShut {NoStop}%
\bibitem [{\citenamefont {Fein}\ \emph {et~al.}(2019)\citenamefont {Fein}, \citenamefont {Geyer}, \citenamefont {Zwick}, \citenamefont {Kiałka}, \citenamefont {Pedalino}, \citenamefont {Mayor}, \citenamefont {Gerlich},\ and\ \citenamefont {Arndt}}]{arndt}%
  \BibitemOpen
  \bibfield  {author} {\bibinfo {author} {\bibfnamefont {Y.}~\bibnamefont {Fein}}, \bibinfo {author} {\bibfnamefont {P.}~\bibnamefont {Geyer}}, \bibinfo {author} {\bibfnamefont {P.}~\bibnamefont {Zwick}}, \bibinfo {author} {\bibfnamefont {F.}~\bibnamefont {Kiałka}}, \bibinfo {author} {\bibfnamefont {S.}~\bibnamefont {Pedalino}}, \bibinfo {author} {\bibfnamefont {M.}~\bibnamefont {Mayor}}, \bibinfo {author} {\bibfnamefont {S.}~\bibnamefont {Gerlich}}, \ and\ \bibinfo {author} {\bibfnamefont {M.}~\bibnamefont {Arndt}},\ }\href {\doibase 10.1038/s41567-019-0663-9} {\bibfield  {journal} {\bibinfo  {journal} {Nature Phys.}\ }\textbf {\bibinfo {volume} {15}},\ \bibinfo {pages} {1242} (\bibinfo {year} {2019})}\BibitemShut {NoStop}%
\bibitem [{\citenamefont {Bonvin}\ \emph {et~al.}(2024)\citenamefont {Bonvin}, \citenamefont {Devaud}, \citenamefont {Rossi}, \citenamefont {Militaru}, \citenamefont {Dania}, \citenamefont {Bykov}, \citenamefont {Romero-Isart}, \citenamefont {Northup}, \citenamefont {Novotny},\ and\ \citenamefont {Frimmer}}]{StateExpansion2024}%
  \BibitemOpen
  \bibfield  {author} {\bibinfo {author} {\bibfnamefont {E.}~\bibnamefont {Bonvin}}, \bibinfo {author} {\bibfnamefont {L.}~\bibnamefont {Devaud}}, \bibinfo {author} {\bibfnamefont {M.}~\bibnamefont {Rossi}}, \bibinfo {author} {\bibfnamefont {A.}~\bibnamefont {Militaru}}, \bibinfo {author} {\bibfnamefont {L.}~\bibnamefont {Dania}}, \bibinfo {author} {\bibfnamefont {D.~S.}\ \bibnamefont {Bykov}}, \bibinfo {author} {\bibfnamefont {O.}~\bibnamefont {Romero-Isart}}, \bibinfo {author} {\bibfnamefont {T.~E.}\ \bibnamefont {Northup}}, \bibinfo {author} {\bibfnamefont {L.}~\bibnamefont {Novotny}}, \ and\ \bibinfo {author} {\bibfnamefont {M.}~\bibnamefont {Frimmer}},\ }\href {\doibase 10.1103/PhysRevLett.132.253602} {\bibfield  {journal} {\bibinfo  {journal} {Phys. Rev. Lett.}\ }\textbf {\bibinfo {volume} {132}},\ \bibinfo {pages} {253602} (\bibinfo {year} {2024})}\BibitemShut {NoStop}%
\bibitem [{\citenamefont {Weiss}\ \emph {et~al.}(2021)\citenamefont {Weiss}, \citenamefont {Roda-Llordes}, \citenamefont {Torrontegui}, \citenamefont {Aspelmeyer},\ and\ \citenamefont {Romero-Isart}}]{StateExpansion2021}%
  \BibitemOpen
  \bibfield  {author} {\bibinfo {author} {\bibfnamefont {T.}~\bibnamefont {Weiss}}, \bibinfo {author} {\bibfnamefont {M.}~\bibnamefont {Roda-Llordes}}, \bibinfo {author} {\bibfnamefont {E.}~\bibnamefont {Torrontegui}}, \bibinfo {author} {\bibfnamefont {M.}~\bibnamefont {Aspelmeyer}}, \ and\ \bibinfo {author} {\bibfnamefont {O.}~\bibnamefont {Romero-Isart}},\ }\href {\doibase 10.1103/PhysRevLett.127.023601} {\bibfield  {journal} {\bibinfo  {journal} {Phys. Rev. Lett.}\ }\textbf {\bibinfo {volume} {127}},\ \bibinfo {pages} {023601} (\bibinfo {year} {2021})}\BibitemShut {NoStop}%
\bibitem [{\citenamefont {Neumeier}\ \emph {et~al.}(2024)\citenamefont {Neumeier}, \citenamefont {Ciampini}, \citenamefont {Romero-Isart}, \citenamefont {Aspelmeyer},\ and\ \citenamefont {Kiesel}}]{KieselPNAS}%
  \BibitemOpen
  \bibfield  {author} {\bibinfo {author} {\bibfnamefont {L.}~\bibnamefont {Neumeier}}, \bibinfo {author} {\bibfnamefont {M.~A.}\ \bibnamefont {Ciampini}}, \bibinfo {author} {\bibfnamefont {O.}~\bibnamefont {Romero-Isart}}, \bibinfo {author} {\bibfnamefont {M.}~\bibnamefont {Aspelmeyer}}, \ and\ \bibinfo {author} {\bibfnamefont {N.}~\bibnamefont {Kiesel}},\ }\href {\doibase 10.1073/pnas.2306953121} {\bibfield  {journal} {\bibinfo  {journal} {Proceedings of the National Academy of Sciences}\ }\textbf {\bibinfo {volume} {121}},\ \bibinfo {pages} {e2306953121} (\bibinfo {year} {2024})}\BibitemShut {NoStop}%
\bibitem [{\citenamefont {Roda-Llordes}\ \emph {et~al.}(2024)\citenamefont {Roda-Llordes}, \citenamefont {Riera-Campeny}, \citenamefont {Candoli}, \citenamefont {Grochowski},\ and\ \citenamefont {Romero-Isart}}]{OriolDoubleWell2024}%
  \BibitemOpen
  \bibfield  {author} {\bibinfo {author} {\bibfnamefont {M.}~\bibnamefont {Roda-Llordes}}, \bibinfo {author} {\bibfnamefont {A.}~\bibnamefont {Riera-Campeny}}, \bibinfo {author} {\bibfnamefont {D.}~\bibnamefont {Candoli}}, \bibinfo {author} {\bibfnamefont {P.~T.}\ \bibnamefont {Grochowski}}, \ and\ \bibinfo {author} {\bibfnamefont {O.}~\bibnamefont {Romero-Isart}},\ }\href {\doibase 10.1103/PhysRevLett.132.023601} {\bibfield  {journal} {\bibinfo  {journal} {Phys. Rev. Lett.}\ }\textbf {\bibinfo {volume} {132}},\ \bibinfo {pages} {023601} (\bibinfo {year} {2024})}\BibitemShut {NoStop}%
\bibitem [{\citenamefont {Bateman}\ \emph {et~al.}(2014)\citenamefont {Bateman}, \citenamefont {Nimmrichter}, \citenamefont {Hornberger},\ and\ \citenamefont {Ulbricht}}]{TalbotLau2014}%
  \BibitemOpen
  \bibfield  {author} {\bibinfo {author} {\bibfnamefont {J.}~\bibnamefont {Bateman}}, \bibinfo {author} {\bibfnamefont {S.}~\bibnamefont {Nimmrichter}}, \bibinfo {author} {\bibfnamefont {K.}~\bibnamefont {Hornberger}}, \ and\ \bibinfo {author} {\bibfnamefont {H.}~\bibnamefont {Ulbricht}},\ }\href {http://dx.doi.org/10.1038/ncomms5788} {\bibfield  {journal} {\bibinfo  {journal} {Nature Comms.}\ }\textbf {\bibinfo {volume} {5}},\ \bibinfo {pages} {4788} (\bibinfo {year} {2014})}\BibitemShut {NoStop}%
\bibitem [{\citenamefont {Datta}\ and\ \citenamefont {Miao}(2021)}]{DattaMiao2021}%
  \BibitemOpen
  \bibfield  {author} {\bibinfo {author} {\bibfnamefont {A.}~\bibnamefont {Datta}}\ and\ \bibinfo {author} {\bibfnamefont {H.}~\bibnamefont {Miao}},\ }\href {\doibase 10.1088/2058-9565/ac1adf} {\bibfield  {journal} {\bibinfo  {journal} {Quantum Science and Technology}\ }\textbf {\bibinfo {volume} {6}},\ \bibinfo {pages} {045014} (\bibinfo {year} {2021})}\BibitemShut {NoStop}%
\bibitem [{\citenamefont {Cosco}\ \emph {et~al.}(2021)\citenamefont {Cosco}, \citenamefont {Pedernales},\ and\ \citenamefont {Plenio}}]{Plenio2021}%
  \BibitemOpen
  \bibfield  {author} {\bibinfo {author} {\bibfnamefont {F.}~\bibnamefont {Cosco}}, \bibinfo {author} {\bibfnamefont {J.~S.}\ \bibnamefont {Pedernales}}, \ and\ \bibinfo {author} {\bibfnamefont {M.~B.}\ \bibnamefont {Plenio}},\ }\href {\doibase 10.1103/PhysRevA.103.L061501} {\bibfield  {journal} {\bibinfo  {journal} {Phys. Rev. A}\ }\textbf {\bibinfo {volume} {103}},\ \bibinfo {pages} {L061501} (\bibinfo {year} {2021})}\BibitemShut {NoStop}%
\bibitem [{\citenamefont {Kuhn}\ \emph {et~al.}(2017)\citenamefont {Kuhn}, \citenamefont {Kosloff}, \citenamefont {Stickler}, \citenamefont {Patolsky}, \citenamefont {Hornberger}, \citenamefont {Arndt},\ and\ \citenamefont {Millen}}]{KuhnRotation2017}%
  \BibitemOpen
  \bibfield  {author} {\bibinfo {author} {\bibfnamefont {S.}~\bibnamefont {Kuhn}}, \bibinfo {author} {\bibfnamefont {A.}~\bibnamefont {Kosloff}}, \bibinfo {author} {\bibfnamefont {B.~A.}\ \bibnamefont {Stickler}}, \bibinfo {author} {\bibfnamefont {F.}~\bibnamefont {Patolsky}}, \bibinfo {author} {\bibfnamefont {K.}~\bibnamefont {Hornberger}}, \bibinfo {author} {\bibfnamefont {M.}~\bibnamefont {Arndt}}, \ and\ \bibinfo {author} {\bibfnamefont {J.}~\bibnamefont {Millen}},\ }\href {\doibase 10.1364/OPTICA.4.000356} {\bibfield  {journal} {\bibinfo  {journal} {Optica}\ }\textbf {\bibinfo {volume} {4}},\ \bibinfo {pages} {356} (\bibinfo {year} {2017})}\BibitemShut {NoStop}%
\bibitem [{\citenamefont {Rusconi}\ \emph {et~al.}(2022)\citenamefont {Rusconi}, \citenamefont {Perdriat}, \citenamefont {H\'etet}, \citenamefont {Romero-Isart},\ and\ \citenamefont {Stickler}}]{SticklerRotors2022}%
  \BibitemOpen
  \bibfield  {author} {\bibinfo {author} {\bibfnamefont {C.~C.}\ \bibnamefont {Rusconi}}, \bibinfo {author} {\bibfnamefont {M.}~\bibnamefont {Perdriat}}, \bibinfo {author} {\bibfnamefont {G.}~\bibnamefont {H\'etet}}, \bibinfo {author} {\bibfnamefont {O.}~\bibnamefont {Romero-Isart}}, \ and\ \bibinfo {author} {\bibfnamefont {B.~A.}\ \bibnamefont {Stickler}},\ }\href {\doibase 10.1103/PhysRevLett.129.093605} {\bibfield  {journal} {\bibinfo  {journal} {Phys. Rev. Lett.}\ }\textbf {\bibinfo {volume} {129}},\ \bibinfo {pages} {093605} (\bibinfo {year} {2022})}\BibitemShut {NoStop}%
\bibitem [{\citenamefont {Moore}\ \emph {et~al.}(2014)\citenamefont {Moore}, \citenamefont {Rider},\ and\ \citenamefont {Gratta}}]{Moore14_mcp}%
  \BibitemOpen
  \bibfield  {author} {\bibinfo {author} {\bibfnamefont {D.~C.}\ \bibnamefont {Moore}}, \bibinfo {author} {\bibfnamefont {A.~D.}\ \bibnamefont {Rider}}, \ and\ \bibinfo {author} {\bibfnamefont {G.}~\bibnamefont {Gratta}},\ }\href {\doibase 10.1103/PhysRevLett.113.251801} {\bibfield  {journal} {\bibinfo  {journal} {Phys. Rev. Lett.}\ }\textbf {\bibinfo {volume} {113}},\ \bibinfo {pages} {251801} (\bibinfo {year} {2014})}\BibitemShut {NoStop}%
\bibitem [{\citenamefont {van~de Kamp}\ \emph {et~al.}(2020)\citenamefont {van~de Kamp}, \citenamefont {Marshman}, \citenamefont {Bose},\ and\ \citenamefont {Mazumdar}}]{vandeKamp:2020rqh}%
  \BibitemOpen
  \bibfield  {author} {\bibinfo {author} {\bibfnamefont {T.~W.}\ \bibnamefont {van~de Kamp}}, \bibinfo {author} {\bibfnamefont {R.~J.}\ \bibnamefont {Marshman}}, \bibinfo {author} {\bibfnamefont {S.}~\bibnamefont {Bose}}, \ and\ \bibinfo {author} {\bibfnamefont {A.}~\bibnamefont {Mazumdar}},\ }\href {\doibase 10.1103/PhysRevA.102.062807} {\bibfield  {journal} {\bibinfo  {journal} {Phys. Rev. A}\ }\textbf {\bibinfo {volume} {102}},\ \bibinfo {pages} {062807} (\bibinfo {year} {2020})}\BibitemShut {NoStop}%
\bibitem [{\citenamefont {Neukirch}\ \emph {et~al.}(2015)\citenamefont {Neukirch}, \citenamefont {von Haartman}, \citenamefont {Rosenholm},\ and\ \citenamefont {Vamivakas}}]{VamivakasNV}%
  \BibitemOpen
  \bibfield  {author} {\bibinfo {author} {\bibfnamefont {L.~P.}\ \bibnamefont {Neukirch}}, \bibinfo {author} {\bibfnamefont {E.}~\bibnamefont {von Haartman}}, \bibinfo {author} {\bibfnamefont {J.~M.}\ \bibnamefont {Rosenholm}}, \ and\ \bibinfo {author} {\bibfnamefont {A.~N.}\ \bibnamefont {Vamivakas}},\ }\href {\doibase 10.1038/nphoton.2015.162} {\bibfield  {journal} {\bibinfo  {journal} {Nat. Photon.}\ }\textbf {\bibinfo {volume} {9}},\ \bibinfo {pages} {653} (\bibinfo {year} {2015})}\BibitemShut {NoStop}%
\bibitem [{\citenamefont {Hoang}\ \emph {et~al.}(2016)\citenamefont {Hoang}, \citenamefont {Ahn}, \citenamefont {Bang},\ and\ \citenamefont {Li}}]{TongcangLiNV2016}%
  \BibitemOpen
  \bibfield  {author} {\bibinfo {author} {\bibfnamefont {T.~M.}\ \bibnamefont {Hoang}}, \bibinfo {author} {\bibfnamefont {J.}~\bibnamefont {Ahn}}, \bibinfo {author} {\bibfnamefont {J.}~\bibnamefont {Bang}}, \ and\ \bibinfo {author} {\bibfnamefont {T.}~\bibnamefont {Li}},\ }\href {\doibase 10.1038/ncomms12250} {\bibfield  {journal} {\bibinfo  {journal} {Nature Communications}\ }\textbf {\bibinfo {volume} {7}},\ \bibinfo {pages} {12250} (\bibinfo {year} {2016})}\BibitemShut {NoStop}%
\bibitem [{\citenamefont {Kuhlicke}\ \emph {et~al.}(2014)\citenamefont {Kuhlicke}, \citenamefont {Schell}, \citenamefont {Zoll},\ and\ \citenamefont {Benson}}]{BensonNV2014}%
  \BibitemOpen
  \bibfield  {author} {\bibinfo {author} {\bibfnamefont {A.}~\bibnamefont {Kuhlicke}}, \bibinfo {author} {\bibfnamefont {A.~W.}\ \bibnamefont {Schell}}, \bibinfo {author} {\bibfnamefont {J.}~\bibnamefont {Zoll}}, \ and\ \bibinfo {author} {\bibfnamefont {O.}~\bibnamefont {Benson}},\ }\href {\doibase 10.1063/1.4893575} {\bibfield  {journal} {\bibinfo  {journal} {Applied Physics Letters}\ }\textbf {\bibinfo {volume} {105}},\ \bibinfo {pages} {073101} (\bibinfo {year} {2014})},\ \Eprint {http://arxiv.org/abs/https://pubs.aip.org/aip/apl/article-pdf/doi/10.1063/1.4893575/14316710/073101\_1\_online.pdf} {https://pubs.aip.org/aip/apl/article-pdf/doi/10.1063/1.4893575/14316710/073101\_1\_online.pdf} \BibitemShut {NoStop}%
\bibitem [{\citenamefont {Delord}\ \emph {et~al.}(2018)\citenamefont {Delord}, \citenamefont {Huillery}, \citenamefont {Schwab}, \citenamefont {Nicolas}, \citenamefont {Lecordier},\ and\ \citenamefont {H\'etet}}]{DelordNV2018}%
  \BibitemOpen
  \bibfield  {author} {\bibinfo {author} {\bibfnamefont {T.}~\bibnamefont {Delord}}, \bibinfo {author} {\bibfnamefont {P.}~\bibnamefont {Huillery}}, \bibinfo {author} {\bibfnamefont {L.}~\bibnamefont {Schwab}}, \bibinfo {author} {\bibfnamefont {L.}~\bibnamefont {Nicolas}}, \bibinfo {author} {\bibfnamefont {L.}~\bibnamefont {Lecordier}}, \ and\ \bibinfo {author} {\bibfnamefont {G.}~\bibnamefont {H\'etet}},\ }\href {\doibase 10.1103/PhysRevLett.121.053602} {\bibfield  {journal} {\bibinfo  {journal} {Phys. Rev. Lett.}\ }\textbf {\bibinfo {volume} {121}},\ \bibinfo {pages} {053602} (\bibinfo {year} {2018})}\BibitemShut {NoStop}%
\bibitem [{\citenamefont {Delord}\ \emph {et~al.}(2020)\citenamefont {Delord}, \citenamefont {Huillery}, \citenamefont {Nicolas},\ and\ \citenamefont {Hétet}}]{DelordNature2020}%
  \BibitemOpen
  \bibfield  {author} {\bibinfo {author} {\bibfnamefont {T.}~\bibnamefont {Delord}}, \bibinfo {author} {\bibfnamefont {P.}~\bibnamefont {Huillery}}, \bibinfo {author} {\bibfnamefont {L.}~\bibnamefont {Nicolas}}, \ and\ \bibinfo {author} {\bibfnamefont {G.}~\bibnamefont {Hétet}},\ }\href {\doibase 10.1038/s41586-020-2133-z} {\bibfield  {journal} {\bibinfo  {journal} {Nature}\ }\textbf {\bibinfo {volume} {580}},\ \bibinfo {pages} {56} (\bibinfo {year} {2020})}\BibitemShut {NoStop}%
\bibitem [{\citenamefont {Jin}\ \emph {et~al.}(2024)\citenamefont {Jin}, \citenamefont {Shen}, \citenamefont {Ju}, \citenamefont {Gao}, \citenamefont {Zu}, \citenamefont {Grine},\ and\ \citenamefont {Li}}]{TongcangLiNV2024}%
  \BibitemOpen
  \bibfield  {author} {\bibinfo {author} {\bibfnamefont {Y.}~\bibnamefont {Jin}}, \bibinfo {author} {\bibfnamefont {K.}~\bibnamefont {Shen}}, \bibinfo {author} {\bibfnamefont {P.}~\bibnamefont {Ju}}, \bibinfo {author} {\bibfnamefont {X.}~\bibnamefont {Gao}}, \bibinfo {author} {\bibfnamefont {C.}~\bibnamefont {Zu}}, \bibinfo {author} {\bibfnamefont {A.~J.}\ \bibnamefont {Grine}}, \ and\ \bibinfo {author} {\bibfnamefont {T.}~\bibnamefont {Li}},\ }\href {\doibase 10.1038/s41467-024-49175-3} {\bibfield  {journal} {\bibinfo  {journal} {Nature Communications}\ }\textbf {\bibinfo {volume} {15}},\ \bibinfo {pages} {5063} (\bibinfo {year} {2024})}\BibitemShut {NoStop}%
\bibitem [{\citenamefont {Conangla}\ \emph {et~al.}(2018)\citenamefont {Conangla}, \citenamefont {Schell}, \citenamefont {Rica},\ and\ \citenamefont {Quidant}}]{Conangla2018}%
  \BibitemOpen
  \bibfield  {author} {\bibinfo {author} {\bibfnamefont {G.~P.}\ \bibnamefont {Conangla}}, \bibinfo {author} {\bibfnamefont {A.~W.}\ \bibnamefont {Schell}}, \bibinfo {author} {\bibfnamefont {R.~A.}\ \bibnamefont {Rica}}, \ and\ \bibinfo {author} {\bibfnamefont {R.}~\bibnamefont {Quidant}},\ }\href@noop {} {\bibfield  {journal} {\bibinfo  {journal} {Nano Lett.}\ }\textbf {\bibinfo {volume} {18}},\ \bibinfo {pages} {3956} (\bibinfo {year} {2018})}\BibitemShut {NoStop}%
\bibitem [{\citenamefont {Rahman}\ \emph {et~al.}(2016)\citenamefont {Rahman}, \citenamefont {Frangeskou}, \citenamefont {Kim}, \citenamefont {Bose}, \citenamefont {Morley},\ and\ \citenamefont {Barker}}]{RahmanBurning2016}%
  \BibitemOpen
  \bibfield  {author} {\bibinfo {author} {\bibfnamefont {A.~T. M.~A.}\ \bibnamefont {Rahman}}, \bibinfo {author} {\bibfnamefont {A.~C.}\ \bibnamefont {Frangeskou}}, \bibinfo {author} {\bibfnamefont {M.~S.}\ \bibnamefont {Kim}}, \bibinfo {author} {\bibfnamefont {S.}~\bibnamefont {Bose}}, \bibinfo {author} {\bibfnamefont {G.~W.}\ \bibnamefont {Morley}}, \ and\ \bibinfo {author} {\bibfnamefont {P.~F.}\ \bibnamefont {Barker}},\ }\href {\doibase 10.1038/srep21633} {\bibfield  {journal} {\bibinfo  {journal} {Scientific Reports}\ }\textbf {\bibinfo {volume} {6}},\ \bibinfo {pages} {21633} (\bibinfo {year} {2016})}\BibitemShut {NoStop}%
\bibitem [{\citenamefont {Frangeskou}\ \emph {et~al.}(2018)\citenamefont {Frangeskou}, \citenamefont {Rahman}, \citenamefont {Gines}, \citenamefont {Mandal}, \citenamefont {Williams}, \citenamefont {Barker},\ and\ \citenamefont {Morley}}]{Frangeskou18_GM}%
  \BibitemOpen
  \bibfield  {author} {\bibinfo {author} {\bibfnamefont {A.~C.}\ \bibnamefont {Frangeskou}}, \bibinfo {author} {\bibfnamefont {A.~T. M.~A.}\ \bibnamefont {Rahman}}, \bibinfo {author} {\bibfnamefont {L.}~\bibnamefont {Gines}}, \bibinfo {author} {\bibfnamefont {S.}~\bibnamefont {Mandal}}, \bibinfo {author} {\bibfnamefont {O.~A.}\ \bibnamefont {Williams}}, \bibinfo {author} {\bibfnamefont {P.~F.}\ \bibnamefont {Barker}}, \ and\ \bibinfo {author} {\bibfnamefont {G.~W.}\ \bibnamefont {Morley}},\ }\href {\doibase 10.1088/1367-2630/aab700} {\bibfield  {journal} {\bibinfo  {journal} {New Journal of Physics}\ }\textbf {\bibinfo {volume} {20}},\ \bibinfo {pages} {043016} (\bibinfo {year} {2018})}\BibitemShut {NoStop}%
\bibitem [{\citenamefont {Di\'osi}\ and\ \citenamefont {Halliwell}(1998)}]{diosi1998coupling}%
  \BibitemOpen
  \bibfield  {author} {\bibinfo {author} {\bibfnamefont {L.}~\bibnamefont {Di\'osi}}\ and\ \bibinfo {author} {\bibfnamefont {J.~J.}\ \bibnamefont {Halliwell}},\ }\href {\doibase 10.1103/PhysRevLett.81.2846} {\bibfield  {journal} {\bibinfo  {journal} {Phys. Rev. Lett.}\ }\textbf {\bibinfo {volume} {81}},\ \bibinfo {pages} {2846} (\bibinfo {year} {1998})}\BibitemShut {NoStop}%
\bibitem [{\citenamefont {Kafri}\ \emph {et~al.}(2014)\citenamefont {Kafri}, \citenamefont {Taylor},\ and\ \citenamefont {Milburn}}]{kafri2014classical}%
  \BibitemOpen
  \bibfield  {author} {\bibinfo {author} {\bibfnamefont {D.}~\bibnamefont {Kafri}}, \bibinfo {author} {\bibfnamefont {J.~M.}\ \bibnamefont {Taylor}}, \ and\ \bibinfo {author} {\bibfnamefont {G.~J.}\ \bibnamefont {Milburn}},\ }\href {\doibase 10.1088/1367-2630/16/6/065020} {\bibfield  {journal} {\bibinfo  {journal} {New Journal of Physics}\ }\textbf {\bibinfo {volume} {16}},\ \bibinfo {pages} {065020} (\bibinfo {year} {2014})}\BibitemShut {NoStop}%
\bibitem [{\citenamefont {Tilloy}\ and\ \citenamefont {Di{\'o}si}(2016)}]{tilloy2016sourcing}%
  \BibitemOpen
  \bibfield  {author} {\bibinfo {author} {\bibfnamefont {A.}~\bibnamefont {Tilloy}}\ and\ \bibinfo {author} {\bibfnamefont {L.}~\bibnamefont {Di{\'o}si}},\ }\href {https://doi.org/10.1103/PhysRevD.93.024026} {\bibfield  {journal} {\bibinfo  {journal} {Physical Review D}\ }\textbf {\bibinfo {volume} {93}},\ \bibinfo {pages} {024026} (\bibinfo {year} {2016})}\BibitemShut {NoStop}%
\bibitem [{\citenamefont {Oppenheim}\ and\ \citenamefont {Weller-Davies}(2022)}]{oppenheim2022constraints}%
  \BibitemOpen
  \bibfield  {author} {\bibinfo {author} {\bibfnamefont {J.}~\bibnamefont {Oppenheim}}\ and\ \bibinfo {author} {\bibfnamefont {Z.}~\bibnamefont {Weller-Davies}},\ }\href {https://doi.org/10.1007/JHEP02(2022)080} {\bibfield  {journal} {\bibinfo  {journal} {Journal of High Energy Physics}\ }\textbf {\bibinfo {volume} {2022}},\ \bibinfo {pages} {1} (\bibinfo {year} {2022})}\BibitemShut {NoStop}%
\bibitem [{\citenamefont {Oppenheim}(2023)}]{OppenheimPRX_GM}%
  \BibitemOpen
  \bibfield  {author} {\bibinfo {author} {\bibfnamefont {J.}~\bibnamefont {Oppenheim}},\ }\href {\doibase 10.1103/PhysRevX.13.041040} {\bibfield  {journal} {\bibinfo  {journal} {Phys. Rev. X}\ }\textbf {\bibinfo {volume} {13}},\ \bibinfo {pages} {041040} (\bibinfo {year} {2023})}\BibitemShut {NoStop}%
\bibitem [{\citenamefont {Christodoulou}\ \emph {et~al.}(2023{\natexlab{b}})\citenamefont {Christodoulou}, \citenamefont {Di~Biagio}, \citenamefont {Howl},\ and\ \citenamefont {Rovelli}}]{Christodoulou2023}%
  \BibitemOpen
  \bibfield  {author} {\bibinfo {author} {\bibfnamefont {M.}~\bibnamefont {Christodoulou}}, \bibinfo {author} {\bibfnamefont {A.}~\bibnamefont {Di~Biagio}}, \bibinfo {author} {\bibfnamefont {R.}~\bibnamefont {Howl}}, \ and\ \bibinfo {author} {\bibfnamefont {C.}~\bibnamefont {Rovelli}},\ }\href {\doibase 10.1088/1361-6382/acb0aa} {\bibfield  {journal} {\bibinfo  {journal} {Classical and Quantum Gravity}\ }\textbf {\bibinfo {volume} {40}},\ \bibinfo {pages} {047001} (\bibinfo {year} {2023}{\natexlab{b}})}\BibitemShut {NoStop}%
\bibitem [{\citenamefont {Ghirardi}\ \emph {et~al.}(1986)\citenamefont {Ghirardi}, \citenamefont {Rimini},\ and\ \citenamefont {Weber}}]{GRW86}%
  \BibitemOpen
  \bibfield  {author} {\bibinfo {author} {\bibfnamefont {G.~C.}\ \bibnamefont {Ghirardi}}, \bibinfo {author} {\bibfnamefont {A.}~\bibnamefont {Rimini}}, \ and\ \bibinfo {author} {\bibfnamefont {T.}~\bibnamefont {Weber}},\ }\href {\doibase 10.1103/PhysRevD.34.470} {\bibfield  {journal} {\bibinfo  {journal} {Phys. Rev. D}\ }\textbf {\bibinfo {volume} {34}},\ \bibinfo {pages} {470} (\bibinfo {year} {1986})}\BibitemShut {NoStop}%
\bibitem [{\citenamefont {Diósi}(1987)}]{Diosi1987}%
  \BibitemOpen
  \bibfield  {author} {\bibinfo {author} {\bibfnamefont {L.}~\bibnamefont {Diósi}},\ }\href {\doibase https://doi.org/10.1016/0375-9601(87)90681-5} {\bibfield  {journal} {\bibinfo  {journal} {Physics Letters A}\ }\textbf {\bibinfo {volume} {120}},\ \bibinfo {pages} {377} (\bibinfo {year} {1987})}\BibitemShut {NoStop}%
\bibitem [{\citenamefont {Penrose}(1996)}]{Penrose96}%
  \BibitemOpen
  \bibfield  {author} {\bibinfo {author} {\bibfnamefont {R.}~\bibnamefont {Penrose}},\ }\href {\doibase https://doi.org/10.1007/BF02105068} {\bibfield  {journal} {\bibinfo  {journal} {Gen. Relat. Gravit.}\ }\textbf {\bibinfo {volume} {28}},\ \bibinfo {pages} {581–600} (\bibinfo {year} {1996})}\BibitemShut {NoStop}%
\bibitem [{\citenamefont {Karolyhazy}(1966)}]{Karolyhazy66}%
  \BibitemOpen
  \bibfield  {author} {\bibinfo {author} {\bibfnamefont {F.}~\bibnamefont {Karolyhazy}},\ }\href {\doibase https://doi.org/10.1007/BF02717926} {\bibfield  {journal} {\bibinfo  {journal} {Nuovo Cimento A (1965-1970)}\ }\textbf {\bibinfo {volume} {42}},\ \bibinfo {pages} {390–402} (\bibinfo {year} {1966})}\BibitemShut {NoStop}%
\bibitem [{\citenamefont {Bassi}\ \emph {et~al.}(2013)\citenamefont {Bassi}, \citenamefont {Lochan}, \citenamefont {Satin}, \citenamefont {Singh},\ and\ \citenamefont {Ulbricht}}]{bassireview}%
  \BibitemOpen
  \bibfield  {author} {\bibinfo {author} {\bibfnamefont {A.}~\bibnamefont {Bassi}}, \bibinfo {author} {\bibfnamefont {K.}~\bibnamefont {Lochan}}, \bibinfo {author} {\bibfnamefont {S.}~\bibnamefont {Satin}}, \bibinfo {author} {\bibfnamefont {T.~P.}\ \bibnamefont {Singh}}, \ and\ \bibinfo {author} {\bibfnamefont {H.}~\bibnamefont {Ulbricht}},\ }\href {\doibase 10.1103/RevModPhys.85.471} {\bibfield  {journal} {\bibinfo  {journal} {Rev. Mod. Phys.}\ }\textbf {\bibinfo {volume} {85}},\ \bibinfo {pages} {471} (\bibinfo {year} {2013})}\BibitemShut {NoStop}%
\bibitem [{\citenamefont {Fr\"owis}\ \emph {et~al.}(2018)\citenamefont {Fr\"owis}, \citenamefont {Sekatski}, \citenamefont {D\"ur}, \citenamefont {Gisin},\ and\ \citenamefont {Sangouard}}]{MacroscopicRMP2018}%
  \BibitemOpen
  \bibfield  {author} {\bibinfo {author} {\bibfnamefont {F.}~\bibnamefont {Fr\"owis}}, \bibinfo {author} {\bibfnamefont {P.}~\bibnamefont {Sekatski}}, \bibinfo {author} {\bibfnamefont {W.}~\bibnamefont {D\"ur}}, \bibinfo {author} {\bibfnamefont {N.}~\bibnamefont {Gisin}}, \ and\ \bibinfo {author} {\bibfnamefont {N.}~\bibnamefont {Sangouard}},\ }\href {\doibase 10.1103/RevModPhys.90.025004} {\bibfield  {journal} {\bibinfo  {journal} {Rev. Mod. Phys.}\ }\textbf {\bibinfo {volume} {90}},\ \bibinfo {pages} {025004} (\bibinfo {year} {2018})}\BibitemShut {NoStop}%
\bibitem [{\citenamefont {Howl}\ \emph {et~al.}(2019)\citenamefont {Howl}, \citenamefont {Penrose},\ and\ \citenamefont {Fuentes}}]{howl2019exploring}%
  \BibitemOpen
  \bibfield  {author} {\bibinfo {author} {\bibfnamefont {R.}~\bibnamefont {Howl}}, \bibinfo {author} {\bibfnamefont {R.}~\bibnamefont {Penrose}}, \ and\ \bibinfo {author} {\bibfnamefont {I.}~\bibnamefont {Fuentes}},\ }\href {\doibase 10.1088/1367-2630/ab104a} {\bibfield  {journal} {\bibinfo  {journal} {New Journal of Physics}\ }\textbf {\bibinfo {volume} {21}},\ \bibinfo {pages} {043047} (\bibinfo {year} {2019})}\BibitemShut {NoStop}%
\bibitem [{\citenamefont {Donadi}\ \emph {et~al.}(2021)\citenamefont {Donadi}, \citenamefont {Piscicchia}, \citenamefont {Curceanu}, \citenamefont {Diósi}, \citenamefont {Laubenstein},\ and\ \citenamefont {Bassi}}]{Donadi2021}%
  \BibitemOpen
  \bibfield  {author} {\bibinfo {author} {\bibfnamefont {S.}~\bibnamefont {Donadi}}, \bibinfo {author} {\bibfnamefont {K.}~\bibnamefont {Piscicchia}}, \bibinfo {author} {\bibfnamefont {C.}~\bibnamefont {Curceanu}}, \bibinfo {author} {\bibfnamefont {L.}~\bibnamefont {Diósi}}, \bibinfo {author} {\bibfnamefont {M.}~\bibnamefont {Laubenstein}}, \ and\ \bibinfo {author} {\bibfnamefont {A.}~\bibnamefont {Bassi}},\ }\href {https://doi.org/10.1038/s41567-020-1008-4} {\bibfield  {journal} {\bibinfo  {journal} {Nature Physics}\ }\textbf {\bibinfo {volume} {17}},\ \bibinfo {pages} {74} (\bibinfo {year} {2021})}\BibitemShut {NoStop}%
\bibitem [{\citenamefont {Penrose}(2022)}]{Penrose2022}%
  \BibitemOpen
  \bibfield  {author} {\bibinfo {author} {\bibfnamefont {R.}~\bibnamefont {Penrose}},\ }in\ \href {https://doi.org/10.1093/oso/9780197501665.001.0001} {\emph {\bibinfo {booktitle} {Consciousness and Quantum Mechanics}}},\ \bibinfo {editor} {edited by\ \bibinfo {editor} {\bibfnamefont {S.}~\bibnamefont {Gao}}}\ (\bibinfo  {publisher} {Oxford University Press},\ \bibinfo {year} {2022})\BibitemShut {NoStop}%
\bibitem [{\citenamefont {Zhou}\ \emph {et~al.}(2025)\citenamefont {Zhou}, \citenamefont {Bose},\ and\ \citenamefont {Mazumdar}}]{Zhou2025}%
  \BibitemOpen
  \bibfield  {author} {\bibinfo {author} {\bibfnamefont {T.}~\bibnamefont {Zhou}}, \bibinfo {author} {\bibfnamefont {S.}~\bibnamefont {Bose}}, \ and\ \bibinfo {author} {\bibfnamefont {A.}~\bibnamefont {Mazumdar}},\ }\href@noop {} {\bibfield  {journal} {\bibinfo  {journal} {Phys. Rev. A}\ }\textbf {\bibinfo {volume} {112}},\ \bibinfo {pages} {013315} (\bibinfo {year} {2025})}\BibitemShut {NoStop}%
\bibitem [{\citenamefont {Rizaldy}\ \emph {et~al.}(2024)\citenamefont {Rizaldy}, \citenamefont {Zhou}, \citenamefont {Bose},\ and\ \citenamefont {Mazumdar}}]{Rizaldy2024}%
  \BibitemOpen
  \bibfield  {author} {\bibinfo {author} {\bibfnamefont {R.}~\bibnamefont {Rizaldy}}, \bibinfo {author} {\bibfnamefont {T.}~\bibnamefont {Zhou}}, \bibinfo {author} {\bibfnamefont {S.}~\bibnamefont {Bose}}, \ and\ \bibinfo {author} {\bibfnamefont {A.}~\bibnamefont {Mazumdar}},\ }\href@noop {} {\  (\bibinfo {year} {2024})},\ \Eprint {http://arxiv.org/abs/2412.15335} {2412.15335 [quant-ph]} \BibitemShut {NoStop}%
\bibitem [{\citenamefont {Magrini}\ \emph {et~al.}(2021)\citenamefont {Magrini}, \citenamefont {Rosenzweig}, \citenamefont {Bach}, \citenamefont {Deutschmann-Olek}, \citenamefont {Hofer}, \citenamefont {Hong}, \citenamefont {Kiesel}, \citenamefont {A.},\ and\ \citenamefont {Aspelmeyer}}]{MagriniCooling2021}%
  \BibitemOpen
  \bibfield  {author} {\bibinfo {author} {\bibfnamefont {L.}~\bibnamefont {Magrini}}, \bibinfo {author} {\bibfnamefont {P.}~\bibnamefont {Rosenzweig}}, \bibinfo {author} {\bibfnamefont {C.}~\bibnamefont {Bach}}, \bibinfo {author} {\bibfnamefont {A.}~\bibnamefont {Deutschmann-Olek}}, \bibinfo {author} {\bibfnamefont {S.~G.}\ \bibnamefont {Hofer}}, \bibinfo {author} {\bibfnamefont {S.}~\bibnamefont {Hong}}, \bibinfo {author} {\bibfnamefont {N.}~\bibnamefont {Kiesel}}, \bibinfo {author} {\bibfnamefont {K.}~\bibnamefont {A.}}, \ and\ \bibinfo {author} {\bibfnamefont {M.}~\bibnamefont {Aspelmeyer}},\ }\href {https://doi.org/10.1038/s41586-021-03602-3} {\bibfield  {journal} {\bibinfo  {journal} {Nature}\ }\textbf {\bibinfo {volume} {595}},\ \bibinfo {pages} {373} (\bibinfo {year} {2021})}\BibitemShut {NoStop}%
\bibitem [{\citenamefont {Tebbenjohanns}\ \emph {et~al.}(2021)\citenamefont {Tebbenjohanns}, \citenamefont {Mattana}, \citenamefont {Rossi}, \citenamefont {M.},\ and\ \citenamefont {Novotny}}]{TebbenjohannsCooling2021}%
  \BibitemOpen
  \bibfield  {author} {\bibinfo {author} {\bibfnamefont {F.}~\bibnamefont {Tebbenjohanns}}, \bibinfo {author} {\bibfnamefont {M.~L.}\ \bibnamefont {Mattana}}, \bibinfo {author} {\bibfnamefont {M.}~\bibnamefont {Rossi}}, \bibinfo {author} {\bibfnamefont {F.}~\bibnamefont {M.}}, \ and\ \bibinfo {author} {\bibfnamefont {L.}~\bibnamefont {Novotny}},\ }\href {https://doi.org/10.1038/s41586-021-03617-w} {\bibfield  {journal} {\bibinfo  {journal} {Nature}\ }\textbf {\bibinfo {volume} {595}},\ \bibinfo {pages} {378} (\bibinfo {year} {2021})}\BibitemShut {NoStop}%
\bibitem [{\citenamefont {Kamba}\ \emph {et~al.}(2022)\citenamefont {Kamba}, \citenamefont {Shimizu},\ and\ \citenamefont {Aikawa}}]{Kamba2022}%
  \BibitemOpen
  \bibfield  {author} {\bibinfo {author} {\bibfnamefont {M.}~\bibnamefont {Kamba}}, \bibinfo {author} {\bibfnamefont {R.}~\bibnamefont {Shimizu}}, \ and\ \bibinfo {author} {\bibfnamefont {K.}~\bibnamefont {Aikawa}},\ }\href {\doibase 10.1364/OE.462921} {\bibfield  {journal} {\bibinfo  {journal} {Opt. Express}\ }\textbf {\bibinfo {volume} {30}},\ \bibinfo {pages} {26716} (\bibinfo {year} {2022})}\BibitemShut {NoStop}%
\bibitem [{\citenamefont {Piotrowski}\ \emph {et~al.}(2023)\citenamefont {Piotrowski}, \citenamefont {Windey}, \citenamefont {Vijayan}, \citenamefont {Gonzalez-Ballestero}, \citenamefont {de~los R{\'\i}os~Sommer}, \citenamefont {Meyer}, \citenamefont {Quidant}, \citenamefont {Romero-Isart}, \citenamefont {Reimann},\ and\ \citenamefont {Novotny}}]{Piotrowski2023}%
  \BibitemOpen
  \bibfield  {author} {\bibinfo {author} {\bibfnamefont {J.}~\bibnamefont {Piotrowski}}, \bibinfo {author} {\bibfnamefont {D.}~\bibnamefont {Windey}}, \bibinfo {author} {\bibfnamefont {J.}~\bibnamefont {Vijayan}}, \bibinfo {author} {\bibfnamefont {C.}~\bibnamefont {Gonzalez-Ballestero}}, \bibinfo {author} {\bibfnamefont {A.}~\bibnamefont {de~los R{\'\i}os~Sommer}}, \bibinfo {author} {\bibfnamefont {N.}~\bibnamefont {Meyer}}, \bibinfo {author} {\bibfnamefont {R.}~\bibnamefont {Quidant}}, \bibinfo {author} {\bibfnamefont {O.}~\bibnamefont {Romero-Isart}}, \bibinfo {author} {\bibfnamefont {R.}~\bibnamefont {Reimann}}, \ and\ \bibinfo {author} {\bibfnamefont {L.}~\bibnamefont {Novotny}},\ }\href@noop {} {\bibfield  {journal} {\bibinfo  {journal} {Nat. Phys.}\ } (\bibinfo {year} {2023})}\BibitemShut {NoStop}%
\bibitem [{\citenamefont {Dania}\ \emph {et~al.}(2025)\citenamefont {Dania}, \citenamefont {Kremer}, \citenamefont {Piotrowski}, \citenamefont {Candoli}, \citenamefont {Vijayan}, \citenamefont {Romero-Isart}, \citenamefont {Gonzalez-Ballestero}, \citenamefont {Novotny},\ and\ \citenamefont {Frimmer}}]{Dania2025}%
  \BibitemOpen
  \bibfield  {author} {\bibinfo {author} {\bibfnamefont {L.}~\bibnamefont {Dania}}, \bibinfo {author} {\bibfnamefont {O.~S.}\ \bibnamefont {Kremer}}, \bibinfo {author} {\bibfnamefont {J.}~\bibnamefont {Piotrowski}}, \bibinfo {author} {\bibfnamefont {D.}~\bibnamefont {Candoli}}, \bibinfo {author} {\bibfnamefont {J.}~\bibnamefont {Vijayan}}, \bibinfo {author} {\bibfnamefont {O.}~\bibnamefont {Romero-Isart}}, \bibinfo {author} {\bibfnamefont {C.}~\bibnamefont {Gonzalez-Ballestero}}, \bibinfo {author} {\bibfnamefont {L.}~\bibnamefont {Novotny}}, \ and\ \bibinfo {author} {\bibfnamefont {M.}~\bibnamefont {Frimmer}},\ }\href@noop {} {\bibfield  {journal} {\bibinfo  {journal} {Nat. Phys.}\ } (\bibinfo {year} {2025})}\BibitemShut {NoStop}%
\bibitem [{\citenamefont {Schut}\ \emph {et~al.}(2022)\citenamefont {Schut}, \citenamefont {Tilly}, \citenamefont {Marshman}, \citenamefont {Bose},\ and\ \citenamefont {Mazumdar}}]{Schut:2021svd}%
  \BibitemOpen
  \bibfield  {author} {\bibinfo {author} {\bibfnamefont {M.}~\bibnamefont {Schut}}, \bibinfo {author} {\bibfnamefont {J.}~\bibnamefont {Tilly}}, \bibinfo {author} {\bibfnamefont {R.~J.}\ \bibnamefont {Marshman}}, \bibinfo {author} {\bibfnamefont {S.}~\bibnamefont {Bose}}, \ and\ \bibinfo {author} {\bibfnamefont {A.}~\bibnamefont {Mazumdar}},\ }\href {\doibase 10.1103/PhysRevA.105.032411} {\bibfield  {journal} {\bibinfo  {journal} {Phys. Rev. A}\ }\textbf {\bibinfo {volume} {105}},\ \bibinfo {pages} {032411} (\bibinfo {year} {2022})}\BibitemShut {NoStop}%
\bibitem [{\citenamefont {Schut}\ \emph {et~al.}(2024{\natexlab{a}})\citenamefont {Schut}, \citenamefont {Geraci}, \citenamefont {Bose},\ and\ \citenamefont {Mazumdar}}]{Schut:2023hsy}%
  \BibitemOpen
  \bibfield  {author} {\bibinfo {author} {\bibfnamefont {M.}~\bibnamefont {Schut}}, \bibinfo {author} {\bibfnamefont {A.}~\bibnamefont {Geraci}}, \bibinfo {author} {\bibfnamefont {S.}~\bibnamefont {Bose}}, \ and\ \bibinfo {author} {\bibfnamefont {A.}~\bibnamefont {Mazumdar}},\ }\href {\doibase 10.1103/PhysRevResearch.6.013199} {\bibfield  {journal} {\bibinfo  {journal} {Phys. Rev. Res.}\ }\textbf {\bibinfo {volume} {6}},\ \bibinfo {pages} {013199} (\bibinfo {year} {2024}{\natexlab{a}})}\BibitemShut {NoStop}%
\bibitem [{\citenamefont {Schut}\ and\ \citenamefont {Mazumdar}(2025)}]{Schut:2025blz}%
  \BibitemOpen
  \bibfield  {author} {\bibinfo {author} {\bibfnamefont {M.}~\bibnamefont {Schut}}\ and\ \bibinfo {author} {\bibfnamefont {A.}~\bibnamefont {Mazumdar}},\ }\href@noop {} {\  (\bibinfo {year} {2025})},\ \Eprint {http://arxiv.org/abs/2502.12474} {arXiv:2502.12474 [quant-ph]} \BibitemShut {NoStop}%
\bibitem [{\citenamefont {Toro\v{s}}\ \emph {et~al.}(2021)\citenamefont {Toro\v{s}}, \citenamefont {Van De~Kamp}, \citenamefont {Marshman}, \citenamefont {Kim}, \citenamefont {Mazumdar},\ and\ \citenamefont {Bose}}]{Toros:2020dbf}%
  \BibitemOpen
  \bibfield  {author} {\bibinfo {author} {\bibfnamefont {M.}~\bibnamefont {Toro\v{s}}}, \bibinfo {author} {\bibfnamefont {T.~W.}\ \bibnamefont {Van De~Kamp}}, \bibinfo {author} {\bibfnamefont {R.~J.}\ \bibnamefont {Marshman}}, \bibinfo {author} {\bibfnamefont {M.~S.}\ \bibnamefont {Kim}}, \bibinfo {author} {\bibfnamefont {A.}~\bibnamefont {Mazumdar}}, \ and\ \bibinfo {author} {\bibfnamefont {S.}~\bibnamefont {Bose}},\ }\href {\doibase 10.1103/PhysRevResearch.3.023178} {\bibfield  {journal} {\bibinfo  {journal} {Phys. Rev. Res.}\ }\textbf {\bibinfo {volume} {3}},\ \bibinfo {pages} {023178} (\bibinfo {year} {2021})}\BibitemShut {NoStop}%
\bibitem [{\citenamefont {Schut}\ \emph {et~al.}(2025)\citenamefont {Schut}, \citenamefont {Andriolo}, \citenamefont {Toro\ifmmode~\check{s}\else \v{s}\fi{}}, \citenamefont {Bose},\ and\ \citenamefont {Mazumdar}}]{Schut2025}%
  \BibitemOpen
  \bibfield  {author} {\bibinfo {author} {\bibfnamefont {M.}~\bibnamefont {Schut}}, \bibinfo {author} {\bibfnamefont {P.}~\bibnamefont {Andriolo}}, \bibinfo {author} {\bibfnamefont {M.}~\bibnamefont {Toro\ifmmode~\check{s}\else \v{s}\fi{}}}, \bibinfo {author} {\bibfnamefont {S.}~\bibnamefont {Bose}}, \ and\ \bibinfo {author} {\bibfnamefont {A.}~\bibnamefont {Mazumdar}},\ }\href {\doibase 10.1103/PhysRevA.111.042211} {\bibfield  {journal} {\bibinfo  {journal} {Phys. Rev. A}\ }\textbf {\bibinfo {volume} {111}},\ \bibinfo {pages} {042211} (\bibinfo {year} {2025})}\BibitemShut {NoStop}%
\bibitem [{\citenamefont {Narasimha~Moorthy}\ \emph {et~al.}(2025)\citenamefont {Narasimha~Moorthy}, \citenamefont {Geraci}, \citenamefont {Bose},\ and\ \citenamefont {Mazumdar}}]{Moorthy2025}%
  \BibitemOpen
  \bibfield  {author} {\bibinfo {author} {\bibfnamefont {S.}~\bibnamefont {Narasimha~Moorthy}}, \bibinfo {author} {\bibfnamefont {A.}~\bibnamefont {Geraci}}, \bibinfo {author} {\bibfnamefont {S.}~\bibnamefont {Bose}}, \ and\ \bibinfo {author} {\bibfnamefont {A.}~\bibnamefont {Mazumdar}},\ }\href {\doibase 10.1103/9n6y-cc7r} {\bibfield  {journal} {\bibinfo  {journal} {Phys. Rev. A}\ }\textbf {\bibinfo {volume} {112}},\ \bibinfo {pages} {022416} (\bibinfo {year} {2025})}\BibitemShut {NoStop}%
\bibitem [{\citenamefont {Romero-Isart}(2011)}]{ORI11_GM}%
  \BibitemOpen
  \bibfield  {author} {\bibinfo {author} {\bibfnamefont {O.}~\bibnamefont {Romero-Isart}},\ }\href {\doibase 10.1103/PhysRevA.84.052121} {\bibfield  {journal} {\bibinfo  {journal} {Phys. Rev. A}\ }\textbf {\bibinfo {volume} {84}},\ \bibinfo {pages} {052121} (\bibinfo {year} {2011})}\BibitemShut {NoStop}%
\bibitem [{\citenamefont {Nair}\ \emph {et~al.}(2023)\citenamefont {Nair}, \citenamefont {Tian}, \citenamefont {Brennen}, \citenamefont {Bose},\ and\ \citenamefont {Twamley}}]{Nair2023}%
  \BibitemOpen
  \bibfield  {author} {\bibinfo {author} {\bibfnamefont {S.~R.}\ \bibnamefont {Nair}}, \bibinfo {author} {\bibfnamefont {S.}~\bibnamefont {Tian}}, \bibinfo {author} {\bibfnamefont {G.~K.}\ \bibnamefont {Brennen}}, \bibinfo {author} {\bibfnamefont {S.}~\bibnamefont {Bose}}, \ and\ \bibinfo {author} {\bibfnamefont {J.}~\bibnamefont {Twamley}},\ }\href@noop {} {\  (\bibinfo {year} {2023})},\ \Eprint {http://arxiv.org/abs/2307.14553} {2307.14553 [quant-ph]} \BibitemShut {NoStop}%
\bibitem [{\citenamefont {Brunner}\ \emph {et~al.}(2014)\citenamefont {Brunner}, \citenamefont {Cavalcanti}, \citenamefont {Pironio}, \citenamefont {Scarani},\ and\ \citenamefont {Wehner}}]{BellReviewRMP2014}%
  \BibitemOpen
  \bibfield  {author} {\bibinfo {author} {\bibfnamefont {N.}~\bibnamefont {Brunner}}, \bibinfo {author} {\bibfnamefont {D.}~\bibnamefont {Cavalcanti}}, \bibinfo {author} {\bibfnamefont {S.}~\bibnamefont {Pironio}}, \bibinfo {author} {\bibfnamefont {V.}~\bibnamefont {Scarani}}, \ and\ \bibinfo {author} {\bibfnamefont {S.}~\bibnamefont {Wehner}},\ }\href {\doibase 10.1103/RevModPhys.86.419} {\bibfield  {journal} {\bibinfo  {journal} {Rev. Mod. Phys.}\ }\textbf {\bibinfo {volume} {86}},\ \bibinfo {pages} {419} (\bibinfo {year} {2014})}\BibitemShut {NoStop}%
\bibitem [{\citenamefont {Kent}\ and\ \citenamefont {Pital\'ua-Garc\'{\i}a}(2021)}]{Kent2021}%
  \BibitemOpen
  \bibfield  {author} {\bibinfo {author} {\bibfnamefont {A.}~\bibnamefont {Kent}}\ and\ \bibinfo {author} {\bibfnamefont {D.}~\bibnamefont {Pital\'ua-Garc\'{\i}a}},\ }\href {\doibase 10.1103/PhysRevD.104.126030} {\bibfield  {journal} {\bibinfo  {journal} {Phys. Rev. D}\ }\textbf {\bibinfo {volume} {104}},\ \bibinfo {pages} {126030} (\bibinfo {year} {2021})}\BibitemShut {NoStop}%
\bibitem [{\citenamefont {Rosskopf}\ \emph {et~al.}(2014)\citenamefont {Rosskopf}, \citenamefont {Dussaux}, \citenamefont {Ohashi}, \citenamefont {Loretz}, \citenamefont {Schirhagl}, \citenamefont {Watanabe}, \citenamefont {Shikata}, \citenamefont {Itoh},\ and\ \citenamefont {Degen}}]{Rosskopf14_GM}%
  \BibitemOpen
  \bibfield  {author} {\bibinfo {author} {\bibfnamefont {T.}~\bibnamefont {Rosskopf}}, \bibinfo {author} {\bibfnamefont {A.}~\bibnamefont {Dussaux}}, \bibinfo {author} {\bibfnamefont {K.}~\bibnamefont {Ohashi}}, \bibinfo {author} {\bibfnamefont {M.}~\bibnamefont {Loretz}}, \bibinfo {author} {\bibfnamefont {R.}~\bibnamefont {Schirhagl}}, \bibinfo {author} {\bibfnamefont {H.}~\bibnamefont {Watanabe}}, \bibinfo {author} {\bibfnamefont {S.}~\bibnamefont {Shikata}}, \bibinfo {author} {\bibfnamefont {K.~M.}\ \bibnamefont {Itoh}}, \ and\ \bibinfo {author} {\bibfnamefont {C.~L.}\ \bibnamefont {Degen}},\ }\href {\doibase 10.1103/PhysRevLett.112.147602} {\bibfield  {journal} {\bibinfo  {journal} {Phys. Rev. Lett.}\ }\textbf {\bibinfo {volume} {112}},\ \bibinfo {pages} {147602} (\bibinfo {year} {2014})}\BibitemShut {NoStop}%
\bibitem [{\citenamefont {Pedernales}\ \emph {et~al.}(2020)\citenamefont {Pedernales}, \citenamefont {Morley},\ and\ \citenamefont {Plenio}}]{Pedernales20_GM}%
  \BibitemOpen
  \bibfield  {author} {\bibinfo {author} {\bibfnamefont {J.~S.}\ \bibnamefont {Pedernales}}, \bibinfo {author} {\bibfnamefont {G.~W.}\ \bibnamefont {Morley}}, \ and\ \bibinfo {author} {\bibfnamefont {M.~B.}\ \bibnamefont {Plenio}},\ }\href {\doibase 10.1103/PhysRevLett.125.023602} {\bibfield  {journal} {\bibinfo  {journal} {Phys. Rev. Lett.}\ }\textbf {\bibinfo {volume} {125}},\ \bibinfo {pages} {023602} (\bibinfo {year} {2020})}\BibitemShut {NoStop}%
\bibitem [{\citenamefont {Wood}\ \emph {et~al.}(2022{\natexlab{a}})\citenamefont {Wood}, \citenamefont {Bose},\ and\ \citenamefont {Morley}}]{WoodPRA22_GM}%
  \BibitemOpen
  \bibfield  {author} {\bibinfo {author} {\bibfnamefont {B.~D.}\ \bibnamefont {Wood}}, \bibinfo {author} {\bibfnamefont {S.}~\bibnamefont {Bose}}, \ and\ \bibinfo {author} {\bibfnamefont {G.~W.}\ \bibnamefont {Morley}},\ }\href {\doibase 10.1103/PhysRevA.105.012824} {\bibfield  {journal} {\bibinfo  {journal} {Phys. Rev. A}\ }\textbf {\bibinfo {volume} {105}},\ \bibinfo {pages} {012824} (\bibinfo {year} {2022}{\natexlab{a}})}\BibitemShut {NoStop}%
\bibitem [{\citenamefont {Zhou}\ \emph {et~al.}(2024{\natexlab{a}})\citenamefont {Zhou}, \citenamefont {Marshman}, \citenamefont {Bose},\ and\ \citenamefont {Mazumdar}}]{Zhou:2022epb}%
  \BibitemOpen
  \bibfield  {author} {\bibinfo {author} {\bibfnamefont {R.}~\bibnamefont {Zhou}}, \bibinfo {author} {\bibfnamefont {R.~J.}\ \bibnamefont {Marshman}}, \bibinfo {author} {\bibfnamefont {S.}~\bibnamefont {Bose}}, \ and\ \bibinfo {author} {\bibfnamefont {A.}~\bibnamefont {Mazumdar}},\ }\href {\doibase 10.1088/1402-4896/ad37df} {\bibfield  {journal} {\bibinfo  {journal} {Physica Scripta}\ }\textbf {\bibinfo {volume} {99}},\ \bibinfo {pages} {055114} (\bibinfo {year} {2024}{\natexlab{a}})}\BibitemShut {NoStop}%
\bibitem [{\citenamefont {Zhou}\ \emph {et~al.}(2022)\citenamefont {Zhou}, \citenamefont {Marshman}, \citenamefont {Bose},\ and\ \citenamefont {Mazumdar}}]{Zhou:2022frl}%
  \BibitemOpen
  \bibfield  {author} {\bibinfo {author} {\bibfnamefont {R.}~\bibnamefont {Zhou}}, \bibinfo {author} {\bibfnamefont {R.~J.}\ \bibnamefont {Marshman}}, \bibinfo {author} {\bibfnamefont {S.}~\bibnamefont {Bose}}, \ and\ \bibinfo {author} {\bibfnamefont {A.}~\bibnamefont {Mazumdar}},\ }\href {\doibase 10.1103/PhysRevResearch.4.043157} {\bibfield  {journal} {\bibinfo  {journal} {Phys. Rev. Res.}\ }\textbf {\bibinfo {volume} {4}},\ \bibinfo {pages} {043157} (\bibinfo {year} {2022})}\BibitemShut {NoStop}%
\bibitem [{\citenamefont {Zhou}\ \emph {et~al.}(2023)\citenamefont {Zhou}, \citenamefont {Marshman}, \citenamefont {Bose},\ and\ \citenamefont {Mazumdar}}]{Zhou:2022jug}%
  \BibitemOpen
  \bibfield  {author} {\bibinfo {author} {\bibfnamefont {R.}~\bibnamefont {Zhou}}, \bibinfo {author} {\bibfnamefont {R.~J.}\ \bibnamefont {Marshman}}, \bibinfo {author} {\bibfnamefont {S.}~\bibnamefont {Bose}}, \ and\ \bibinfo {author} {\bibfnamefont {A.}~\bibnamefont {Mazumdar}},\ }\href {\doibase 10.1103/PhysRevA.107.032212} {\bibfield  {journal} {\bibinfo  {journal} {Phys. Rev. A}\ }\textbf {\bibinfo {volume} {107}},\ \bibinfo {pages} {032212} (\bibinfo {year} {2023})}\BibitemShut {NoStop}%
\bibitem [{\citenamefont {Zhou}\ \emph {et~al.}(2024{\natexlab{b}})\citenamefont {Zhou}, \citenamefont {Xiang},\ and\ \citenamefont {Mazumdar}}]{Zhou:2024voj}%
  \BibitemOpen
  \bibfield  {author} {\bibinfo {author} {\bibfnamefont {R.}~\bibnamefont {Zhou}}, \bibinfo {author} {\bibfnamefont {Q.}~\bibnamefont {Xiang}}, \ and\ \bibinfo {author} {\bibfnamefont {A.}~\bibnamefont {Mazumdar}},\ }\href@noop {} {\  (\bibinfo {year} {2024}{\natexlab{b}})},\ \Eprint {http://arxiv.org/abs/2408.11909} {arXiv:2408.11909 [quant-ph]} \BibitemShut {NoStop}%
\bibitem [{\citenamefont {Braccini}\ \emph {et~al.}(2024)\citenamefont {Braccini}, \citenamefont {Serafini},\ and\ \citenamefont {Bose}}]{Braccini:2024fey}%
  \BibitemOpen
  \bibfield  {author} {\bibinfo {author} {\bibfnamefont {L.}~\bibnamefont {Braccini}}, \bibinfo {author} {\bibfnamefont {A.}~\bibnamefont {Serafini}}, \ and\ \bibinfo {author} {\bibfnamefont {S.}~\bibnamefont {Bose}},\ }\href@noop {} {\  (\bibinfo {year} {2024})},\ \Eprint {http://arxiv.org/abs/2408.11930} {arXiv:2408.11930 [quant-ph]} \BibitemShut {NoStop}%
\bibitem [{\citenamefont {Henkel}\ and\ \citenamefont {Folman}(2022)}]{Henkel:2021wmj}%
  \BibitemOpen
  \bibfield  {author} {\bibinfo {author} {\bibfnamefont {C.}~\bibnamefont {Henkel}}\ and\ \bibinfo {author} {\bibfnamefont {R.}~\bibnamefont {Folman}},\ }\href {\doibase 10.1116/5.0080503} {\bibfield  {journal} {\bibinfo  {journal} {AVS Quantum Sci.}\ }\textbf {\bibinfo {volume} {4}},\ \bibinfo {pages} {025602} (\bibinfo {year} {2022})}\BibitemShut {NoStop}%
\bibitem [{\citenamefont {Japha}\ and\ \citenamefont {Folman}(2023)}]{Japha:2022xyg}%
  \BibitemOpen
  \bibfield  {author} {\bibinfo {author} {\bibfnamefont {Y.}~\bibnamefont {Japha}}\ and\ \bibinfo {author} {\bibfnamefont {R.}~\bibnamefont {Folman}},\ }\href {\doibase 10.1103/PhysRevLett.130.113602} {\bibfield  {journal} {\bibinfo  {journal} {Phys. Rev. Lett.}\ }\textbf {\bibinfo {volume} {130}},\ \bibinfo {pages} {113602} (\bibinfo {year} {2023})}\BibitemShut {NoStop}%
\bibitem [{\citenamefont {Wan}\ \emph {et~al.}(2016)\citenamefont {Wan}, \citenamefont {Scala}, \citenamefont {Morley}, \citenamefont {Rahman}, \citenamefont {Ulbricht}, \citenamefont {Bateman}, \citenamefont {Barker}, \citenamefont {Bose},\ and\ \citenamefont {Kim}}]{Wan16_GM}%
  \BibitemOpen
  \bibfield  {author} {\bibinfo {author} {\bibfnamefont {C.}~\bibnamefont {Wan}}, \bibinfo {author} {\bibfnamefont {M.}~\bibnamefont {Scala}}, \bibinfo {author} {\bibfnamefont {G.}~\bibnamefont {Morley}}, \bibinfo {author} {\bibfnamefont {A.}~\bibnamefont {Rahman}}, \bibinfo {author} {\bibfnamefont {H.}~\bibnamefont {Ulbricht}}, \bibinfo {author} {\bibfnamefont {J.}~\bibnamefont {Bateman}}, \bibinfo {author} {\bibfnamefont {P.}~\bibnamefont {Barker}}, \bibinfo {author} {\bibfnamefont {S.}~\bibnamefont {Bose}}, \ and\ \bibinfo {author} {\bibfnamefont {M.}~\bibnamefont {Kim}},\ }\href {http://link.aps.org/doi/10.1103/PhysRevLett.117.143003} {\bibfield  {journal} {\bibinfo  {journal} {Phys. Rev. Lett.}\ }\textbf {\bibinfo {volume} {117}},\ \bibinfo {pages} {143003} (\bibinfo {year} {2016})}\BibitemShut {NoStop}%
\bibitem [{\citenamefont {Hsu}\ \emph {et~al.}(2016)\citenamefont {Hsu}, \citenamefont {Ji}, \citenamefont {Lewandowski},\ and\ \citenamefont {D’Urso}}]{DUrso16_GM}%
  \BibitemOpen
  \bibfield  {author} {\bibinfo {author} {\bibfnamefont {J.-F.}\ \bibnamefont {Hsu}}, \bibinfo {author} {\bibfnamefont {P.}~\bibnamefont {Ji}}, \bibinfo {author} {\bibfnamefont {C.~W.}\ \bibnamefont {Lewandowski}}, \ and\ \bibinfo {author} {\bibfnamefont {B.}~\bibnamefont {D’Urso}},\ }\href {\doibase 10.1038/srep30125 http://www.nature.com/articles/srep30125#supplementary-information} {\bibfield  {journal} {\bibinfo  {journal} {Sci. Rep.}\ }\textbf {\bibinfo {volume} {6}},\ \bibinfo {pages} {30125} (\bibinfo {year} {2016})}\BibitemShut {NoStop}%
\bibitem [{\citenamefont {Dobkowski}\ \emph {et~al.}(2025)\citenamefont {Dobkowski}, \citenamefont {Trok}, \citenamefont {Skakunenko}, \citenamefont {Japha}, \citenamefont {Groswasser}, \citenamefont {Efremov}, \citenamefont {Marletto}, \citenamefont {Fuentes}, \citenamefont {Penrose}, \citenamefont {Vedral}, \citenamefont {Schleich},\ and\ \citenamefont {Folman}}]{QequivalenceFolman}%
  \BibitemOpen
  \bibfield  {author} {\bibinfo {author} {\bibfnamefont {O.}~\bibnamefont {Dobkowski}}, \bibinfo {author} {\bibfnamefont {B.}~\bibnamefont {Trok}}, \bibinfo {author} {\bibfnamefont {P.}~\bibnamefont {Skakunenko}}, \bibinfo {author} {\bibfnamefont {Y.}~\bibnamefont {Japha}}, \bibinfo {author} {\bibfnamefont {D.}~\bibnamefont {Groswasser}}, \bibinfo {author} {\bibfnamefont {M.}~\bibnamefont {Efremov}}, \bibinfo {author} {\bibfnamefont {C.}~\bibnamefont {Marletto}}, \bibinfo {author} {\bibfnamefont {I.}~\bibnamefont {Fuentes}}, \bibinfo {author} {\bibfnamefont {R.}~\bibnamefont {Penrose}}, \bibinfo {author} {\bibfnamefont {V.}~\bibnamefont {Vedral}}, \bibinfo {author} {\bibfnamefont {W.~P.}\ \bibnamefont {Schleich}}, \ and\ \bibinfo {author} {\bibfnamefont {R.}~\bibnamefont {Folman}},\ }\href {https://arxiv.org/abs/2502.14535} {\  (\bibinfo {year} {2025})},\ \Eprint {http://arxiv.org/abs/2502.14535} {arXiv:2502.14535 [quant-ph]} \BibitemShut {NoStop}%
\bibitem [{\citenamefont {Keil}\ \emph {et~al.}(2016)\citenamefont {Keil}, \citenamefont {Amit}, \citenamefont {Zhou}, \citenamefont {Groswasser}, \citenamefont {Japha},\ and\ \citenamefont {Folman}}]{AtomChipReview2016}%
  \BibitemOpen
  \bibfield  {author} {\bibinfo {author} {\bibfnamefont {M.}~\bibnamefont {Keil}}, \bibinfo {author} {\bibfnamefont {O.}~\bibnamefont {Amit}}, \bibinfo {author} {\bibfnamefont {S.}~\bibnamefont {Zhou}}, \bibinfo {author} {\bibfnamefont {D.}~\bibnamefont {Groswasser}}, \bibinfo {author} {\bibfnamefont {Y.}~\bibnamefont {Japha}}, \ and\ \bibinfo {author} {\bibfnamefont {R.}~\bibnamefont {Folman}},\ }\href {\doibase 10.1080/09500340.2016.1178820} {\bibfield  {journal} {\bibinfo  {journal} {Journal of Modern Optics}\ }\textbf {\bibinfo {volume} {63}},\ \bibinfo {pages} {1840} (\bibinfo {year} {2016})}\BibitemShut {NoStop}%
\bibitem [{\citenamefont {O'Brien}\ \emph {et~al.}(2019)\citenamefont {O'Brien}, \citenamefont {Dunn}, \citenamefont {Downes},\ and\ \citenamefont {Twamley}}]{Twamley19_GM}%
  \BibitemOpen
  \bibfield  {author} {\bibinfo {author} {\bibfnamefont {M.~C.}\ \bibnamefont {O'Brien}}, \bibinfo {author} {\bibfnamefont {S.}~\bibnamefont {Dunn}}, \bibinfo {author} {\bibfnamefont {J.~E.}\ \bibnamefont {Downes}}, \ and\ \bibinfo {author} {\bibfnamefont {J.}~\bibnamefont {Twamley}},\ }\href {\doibase 10.1063/1.5066065} {\bibfield  {journal} {\bibinfo  {journal} {App. Phys. Lett.}\ }\textbf {\bibinfo {volume} {114}},\ \bibinfo {pages} {053103} (\bibinfo {year} {2019})}\BibitemShut {NoStop}%
\bibitem [{\citenamefont {Afek}\ \emph {et~al.}(2021)\citenamefont {Afek}, \citenamefont {Monteiro}, \citenamefont {Siegel}, \citenamefont {Wang}, \citenamefont {Dickson}, \citenamefont {Recoaro}, \citenamefont {Watts},\ and\ \citenamefont {Moore}}]{Afek21_dipoles}%
  \BibitemOpen
  \bibfield  {author} {\bibinfo {author} {\bibfnamefont {G.}~\bibnamefont {Afek}}, \bibinfo {author} {\bibfnamefont {F.}~\bibnamefont {Monteiro}}, \bibinfo {author} {\bibfnamefont {B.}~\bibnamefont {Siegel}}, \bibinfo {author} {\bibfnamefont {J.}~\bibnamefont {Wang}}, \bibinfo {author} {\bibfnamefont {S.}~\bibnamefont {Dickson}}, \bibinfo {author} {\bibfnamefont {J.}~\bibnamefont {Recoaro}}, \bibinfo {author} {\bibfnamefont {M.}~\bibnamefont {Watts}}, \ and\ \bibinfo {author} {\bibfnamefont {D.~C.}\ \bibnamefont {Moore}},\ }\href {\doibase 10.1103/PhysRevA.104.053512} {\bibfield  {journal} {\bibinfo  {journal} {Phys. Rev. A}\ }\textbf {\bibinfo {volume} {104}},\ \bibinfo {pages} {053512} (\bibinfo {year} {2021})}\BibitemShut {NoStop}%
\bibitem [{\citenamefont {Rider}\ \emph {et~al.}(2016)\citenamefont {Rider}, \citenamefont {Moore}, \citenamefont {Blakemore}, \citenamefont {Louis}, \citenamefont {Lu},\ and\ \citenamefont {Gratta}}]{Rider16_chameleons}%
  \BibitemOpen
  \bibfield  {author} {\bibinfo {author} {\bibfnamefont {A.~D.}\ \bibnamefont {Rider}}, \bibinfo {author} {\bibfnamefont {D.~C.}\ \bibnamefont {Moore}}, \bibinfo {author} {\bibfnamefont {C.~P.}\ \bibnamefont {Blakemore}}, \bibinfo {author} {\bibfnamefont {M.}~\bibnamefont {Louis}}, \bibinfo {author} {\bibfnamefont {M.}~\bibnamefont {Lu}}, \ and\ \bibinfo {author} {\bibfnamefont {G.}~\bibnamefont {Gratta}},\ }\href {\doibase 10.1103/PhysRevLett.117.101101} {\bibfield  {journal} {\bibinfo  {journal} {Phys. Rev. Lett.}\ }\textbf {\bibinfo {volume} {117}},\ \bibinfo {pages} {101101} (\bibinfo {year} {2016})}\BibitemShut {NoStop}%
\bibitem [{\citenamefont {Fragolino}\ \emph {et~al.}(2024)\citenamefont {Fragolino}, \citenamefont {Schut}, \citenamefont {Toro\v{s}}, \citenamefont {Bose},\ and\ \citenamefont {Mazumdar}}]{Fragolino:2023agd}%
  \BibitemOpen
  \bibfield  {author} {\bibinfo {author} {\bibfnamefont {P.}~\bibnamefont {Fragolino}}, \bibinfo {author} {\bibfnamefont {M.}~\bibnamefont {Schut}}, \bibinfo {author} {\bibfnamefont {M.}~\bibnamefont {Toro\v{s}}}, \bibinfo {author} {\bibfnamefont {S.}~\bibnamefont {Bose}}, \ and\ \bibinfo {author} {\bibfnamefont {A.}~\bibnamefont {Mazumdar}},\ }\href {\doibase 10.1103/PhysRevA.109.033301} {\bibfield  {journal} {\bibinfo  {journal} {Phys. Rev. A}\ }\textbf {\bibinfo {volume} {109}},\ \bibinfo {pages} {033301} (\bibinfo {year} {2024})}\BibitemShut {NoStop}%
\bibitem [{\citenamefont {Schut}\ \emph {et~al.}(2024{\natexlab{b}})\citenamefont {Schut}, \citenamefont {Bosma}, \citenamefont {Wu}, \citenamefont {Toro\v{s}}, \citenamefont {Bose},\ and\ \citenamefont {Mazumdar}}]{Schut:2023tce}%
  \BibitemOpen
  \bibfield  {author} {\bibinfo {author} {\bibfnamefont {M.}~\bibnamefont {Schut}}, \bibinfo {author} {\bibfnamefont {H.}~\bibnamefont {Bosma}}, \bibinfo {author} {\bibfnamefont {M.}~\bibnamefont {Wu}}, \bibinfo {author} {\bibfnamefont {M.}~\bibnamefont {Toro\v{s}}}, \bibinfo {author} {\bibfnamefont {S.}~\bibnamefont {Bose}}, \ and\ \bibinfo {author} {\bibfnamefont {A.}~\bibnamefont {Mazumdar}},\ }\href {\doibase 10.1103/PhysRevA.110.022412} {\bibfield  {journal} {\bibinfo  {journal} {Phys. Rev. A}\ }\textbf {\bibinfo {volume} {110}},\ \bibinfo {pages} {022412} (\bibinfo {year} {2024}{\natexlab{b}})}\BibitemShut {NoStop}%
\bibitem [{\citenamefont {Garrett}\ \emph {et~al.}(2020)\citenamefont {Garrett}, \citenamefont {Kim},\ and\ \citenamefont {Munday}}]{Garrett20_patches}%
  \BibitemOpen
  \bibfield  {author} {\bibinfo {author} {\bibfnamefont {J.~L.}\ \bibnamefont {Garrett}}, \bibinfo {author} {\bibfnamefont {J.}~\bibnamefont {Kim}}, \ and\ \bibinfo {author} {\bibfnamefont {J.~N.}\ \bibnamefont {Munday}},\ }\href {\doibase 10.1103/PhysRevResearch.2.023355} {\bibfield  {journal} {\bibinfo  {journal} {Phys. Rev. Res.}\ }\textbf {\bibinfo {volume} {2}},\ \bibinfo {pages} {023355} (\bibinfo {year} {2020})}\BibitemShut {NoStop}%
\bibitem [{\citenamefont {Ranjit}\ \emph {et~al.}(2016)\citenamefont {Ranjit}, \citenamefont {Cunningham}, \citenamefont {Casey},\ and\ \citenamefont {Geraci}}]{Ranjit:2016}%
  \BibitemOpen
  \bibfield  {author} {\bibinfo {author} {\bibfnamefont {G.}~\bibnamefont {Ranjit}}, \bibinfo {author} {\bibfnamefont {M.}~\bibnamefont {Cunningham}}, \bibinfo {author} {\bibfnamefont {K.}~\bibnamefont {Casey}}, \ and\ \bibinfo {author} {\bibfnamefont {A.~A.}\ \bibnamefont {Geraci}},\ }\href {\doibase 10.1103/PhysRevA.93.053801} {\bibfield  {journal} {\bibinfo  {journal} {Phys. Rev. A}\ }\textbf {\bibinfo {volume} {93}},\ \bibinfo {pages} {053801} (\bibinfo {year} {2016})}\BibitemShut {NoStop}%
\bibitem [{\citenamefont {Montoya}\ \emph {et~al.}(2022)\citenamefont {Montoya}, \citenamefont {Alejandro}, \citenamefont {Eom}, \citenamefont {Grass}, \citenamefont {Clarisse}, \citenamefont {Witherspoon},\ and\ \citenamefont {Geraci}}]{montoya2022}%
  \BibitemOpen
  \bibfield  {author} {\bibinfo {author} {\bibfnamefont {C.}~\bibnamefont {Montoya}}, \bibinfo {author} {\bibfnamefont {E.}~\bibnamefont {Alejandro}}, \bibinfo {author} {\bibfnamefont {W.}~\bibnamefont {Eom}}, \bibinfo {author} {\bibfnamefont {D.}~\bibnamefont {Grass}}, \bibinfo {author} {\bibfnamefont {N.}~\bibnamefont {Clarisse}}, \bibinfo {author} {\bibfnamefont {A.}~\bibnamefont {Witherspoon}}, \ and\ \bibinfo {author} {\bibfnamefont {A.~A.}\ \bibnamefont {Geraci}},\ }\href {\doibase 10.1364/AO.457148} {\bibfield  {journal} {\bibinfo  {journal} {Appl. Opt.}\ }\textbf {\bibinfo {volume} {61}},\ \bibinfo {pages} {3486} (\bibinfo {year} {2022})}\BibitemShut {NoStop}%
\bibitem [{\citenamefont {Casimir}\ and\ \citenamefont {Polder}(1948)}]{casimirpolder}%
  \BibitemOpen
  \bibfield  {author} {\bibinfo {author} {\bibfnamefont {H.~B.~G.}\ \bibnamefont {Casimir}}\ and\ \bibinfo {author} {\bibfnamefont {D.}~\bibnamefont {Polder}},\ }\href {\doibase 10.1103/PhysRev.73.360} {\bibfield  {journal} {\bibinfo  {journal} {Phys. Rev.}\ }\textbf {\bibinfo {volume} {73}},\ \bibinfo {pages} {360} (\bibinfo {year} {1948})}\BibitemShut {NoStop}%
\bibitem [{\citenamefont {Lambrecht}\ and\ \citenamefont {Reynaud}(2000)}]{lambrecht}%
  \BibitemOpen
  \bibfield  {author} {\bibinfo {author} {\bibfnamefont {A.}~\bibnamefont {Lambrecht}}\ and\ \bibinfo {author} {\bibfnamefont {S.}~\bibnamefont {Reynaud}},\ }\href {\doibase 10.1007/s100530050041} {\bibfield  {journal} {\bibinfo  {journal} {The European Physical Journal D}\ }\textbf {\bibinfo {volume} {8}},\ \bibinfo {pages} {309} (\bibinfo {year} {2000})}\BibitemShut {NoStop}%
\bibitem [{\citenamefont {Harber}\ \emph {et~al.}(2005)\citenamefont {Harber}, \citenamefont {Obrecht}, \citenamefont {McGuirk},\ and\ \citenamefont {Cornell}}]{harber}%
  \BibitemOpen
  \bibfield  {author} {\bibinfo {author} {\bibfnamefont {D.~M.}\ \bibnamefont {Harber}}, \bibinfo {author} {\bibfnamefont {J.~M.}\ \bibnamefont {Obrecht}}, \bibinfo {author} {\bibfnamefont {J.~M.}\ \bibnamefont {McGuirk}}, \ and\ \bibinfo {author} {\bibfnamefont {E.~A.}\ \bibnamefont {Cornell}},\ }\href {\doibase 10.1103/PhysRevA.72.033610} {\bibfield  {journal} {\bibinfo  {journal} {Phys. Rev. A}\ }\textbf {\bibinfo {volume} {72}},\ \bibinfo {pages} {033610} (\bibinfo {year} {2005})}\BibitemShut {NoStop}%
\bibitem [{\citenamefont {Elahi}\ \emph {et~al.}(2024)\citenamefont {Elahi}, \citenamefont {Schut}, \citenamefont {Dana}, \citenamefont {Grinin}, \citenamefont {Bose}, \citenamefont {Mazumdar},\ and\ \citenamefont {Geraci}}]{Elahi:2024dbb}%
  \BibitemOpen
  \bibfield  {author} {\bibinfo {author} {\bibfnamefont {S.~G.}\ \bibnamefont {Elahi}}, \bibinfo {author} {\bibfnamefont {M.}~\bibnamefont {Schut}}, \bibinfo {author} {\bibfnamefont {A.}~\bibnamefont {Dana}}, \bibinfo {author} {\bibfnamefont {A.}~\bibnamefont {Grinin}}, \bibinfo {author} {\bibfnamefont {S.}~\bibnamefont {Bose}}, \bibinfo {author} {\bibfnamefont {A.}~\bibnamefont {Mazumdar}}, \ and\ \bibinfo {author} {\bibfnamefont {A.}~\bibnamefont {Geraci}},\ }\href@noop {} {\  (\bibinfo {year} {2024})},\ \Eprint {http://arxiv.org/abs/2411.02325} {arXiv:2411.02325 [quant-ph]} \BibitemShut {NoStop}%
\bibitem [{\citenamefont {Schut}\ \emph {et~al.}(2023)\citenamefont {Schut}, \citenamefont {Grinin}, \citenamefont {Dana}, \citenamefont {Bose}, \citenamefont {Geraci},\ and\ \citenamefont {Mazumdar}}]{Schut:2023eux}%
  \BibitemOpen
  \bibfield  {author} {\bibinfo {author} {\bibfnamefont {M.}~\bibnamefont {Schut}}, \bibinfo {author} {\bibfnamefont {A.}~\bibnamefont {Grinin}}, \bibinfo {author} {\bibfnamefont {A.}~\bibnamefont {Dana}}, \bibinfo {author} {\bibfnamefont {S.}~\bibnamefont {Bose}}, \bibinfo {author} {\bibfnamefont {A.}~\bibnamefont {Geraci}}, \ and\ \bibinfo {author} {\bibfnamefont {A.}~\bibnamefont {Mazumdar}},\ }\href {\doibase 10.1103/PhysRevResearch.5.043170} {\bibfield  {journal} {\bibinfo  {journal} {Phys. Rev. Res.}\ }\textbf {\bibinfo {volume} {5}},\ \bibinfo {pages} {043170} (\bibinfo {year} {2023})}\BibitemShut {NoStop}%
\bibitem [{\citenamefont {Marshman}\ \emph {et~al.}(2022)\citenamefont {Marshman}, \citenamefont {Mazumdar}, \citenamefont {Folman},\ and\ \citenamefont {Bose}}]{Marshman:2021wyk}%
  \BibitemOpen
  \bibfield  {author} {\bibinfo {author} {\bibfnamefont {R.~J.}\ \bibnamefont {Marshman}}, \bibinfo {author} {\bibfnamefont {A.}~\bibnamefont {Mazumdar}}, \bibinfo {author} {\bibfnamefont {R.}~\bibnamefont {Folman}}, \ and\ \bibinfo {author} {\bibfnamefont {S.}~\bibnamefont {Bose}},\ }\href {\doibase 10.1103/PhysRevResearch.4.023087} {\bibfield  {journal} {\bibinfo  {journal} {Phys. Rev. Res.}\ }\textbf {\bibinfo {volume} {4}},\ \bibinfo {pages} {023087} (\bibinfo {year} {2022})}\BibitemShut {NoStop}%
\bibitem [{\citenamefont {Geraci}\ \emph {et~al.}(2008)\citenamefont {Geraci}, \citenamefont {Smullin}, \citenamefont {Weld}, \citenamefont {Chiaverini},\ and\ \citenamefont {Kapitulnik}}]{Geraci:2008}%
  \BibitemOpen
  \bibfield  {author} {\bibinfo {author} {\bibfnamefont {A.~A.}\ \bibnamefont {Geraci}}, \bibinfo {author} {\bibfnamefont {S.~J.}\ \bibnamefont {Smullin}}, \bibinfo {author} {\bibfnamefont {D.~M.}\ \bibnamefont {Weld}}, \bibinfo {author} {\bibfnamefont {J.}~\bibnamefont {Chiaverini}}, \ and\ \bibinfo {author} {\bibfnamefont {A.}~\bibnamefont {Kapitulnik}},\ }\href {\doibase 10.1103/PhysRevD.78.022002} {\bibfield  {journal} {\bibinfo  {journal} {Phys. Rev. D}\ }\textbf {\bibinfo {volume} {78}},\ \bibinfo {pages} {022002} (\bibinfo {year} {2008})}\BibitemShut {NoStop}%
\bibitem [{\citenamefont {Barker}\ \emph {et~al.}(2022)\citenamefont {Barker}, \citenamefont {Bose}, \citenamefont {Marshman},\ and\ \citenamefont {Mazumdar}}]{Barker:2022mdz}%
  \BibitemOpen
  \bibfield  {author} {\bibinfo {author} {\bibfnamefont {P.~F.}\ \bibnamefont {Barker}}, \bibinfo {author} {\bibfnamefont {S.}~\bibnamefont {Bose}}, \bibinfo {author} {\bibfnamefont {R.~J.}\ \bibnamefont {Marshman}}, \ and\ \bibinfo {author} {\bibfnamefont {A.}~\bibnamefont {Mazumdar}},\ }\href {\doibase 10.1103/PhysRevD.106.L041901} {\bibfield  {journal} {\bibinfo  {journal} {Phys. Rev. D}\ }\textbf {\bibinfo {volume} {106}},\ \bibinfo {pages} {L041901} (\bibinfo {year} {2022})}\BibitemShut {NoStop}%
\bibitem [{\citenamefont {Wood}\ \emph {et~al.}(2022{\natexlab{b}})\citenamefont {Wood}, \citenamefont {Stimpson}, \citenamefont {March}, \citenamefont {Lekhai}, \citenamefont {Stephen}, \citenamefont {Green}, \citenamefont {Frangeskou}, \citenamefont {Gin\'es}, \citenamefont {Mandal}, \citenamefont {Williams},\ and\ \citenamefont {Morley}}]{WoodPRB22_GM}%
  \BibitemOpen
  \bibfield  {author} {\bibinfo {author} {\bibfnamefont {B.~D.}\ \bibnamefont {Wood}}, \bibinfo {author} {\bibfnamefont {G.~A.}\ \bibnamefont {Stimpson}}, \bibinfo {author} {\bibfnamefont {J.~E.}\ \bibnamefont {March}}, \bibinfo {author} {\bibfnamefont {Y.~N.~D.}\ \bibnamefont {Lekhai}}, \bibinfo {author} {\bibfnamefont {C.~J.}\ \bibnamefont {Stephen}}, \bibinfo {author} {\bibfnamefont {B.~L.}\ \bibnamefont {Green}}, \bibinfo {author} {\bibfnamefont {A.~C.}\ \bibnamefont {Frangeskou}}, \bibinfo {author} {\bibfnamefont {L.}~\bibnamefont {Gin\'es}}, \bibinfo {author} {\bibfnamefont {S.}~\bibnamefont {Mandal}}, \bibinfo {author} {\bibfnamefont {O.~A.}\ \bibnamefont {Williams}}, \ and\ \bibinfo {author} {\bibfnamefont {G.~W.}\ \bibnamefont {Morley}},\ }\href {\doibase 10.1103/PhysRevB.105.205401} {\bibfield  {journal} {\bibinfo  {journal} {Phys. Rev. B}\ }\textbf {\bibinfo {volume} {105}},\ \bibinfo {pages} {205401} (\bibinfo {year} {2022}{\natexlab{b}})}\BibitemShut {NoStop}%
\bibitem [{\citenamefont {Andrich}\ \emph {et~al.}(2014)\citenamefont {Andrich}, \citenamefont {Alemán}, \citenamefont {Lee}, \citenamefont {Ohno}, \citenamefont {de~las Casas}, \citenamefont {Heremans}, \citenamefont {Hu},\ and\ \citenamefont {Awschalom}}]{Awschalom}%
  \BibitemOpen
  \bibfield  {author} {\bibinfo {author} {\bibfnamefont {P.}~\bibnamefont {Andrich}}, \bibinfo {author} {\bibfnamefont {B.~J.}\ \bibnamefont {Alemán}}, \bibinfo {author} {\bibfnamefont {J.~C.}\ \bibnamefont {Lee}}, \bibinfo {author} {\bibfnamefont {K.}~\bibnamefont {Ohno}}, \bibinfo {author} {\bibfnamefont {C.~F.}\ \bibnamefont {de~las Casas}}, \bibinfo {author} {\bibfnamefont {F.~J.}\ \bibnamefont {Heremans}}, \bibinfo {author} {\bibfnamefont {E.~L.}\ \bibnamefont {Hu}}, \ and\ \bibinfo {author} {\bibfnamefont {D.~D.}\ \bibnamefont {Awschalom}},\ }\href {\doibase 10.1021/nl501208s} {\bibfield  {journal} {\bibinfo  {journal} {Nano Lett.}\ }\textbf {\bibinfo {volume} {14}},\ \bibinfo {pages} {4959} (\bibinfo {year} {2014})}\BibitemShut {NoStop}%
\bibitem [{\citenamefont {Trusheim}\ \emph {et~al.}(2013)\citenamefont {Trusheim}, \citenamefont {Li}, \citenamefont {Laraoui}, \citenamefont {Chen}, \citenamefont {Bakhru}, \citenamefont {Schr{\"o}der}, \citenamefont {Gaathon}, \citenamefont {Meriles},\ and\ \citenamefont {Englund}}]{Englund}%
  \BibitemOpen
  \bibfield  {author} {\bibinfo {author} {\bibfnamefont {M.~E.}\ \bibnamefont {Trusheim}}, \bibinfo {author} {\bibfnamefont {L.}~\bibnamefont {Li}}, \bibinfo {author} {\bibfnamefont {A.}~\bibnamefont {Laraoui}}, \bibinfo {author} {\bibfnamefont {E.~H.}\ \bibnamefont {Chen}}, \bibinfo {author} {\bibfnamefont {H.}~\bibnamefont {Bakhru}}, \bibinfo {author} {\bibfnamefont {T.}~\bibnamefont {Schr{\"o}der}}, \bibinfo {author} {\bibfnamefont {O.}~\bibnamefont {Gaathon}}, \bibinfo {author} {\bibfnamefont {C.~A.}\ \bibnamefont {Meriles}}, \ and\ \bibinfo {author} {\bibfnamefont {D.}~\bibnamefont {Englund}},\ }\href {\doibase 10.1021/nl402799u} {\bibfield  {journal} {\bibinfo  {journal} {Nano Lett.}\ }\textbf {\bibinfo {volume} {14}},\ \bibinfo {pages} {32} (\bibinfo {year} {2013})}\BibitemShut {NoStop}%
\bibitem [{\citenamefont {Stephen}\ \emph {et~al.}(2019)\citenamefont {Stephen}, \citenamefont {Green}, \citenamefont {Lekhai}, \citenamefont {Weng}, \citenamefont {Hill}, \citenamefont {Johnson}, \citenamefont {Frangeskou}, \citenamefont {Diggle}, \citenamefont {Chen}, \citenamefont {Strain}, \citenamefont {Gu}, \citenamefont {Newton}, \citenamefont {Smith}, \citenamefont {Salter},\ and\ \citenamefont {Morley}}]{Stephen19_GM}%
  \BibitemOpen
  \bibfield  {author} {\bibinfo {author} {\bibfnamefont {C.~J.}\ \bibnamefont {Stephen}}, \bibinfo {author} {\bibfnamefont {B.~L.}\ \bibnamefont {Green}}, \bibinfo {author} {\bibfnamefont {Y.~N.~D.}\ \bibnamefont {Lekhai}}, \bibinfo {author} {\bibfnamefont {L.}~\bibnamefont {Weng}}, \bibinfo {author} {\bibfnamefont {P.}~\bibnamefont {Hill}}, \bibinfo {author} {\bibfnamefont {S.}~\bibnamefont {Johnson}}, \bibinfo {author} {\bibfnamefont {A.~C.}\ \bibnamefont {Frangeskou}}, \bibinfo {author} {\bibfnamefont {P.~L.}\ \bibnamefont {Diggle}}, \bibinfo {author} {\bibfnamefont {Y.-C.}\ \bibnamefont {Chen}}, \bibinfo {author} {\bibfnamefont {M.~J.}\ \bibnamefont {Strain}}, \bibinfo {author} {\bibfnamefont {E.}~\bibnamefont {Gu}}, \bibinfo {author} {\bibfnamefont {M.~E.}\ \bibnamefont {Newton}}, \bibinfo {author} {\bibfnamefont {J.~M.}\ \bibnamefont {Smith}}, \bibinfo {author} {\bibfnamefont {P.~S.}\ \bibnamefont {Salter}}, \ and\ \bibinfo {author} {\bibfnamefont {G.~W.}\ \bibnamefont {Morley}},\ }\href {\doibase
  10.1103/PhysRevApplied.12.064005} {\bibfield  {journal} {\bibinfo  {journal} {Phys. Rev. Applied}\ }\textbf {\bibinfo {volume} {12}},\ \bibinfo {pages} {064005} (\bibinfo {year} {2019})}\BibitemShut {NoStop}%
\bibitem [{\citenamefont {Tyler}\ \emph {et~al.}(2025)\citenamefont {Tyler}, \citenamefont {Newland}, \citenamefont {Hepworth}, \citenamefont {Wijesekara}, \citenamefont {Gullick}, \citenamefont {Markham}, \citenamefont {Newton},\ and\ \citenamefont {Green}}]{TylerCMP2025}%
  \BibitemOpen
  \bibfield  {author} {\bibinfo {author} {\bibfnamefont {S.}~\bibnamefont {Tyler}}, \bibinfo {author} {\bibfnamefont {J.}~\bibnamefont {Newland}}, \bibinfo {author} {\bibfnamefont {P.}~\bibnamefont {Hepworth}}, \bibinfo {author} {\bibfnamefont {A.}~\bibnamefont {Wijesekara}}, \bibinfo {author} {\bibfnamefont {I.~R.}\ \bibnamefont {Gullick}}, \bibinfo {author} {\bibfnamefont {M.~L.}\ \bibnamefont {Markham}}, \bibinfo {author} {\bibfnamefont {M.~E.}\ \bibnamefont {Newton}}, \ and\ \bibinfo {author} {\bibfnamefont {B.~L.}\ \bibnamefont {Green}},\ }\href {\doibase 10.1063/5.0244913} {\bibfield  {journal} {\bibinfo  {journal} {Applied Physics Letters}\ }\textbf {\bibinfo {volume} {126}},\ \bibinfo {pages} {054001} (\bibinfo {year} {2025})}\BibitemShut {NoStop}%
\bibitem [{\citenamefont {Steiner}\ \emph {et~al.}(2024)\citenamefont {Steiner}, \citenamefont {Pedernales},\ and\ \citenamefont {Plenio}}]{SteinerPentacene}%
  \BibitemOpen
  \bibfield  {author} {\bibinfo {author} {\bibfnamefont {M.~O.~E.}\ \bibnamefont {Steiner}}, \bibinfo {author} {\bibfnamefont {J.~S.}\ \bibnamefont {Pedernales}}, \ and\ \bibinfo {author} {\bibfnamefont {M.~B.}\ \bibnamefont {Plenio}},\ }\href@noop {} {\enquote {\bibinfo {title} {Pentacene-doped naphthalene for levitated optomechanics},}\ } (\bibinfo {year} {2024}),\ \Eprint {http://arxiv.org/abs/2405.13869} {arXiv:2405.13869 [quant-ph]} \BibitemShut {NoStop}%
\bibitem [{\citenamefont {Bradley}\ \emph {et~al.}(2019)\citenamefont {Bradley}, \citenamefont {Randall}, \citenamefont {Abobeih}, \citenamefont {Berrevoets}, \citenamefont {Degen}, \citenamefont {Bakker}, \citenamefont {Markham}, \citenamefont {Twitchen},\ and\ \citenamefont {Taminiau}}]{Taminiau_GM}%
  \BibitemOpen
  \bibfield  {author} {\bibinfo {author} {\bibfnamefont {C.~E.}\ \bibnamefont {Bradley}}, \bibinfo {author} {\bibfnamefont {J.}~\bibnamefont {Randall}}, \bibinfo {author} {\bibfnamefont {M.~H.}\ \bibnamefont {Abobeih}}, \bibinfo {author} {\bibfnamefont {R.~C.}\ \bibnamefont {Berrevoets}}, \bibinfo {author} {\bibfnamefont {M.~J.}\ \bibnamefont {Degen}}, \bibinfo {author} {\bibfnamefont {M.~A.}\ \bibnamefont {Bakker}}, \bibinfo {author} {\bibfnamefont {M.}~\bibnamefont {Markham}}, \bibinfo {author} {\bibfnamefont {D.~J.}\ \bibnamefont {Twitchen}}, \ and\ \bibinfo {author} {\bibfnamefont {T.~H.}\ \bibnamefont {Taminiau}},\ }\href {\doibase 10.1103/PhysRevX.9.031045} {\bibfield  {journal} {\bibinfo  {journal} {Phys. Rev. X}\ }\textbf {\bibinfo {volume} {9}},\ \bibinfo {pages} {031045} (\bibinfo {year} {2019})}\BibitemShut {NoStop}%
\bibitem [{\citenamefont {Geraci}\ \emph {et~al.}(2010)\citenamefont {Geraci}, \citenamefont {Papp},\ and\ \citenamefont {Kitching}}]{Geraci:2010}%
  \BibitemOpen
  \bibfield  {author} {\bibinfo {author} {\bibfnamefont {A.~A.}\ \bibnamefont {Geraci}}, \bibinfo {author} {\bibfnamefont {S.~B.}\ \bibnamefont {Papp}}, \ and\ \bibinfo {author} {\bibfnamefont {J.}~\bibnamefont {Kitching}},\ }\href {\doibase 10.1103/PhysRevLett.105.101101} {\bibfield  {journal} {\bibinfo  {journal} {Phys. Rev. Lett.}\ }\textbf {\bibinfo {volume} {105}},\ \bibinfo {pages} {101101} (\bibinfo {year} {2010})}\BibitemShut {NoStop}%
\bibitem [{\citenamefont {Carney}\ \emph {et~al.}(2021)\citenamefont {Carney}, \citenamefont {Krnjaic}, \citenamefont {Moore}, \citenamefont {Regal}, \citenamefont {Afek}, \citenamefont {Bhave}, \citenamefont {Brubaker}, \citenamefont {Corbitt}, \citenamefont {Cripe}, \citenamefont {Crisosto}, \citenamefont {Geraci}, \citenamefont {Ghosh}, \citenamefont {Harris}, \citenamefont {Hook}, \citenamefont {Kolb}, \citenamefont {Kunjummen}, \citenamefont {Lang}, \citenamefont {Li}, \citenamefont {Lin}, \citenamefont {Liu}, \citenamefont {Lykken}, \citenamefont {Magrini}, \citenamefont {Manley}, \citenamefont {Matsumoto}, \citenamefont {Monte}, \citenamefont {Monteiro}, \citenamefont {Purdy}, \citenamefont {Riedel}, \citenamefont {Singh}, \citenamefont {Singh}, \citenamefont {Sinha}, \citenamefont {Taylor}, \citenamefont {Qin}, \citenamefont {Wilson},\ and\ \citenamefont {Zhao}}]{Carney20_DM}%
  \BibitemOpen
  \bibfield  {author} {\bibinfo {author} {\bibfnamefont {D.}~\bibnamefont {Carney}}, \bibinfo {author} {\bibfnamefont {G.}~\bibnamefont {Krnjaic}}, \bibinfo {author} {\bibfnamefont {D.~C.}\ \bibnamefont {Moore}}, \bibinfo {author} {\bibfnamefont {C.~A.}\ \bibnamefont {Regal}}, \bibinfo {author} {\bibfnamefont {G.}~\bibnamefont {Afek}}, \bibinfo {author} {\bibfnamefont {S.}~\bibnamefont {Bhave}}, \bibinfo {author} {\bibfnamefont {B.}~\bibnamefont {Brubaker}}, \bibinfo {author} {\bibfnamefont {T.}~\bibnamefont {Corbitt}}, \bibinfo {author} {\bibfnamefont {J.}~\bibnamefont {Cripe}}, \bibinfo {author} {\bibfnamefont {N.}~\bibnamefont {Crisosto}}, \bibinfo {author} {\bibfnamefont {A.}~\bibnamefont {Geraci}}, \bibinfo {author} {\bibfnamefont {S.}~\bibnamefont {Ghosh}}, \bibinfo {author} {\bibfnamefont {J.~G.~E.}\ \bibnamefont {Harris}}, \bibinfo {author} {\bibfnamefont {A.}~\bibnamefont {Hook}}, \bibinfo {author} {\bibfnamefont {E.~W.}\ \bibnamefont {Kolb}}, \bibinfo {author} {\bibfnamefont {J.}~\bibnamefont
  {Kunjummen}}, \bibinfo {author} {\bibfnamefont {R.~F.}\ \bibnamefont {Lang}}, \bibinfo {author} {\bibfnamefont {T.}~\bibnamefont {Li}}, \bibinfo {author} {\bibfnamefont {T.}~\bibnamefont {Lin}}, \bibinfo {author} {\bibfnamefont {Z.}~\bibnamefont {Liu}}, \bibinfo {author} {\bibfnamefont {J.}~\bibnamefont {Lykken}}, \bibinfo {author} {\bibfnamefont {L.}~\bibnamefont {Magrini}}, \bibinfo {author} {\bibfnamefont {J.}~\bibnamefont {Manley}}, \bibinfo {author} {\bibfnamefont {N.}~\bibnamefont {Matsumoto}}, \bibinfo {author} {\bibfnamefont {A.}~\bibnamefont {Monte}}, \bibinfo {author} {\bibfnamefont {F.}~\bibnamefont {Monteiro}}, \bibinfo {author} {\bibfnamefont {T.}~\bibnamefont {Purdy}}, \bibinfo {author} {\bibfnamefont {C.~J.}\ \bibnamefont {Riedel}}, \bibinfo {author} {\bibfnamefont {R.}~\bibnamefont {Singh}}, \bibinfo {author} {\bibfnamefont {S.}~\bibnamefont {Singh}}, \bibinfo {author} {\bibfnamefont {K.}~\bibnamefont {Sinha}}, \bibinfo {author} {\bibfnamefont {J.~M.}\ \bibnamefont {Taylor}}, \bibinfo
  {author} {\bibfnamefont {J.}~\bibnamefont {Qin}}, \bibinfo {author} {\bibfnamefont {D.~J.}\ \bibnamefont {Wilson}}, \ and\ \bibinfo {author} {\bibfnamefont {Y.}~\bibnamefont {Zhao}},\ }\href {\doibase 10.1088/2058-9565/abcfcd} {\bibfield  {journal} {\bibinfo  {journal} {Quantum Sci. Technol.}\ }\textbf {\bibinfo {volume} {6}},\ \bibinfo {pages} {024002} (\bibinfo {year} {2021})}\BibitemShut {NoStop}%
\bibitem [{\citenamefont {Afek}\ \emph {et~al.}(2022)\citenamefont {Afek}, \citenamefont {Carney},\ and\ \citenamefont {Moore}}]{Afek22_DM}%
  \BibitemOpen
  \bibfield  {author} {\bibinfo {author} {\bibfnamefont {G.}~\bibnamefont {Afek}}, \bibinfo {author} {\bibfnamefont {D.}~\bibnamefont {Carney}}, \ and\ \bibinfo {author} {\bibfnamefont {D.~C.}\ \bibnamefont {Moore}},\ }\href {\doibase 10.1103/PhysRevLett.128.101301} {\bibfield  {journal} {\bibinfo  {journal} {Phys. Rev. Lett.}\ }\textbf {\bibinfo {volume} {128}},\ \bibinfo {pages} {101301} (\bibinfo {year} {2022})}\BibitemShut {NoStop}%
\bibitem [{\citenamefont {Moore}\ and\ \citenamefont {Geraci}(2021)}]{Moore20_BSM}%
  \BibitemOpen
  \bibfield  {author} {\bibinfo {author} {\bibfnamefont {D.~C.}\ \bibnamefont {Moore}}\ and\ \bibinfo {author} {\bibfnamefont {A.~A.}\ \bibnamefont {Geraci}},\ }\href {\doibase 10.1088/2058-9565/abcf8a} {\bibfield  {journal} {\bibinfo  {journal} {Quantum Sci. Technol.}\ }\textbf {\bibinfo {volume} {6}},\ \bibinfo {pages} {014008} (\bibinfo {year} {2021})}\BibitemShut {NoStop}%
\bibitem [{\citenamefont {Riedel}(2013)}]{Riedel13_DM}%
  \BibitemOpen
  \bibfield  {author} {\bibinfo {author} {\bibfnamefont {C.~J.}\ \bibnamefont {Riedel}},\ }\href {\doibase 10.1103/PhysRevD.88.116005} {\bibfield  {journal} {\bibinfo  {journal} {Phys. Rev. D}\ }\textbf {\bibinfo {volume} {88}},\ \bibinfo {pages} {116005} (\bibinfo {year} {2013})}\BibitemShut {NoStop}%
\bibitem [{\citenamefont {Carmona~Rufo}\ \emph {et~al.}(2025)\citenamefont {Carmona~Rufo}, \citenamefont {Kumar}, \citenamefont {Sab\'{\i}n},\ and\ \citenamefont {Mazumdar}}]{Rufo:2025rps}%
  \BibitemOpen
  \bibfield  {author} {\bibinfo {author} {\bibfnamefont {P.~G.}\ \bibnamefont {Carmona~Rufo}}, \bibinfo {author} {\bibfnamefont {A.}~\bibnamefont {Kumar}}, \bibinfo {author} {\bibfnamefont {C.}~\bibnamefont {Sab\'{\i}n}}, \ and\ \bibinfo {author} {\bibfnamefont {A.}~\bibnamefont {Mazumdar}},\ }\href {\doibase 10.1103/14tm-ddnv} {\bibfield  {journal} {\bibinfo  {journal} {Phys. Rev. D}\ }\textbf {\bibinfo {volume} {111}},\ \bibinfo {pages} {115005} (\bibinfo {year} {2025})}\BibitemShut {NoStop}%
\bibitem [{\citenamefont {Liu}\ and\ \citenamefont {Zhu}(2018)}]{Liu2018}%
  \BibitemOpen
  \bibfield  {author} {\bibinfo {author} {\bibfnamefont {J.}~\bibnamefont {Liu}}\ and\ \bibinfo {author} {\bibfnamefont {K.-D.}\ \bibnamefont {Zhu}},\ }\href {\doibase 10.1088/1361-6382/aae062} {\bibfield  {journal} {\bibinfo  {journal} {Classical and Quantum Gravity}\ }\textbf {\bibinfo {volume} {35}},\ \bibinfo {pages} {205010} (\bibinfo {year} {2018})}\BibitemShut {NoStop}%
\bibitem [{\citenamefont {Monteiro}\ \emph {et~al.}(2020{\natexlab{a}})\citenamefont {Monteiro}, \citenamefont {Afek}, \citenamefont {Carney}, \citenamefont {Krnjaic}, \citenamefont {Wang},\ and\ \citenamefont {Moore}}]{Monteiro20_DM}%
  \BibitemOpen
  \bibfield  {author} {\bibinfo {author} {\bibfnamefont {F.}~\bibnamefont {Monteiro}}, \bibinfo {author} {\bibfnamefont {G.}~\bibnamefont {Afek}}, \bibinfo {author} {\bibfnamefont {D.}~\bibnamefont {Carney}}, \bibinfo {author} {\bibfnamefont {G.}~\bibnamefont {Krnjaic}}, \bibinfo {author} {\bibfnamefont {J.}~\bibnamefont {Wang}}, \ and\ \bibinfo {author} {\bibfnamefont {D.~C.}\ \bibnamefont {Moore}},\ }\href {\doibase 10.1103/PhysRevLett.125.181102} {\bibfield  {journal} {\bibinfo  {journal} {Phys. Rev. Lett.}\ }\textbf {\bibinfo {volume} {125}},\ \bibinfo {pages} {181102} (\bibinfo {year} {2020}{\natexlab{a}})}\BibitemShut {NoStop}%
\bibitem [{\citenamefont {Kilian}\ \emph {et~al.}(2023)\citenamefont {Kilian}, \citenamefont {Toro{\v{s}}}, \citenamefont {Deppisch}, \citenamefont {Saakyan},\ and\ \citenamefont {Bose}}]{kilian2023requirements}%
  \BibitemOpen
  \bibfield  {author} {\bibinfo {author} {\bibfnamefont {E.}~\bibnamefont {Kilian}}, \bibinfo {author} {\bibfnamefont {M.}~\bibnamefont {Toro{\v{s}}}}, \bibinfo {author} {\bibfnamefont {F.~F.}\ \bibnamefont {Deppisch}}, \bibinfo {author} {\bibfnamefont {R.}~\bibnamefont {Saakyan}}, \ and\ \bibinfo {author} {\bibfnamefont {S.}~\bibnamefont {Bose}},\ }\href {https://doi.org/10.1103/PhysRevResearch.5.023012} {\bibfield  {journal} {\bibinfo  {journal} {Physical Review Research}\ }\textbf {\bibinfo {volume} {5}},\ \bibinfo {pages} {023012} (\bibinfo {year} {2023})}\BibitemShut {NoStop}%
\bibitem [{\citenamefont {Kilian}\ \emph {et~al.}(2024{\natexlab{a}})\citenamefont {Kilian}, \citenamefont {Toro{\v{s}}}, \citenamefont {Barker},\ and\ \citenamefont {Bose}}]{kilian2024optimal}%
  \BibitemOpen
  \bibfield  {author} {\bibinfo {author} {\bibfnamefont {E.}~\bibnamefont {Kilian}}, \bibinfo {author} {\bibfnamefont {M.}~\bibnamefont {Toro{\v{s}}}}, \bibinfo {author} {\bibfnamefont {P.}~\bibnamefont {Barker}}, \ and\ \bibinfo {author} {\bibfnamefont {S.}~\bibnamefont {Bose}},\ }\href {https://doi.org/10.1103/PhysRevResearch.6.023037} {\bibfield  {journal} {\bibinfo  {journal} {Physical Review Research}\ }\textbf {\bibinfo {volume} {6}},\ \bibinfo {pages} {023037} (\bibinfo {year} {2024}{\natexlab{a}})}\BibitemShut {NoStop}%
\bibitem [{\citenamefont {Kilian}\ \emph {et~al.}(2024{\natexlab{b}})\citenamefont {Kilian}, \citenamefont {Rademacher}, \citenamefont {Gosling}, \citenamefont {Iacoponi}, \citenamefont {Alder}, \citenamefont {Toro{\v{s}}}, \citenamefont {Pontin}, \citenamefont {Ghag}, \citenamefont {Bose}, \citenamefont {Monteiro},\ and\ \citenamefont {Barker}}]{kilian2024dark}%
  \BibitemOpen
  \bibfield  {author} {\bibinfo {author} {\bibfnamefont {E.}~\bibnamefont {Kilian}}, \bibinfo {author} {\bibfnamefont {M.}~\bibnamefont {Rademacher}}, \bibinfo {author} {\bibfnamefont {J.~M.}\ \bibnamefont {Gosling}}, \bibinfo {author} {\bibfnamefont {J.~H.}\ \bibnamefont {Iacoponi}}, \bibinfo {author} {\bibfnamefont {F.}~\bibnamefont {Alder}}, \bibinfo {author} {\bibfnamefont {M.}~\bibnamefont {Toro{\v{s}}}}, \bibinfo {author} {\bibfnamefont {A.}~\bibnamefont {Pontin}}, \bibinfo {author} {\bibfnamefont {C.}~\bibnamefont {Ghag}}, \bibinfo {author} {\bibfnamefont {S.}~\bibnamefont {Bose}}, \bibinfo {author} {\bibfnamefont {T.~S.}\ \bibnamefont {Monteiro}}, \ and\ \bibinfo {author} {\bibfnamefont {P.~F.}\ \bibnamefont {Barker}},\ }\href {https://doi.org/10.1116/5.0200916} {\bibfield  {journal} {\bibinfo  {journal} {AVS Quantum Science}\ }\textbf {\bibinfo {volume} {6}} (\bibinfo {year} {2024}{\natexlab{b}})}\BibitemShut {NoStop}%
\bibitem [{\citenamefont {Geraci}\ and\ \citenamefont {Goldman}(2015)}]{andyhart2015}%
  \BibitemOpen
  \bibfield  {author} {\bibinfo {author} {\bibfnamefont {A.}~\bibnamefont {Geraci}}\ and\ \bibinfo {author} {\bibfnamefont {H.}~\bibnamefont {Goldman}},\ }\href {\doibase 10.1103/PhysRevD.92.062002} {\bibfield  {journal} {\bibinfo  {journal} {Phys. Rev. D}\ }\textbf {\bibinfo {volume} {92}},\ \bibinfo {pages} {062002} (\bibinfo {year} {2015})}\BibitemShut {NoStop}%
\bibitem [{\citenamefont {Marshman}\ \emph {et~al.}(2020{\natexlab{b}})\citenamefont {Marshman}, \citenamefont {Mazumdar}, \citenamefont {Morley}, \citenamefont {Barker}, \citenamefont {Hoekstra},\ and\ \citenamefont {Bose}}]{Marshman20_GM}%
  \BibitemOpen
  \bibfield  {author} {\bibinfo {author} {\bibfnamefont {R.~J.}\ \bibnamefont {Marshman}}, \bibinfo {author} {\bibfnamefont {A.}~\bibnamefont {Mazumdar}}, \bibinfo {author} {\bibfnamefont {G.~W.}\ \bibnamefont {Morley}}, \bibinfo {author} {\bibfnamefont {P.~F.}\ \bibnamefont {Barker}}, \bibinfo {author} {\bibfnamefont {S.}~\bibnamefont {Hoekstra}}, \ and\ \bibinfo {author} {\bibfnamefont {S.}~\bibnamefont {Bose}},\ }\href {\doibase 10.1088/1367-2630/ab9f6c} {\bibfield  {journal} {\bibinfo  {journal} {New J. Phys.}\ }\textbf {\bibinfo {volume} {22}},\ \bibinfo {pages} {083012} (\bibinfo {year} {2020}{\natexlab{b}})}\BibitemShut {NoStop}%
\bibitem [{\citenamefont {Wu}\ \emph {et~al.}(2025)\citenamefont {Wu}, \citenamefont {Toro\v{s}}, \citenamefont {Bose},\ and\ \citenamefont {Mazumdar}}]{Wu:2024bzd}%
  \BibitemOpen
  \bibfield  {author} {\bibinfo {author} {\bibfnamefont {M.-Z.}\ \bibnamefont {Wu}}, \bibinfo {author} {\bibfnamefont {M.}~\bibnamefont {Toro\v{s}}}, \bibinfo {author} {\bibfnamefont {S.}~\bibnamefont {Bose}}, \ and\ \bibinfo {author} {\bibfnamefont {A.}~\bibnamefont {Mazumdar}},\ }\href {\doibase 10.1103/PhysRevD.111.064004} {\bibfield  {journal} {\bibinfo  {journal} {Phys. Rev. D}\ }\textbf {\bibinfo {volume} {111}},\ \bibinfo {pages} {064004} (\bibinfo {year} {2025})}\BibitemShut {NoStop}%
\bibitem [{\citenamefont {Rademacher}\ \emph {et~al.}(2019)\citenamefont {Rademacher}, \citenamefont {Millen},\ and\ \citenamefont {Li}}]{Rademacher2019}%
  \BibitemOpen
  \bibfield  {author} {\bibinfo {author} {\bibfnamefont {M.}~\bibnamefont {Rademacher}}, \bibinfo {author} {\bibfnamefont {J.}~\bibnamefont {Millen}}, \ and\ \bibinfo {author} {\bibfnamefont {Y.~L.}\ \bibnamefont {Li}},\ }\href {\doibase 10.1515/aot-2020-0019} {\bibfield  {journal} {\bibinfo  {journal} {Advanced Optical Technologies}\ }\textbf {\bibinfo {volume} {9}},\ \bibinfo {pages} {227–239} (\bibinfo {year} {2019})}\BibitemShut {NoStop}%
\bibitem [{\citenamefont {Monteiro}\ \emph {et~al.}(2020{\natexlab{b}})\citenamefont {Monteiro}, \citenamefont {Li}, \citenamefont {Afek}, \citenamefont {Li}, \citenamefont {Mossman},\ and\ \citenamefont {Moore}}]{Monteiro20_force}%
  \BibitemOpen
  \bibfield  {author} {\bibinfo {author} {\bibfnamefont {F.}~\bibnamefont {Monteiro}}, \bibinfo {author} {\bibfnamefont {W.}~\bibnamefont {Li}}, \bibinfo {author} {\bibfnamefont {G.}~\bibnamefont {Afek}}, \bibinfo {author} {\bibfnamefont {C.-l.}\ \bibnamefont {Li}}, \bibinfo {author} {\bibfnamefont {M.}~\bibnamefont {Mossman}}, \ and\ \bibinfo {author} {\bibfnamefont {D.~C.}\ \bibnamefont {Moore}},\ }\href {\doibase 10.1103/PhysRevA.101.053835} {\bibfield  {journal} {\bibinfo  {journal} {Phys. Rev. A}\ }\textbf {\bibinfo {volume} {101}},\ \bibinfo {pages} {053835} (\bibinfo {year} {2020}{\natexlab{b}})}\BibitemShut {NoStop}%
\bibitem [{\citenamefont {Winstone}\ \emph {et~al.}(2022)\citenamefont {Winstone} \emph {et~al.}}]{winstone_2022}%
  \BibitemOpen
  \bibfield  {author} {\bibinfo {author} {\bibfnamefont {G.}~\bibnamefont {Winstone}} \emph {et~al.} (\bibinfo {collaboration} {LSD Collaboration}),\ }\href {\doibase 10.1103/PhysRevLett.129.053604} {\bibfield  {journal} {\bibinfo  {journal} {Phys. Rev. Lett.}\ }\textbf {\bibinfo {volume} {129}},\ \bibinfo {pages} {053604} (\bibinfo {year} {2022})}\BibitemShut {NoStop}%
\end{thebibliography}%

\end{document}